\documentclass[10pt,journal,twocolumn]{IEEEtran}
\newif\ifCLASSOPTIONromanappendices \CLASSOPTIONromanappendicestrue
 \usepackage{bm}
\usepackage{stfloats}
\usepackage{relsize}
\usepackage{geometry}
\usepackage[utf8x]{inputenc}
\usepackage[T1]{fontenc}
\usepackage{url}
\usepackage{xfrac}
\usepackage{multicol}
\usepackage{xcolor}
\usepackage{ifthen}
\usepackage{amsfonts}
\usepackage{amssymb}
\usepackage{amsmath,hyperref}
\usepackage{sansmath}
\usepackage{siunitx}
\usepackage{mathtools}
\usepackage{nccmath}
\usepackage{cite}
\usepackage{systeme}
\usepackage{spalign}
\usepackage{caption}
\usepackage{algpseudocode}
\usepackage{algorithm}
\algnewcommand\algorithmicforeach{\textbf{for each}}
\algdef{S}[FOR]{ForEach}[1]{\algorithmicforeach\ #1\ \algorithmicdo}

\usepackage{eqparbox}
\newdimen{\algindent}
\algnewcommand\LeftComment[2]{%
\hspace{#1\algindent}$\triangleright$ \eqparbox{COMMENT}{#2} \hfill %
}
\algnewcommand\LeftCommentNoTriangle[2]{%
\hspace{#1\algindent} \eqparbox{COMMENT}{#2} \hfill %
}
\algnewcommand\LeftCommentNoIntent[1]{%
$\triangleright$ \eqparbox{COMMENT}{#1} \hfill %
}
\usepackage{verbatim}
\usepackage{tikz}
\usetikzlibrary{positioning,quotes,calc,fit,shapes,shadows,arrows,decorations.pathreplacing}
\tikzset{block/.style={draw,very thick,text width=2cm,minimum height=4cm,align=center},
         line/.style={-latex}}
\tikzset{blockV/.style={draw,very thick,text width=2cm,minimum height=2cm, minimum width=4cm,align=center},
         line/.style={-latex}}
\tikzset{blockExt/.style={draw,very thick,minimum height=1cm, minimum width=1cm,align=center},
         line/.style={-latex}}
\usepgflibrary{arrows}
\usetikzlibrary{bayesnet}
\newcommand*{\Scale}[2][4]{\scalebox{#1}{$#2$}}
\definecolor{light-gray}{HTML}{E0E0E0}

\newcommand*{\rv}{\fontfamily{cmss}\selectfont}

\makeatletter
\newcommand\notsotiny{\@setfontsize\notsotiny{6.82}{7.5}}
\makeatother

\newcommand\figscale{0.569}

\newcommand\Tstrut{\rule{0pt}{2.6ex}}       
\newcommand\Bstrut{\rule[-2.6ex]{0ex}{0ex}} 

\newcommand\norm[1]{\left\lVert#1\right\rVert}

\makeatletter
\newcommand{\labeltarget}[1]{\Hy@raisedlink{\hypertarget{#1}{}}}
\makeatother

\newcommand*{\myDots}{\ifmmode\mathellipsis\else.\kern1.40em.\kern 1.40em.\fi}
\newcommand*{\yourtightDots}{\ifmmode\mathellipsis\else.\kern0.1em.\kern 0.18em.\kern 0.1em.\fi}
\newcommand*{\mytightDots}{\ifmmode\mathellipsis\else.\kern0.05em.\kern 0.05em.\kern 0.05em\fi}

\usepackage{caption}
\usepackage{subfig}

\usepackage{balance}

\begin{document}
\title{Bilinear Generalized \\Vector Approximate Message Passing}

\author{Mohamed Akrout, Anis Housseini, Faouzi Bellili, \IEEEmembership{Member, IEEE}, and Amine Mezghani, \IEEEmembership{Member, IEEE}
 \vspace{0.3cm}
\\\small E2-390 E.I.T.C,  75 Chancellor's Circle  Winnipeg, MB, Canada, R3T 5V6.
  \vspace{0.1cm}
  \\\small Emails:  \{akroutm, {housseia}\}@myumanitoba.ca, \{faouzi.bellili, amine.mezghani\}@umanitoba.ca.
  \vspace{0.3cm}
\thanks{The authors are with the Department of Electrical and Computer Engineering at the University of Manitoba, Winnipeg, MB, Canada.  This work was supported by the Discovery Grants Program of the Natural Sciences and Engineering Research Council of Canada (NSERC). }}

\maketitle
\begin{abstract}
We introduce the bilinear generalized vector approximate message passing (BiG-VAMP) algorithm which jointly recovers two matrices $\boldsymbol{U}$ and $\boldsymbol{V}$ from their noisy product through a probabilistic  observation model. BiG-VAMP provides computationally efficient approximate implementations of both max-sum and sum-product loopy belief propagation (BP). We show how the proposed BiG-VAMP algorithm recovers different types of structured matrices and overcomes the fundamental limitations of other state-of-the-art approaches to the bilinear recovery problem, such as BiG-AMP, BAd-VAMP and LowRAMP. In essence, BiG-VAMP applies to a broader class of practical applications which involve  a general form of structured matrices.
For the sake of theoretical performance prediction, we also conduct a state evolution (SE) analysis of the proposed algorithm  and show its consistency with the asymptotic empirical  mean-squared error (MSE). Numerical results on various applications such as matrix factorization, dictionary learning, and matrix completion demonstrate unambiguously the effectiveness of the proposed BiG-VAMP algorithm and its superiority over state-of-the-art algorithms. Using the developed SE framework, we also examine (as one example) the phase transition diagrams of the matrix completion problem, thereby unveiling a low detectability region corresponding to the low signal-to-noise ratio (SNR) regime.
\end{abstract}



\section{Introduction}
\subsection{Background and related work}
\IEEEPARstart{W}{e} consider an observation matrix, $\boldsymbol{Y} \in$ $\mathbb{R}^{N \times M}$, obtained from the following generalized bilinear model:
\begin{equation}\label{eq:non-linear-distribution0}
p_{\textrm{\textbf{{\rv{Y}}}} | \textrm{\textbf{{\rv{Z}}}}}(\boldsymbol{Y} | \boldsymbol{Z})~=~\prod_{i=1}^{N} \,\prod_{j=1}^{M} p_{\textrm{{\rv{y}}}_{ij} | \textrm{{\rv{z}}}_{ij}}(y_{ij} | z_{ij}) ~~~\textrm{with}~~~\bm{Z}\,=\,\boldsymbol{U}\boldsymbol{V}^{\mathsf{T}},
\end{equation}
where $\boldsymbol{U}$ and $\boldsymbol{V}$ are two unknown matrices in $\mathbb{R}^{N \times r}$ and $\mathbb{R}^{M \times r}$, respectively. The goal is to recover  $\boldsymbol{U}$ and $\boldsymbol{V}$ based on the knowledge of $\boldsymbol{Y}$ and the model in (\ref{eq:non-linear-distribution0}).
The latter
applies to a myriad of problems ranging from noisy dictionary learning\cite{gribonval2015sparse}, matrix completion \cite{montanari2012graphical}, and sparse PCA \cite{zou2006sparse}, to matrix factorization \cite{koren2009matrix}, low-rank matrix reconstruction \cite{matsushita2013low}, and subgraph estimation \cite{wang2019symmetric}, just to name a few. A special relevant case of the generalized bilinear model in (\ref{eq:non-linear-distribution0}) is:
\begin{equation}
\label{eq:bilinear-model} 
\boldsymbol{Y}~ =~ \phi\left(\boldsymbol{U}\,\boldsymbol{V}^{\top} +\, \boldsymbol{W}\right), 
\end{equation}
in which $\boldsymbol{W}$ $\in$ $\mathbb{R}^{N \times M}$ is an  additive white Gaussian noise matrix whose entries are assumed to be mutually independent with mean zero and variance $\gamma_w^{-1}$, i.e., $w_{i,j}~\,\sim\,\mathcal{N}(w_{ij}; 0, \gamma_w^{-1})$. 
While convex relaxation of the bilinear recovery problem under the observation model in (\ref{eq:bilinear-model}) is possible in some cases using the augmented Lagrange multiplier method (ALMM)\cite{lin2010augmented}, different non-convex formulations have been investigated in some special cases over the last decade based on:
\begin{enumerate}
    \item the alternating direction method of multipliers (ADMM)\cite{boyd2011distributed},
    \item the variational sparse Bayesian learning (VSBL) method \cite{babacan2012sparse},
    \item the approximate message passing (AMP) paradigm which is  discussed hereafter in some depth in order to put our contribution in a proper perspective. 
\end{enumerate}
\par \noindent In fact, AMP-based computational information/data processing  have attracted a lot of interest in different fields since the early introduction of the AMP algorithm in \cite{donoho2009message} within the compressed sensing (CS) framework. More specifically, in a typical CS problem,  one is interested  in recovering an unknown sparse vector, $\boldsymbol{x}\!\in\! \mathbb{R}^{N}$, from its noisy linear measurements/observations:
\begin{equation}\label{linear_model}
\boldsymbol{y}~=~\boldsymbol{A}\bm{x}~+~\bm{w},
\end{equation} 
wherein   $\bm{A}\in\mathbb{R}^{M\times N}$ (with $M\ll N$) is a known sensing matrix. AMP strikes a proper balance between reconstruction performance and  computational complexity as compared to traditional convex optimization-based and iterative soft thresholding algorithms \cite{eldar2012compressed}. The AMP algorithm was later extended in \cite{rangan2011generalized} to generalized linear models of the form:
\begin{eqnarray}\label{eq:non-linear-distribution}
\!\!\!\!p_{\textrm{\textbf{{\rv{y}}}} | \textrm{\textbf{{\rv{z}}}}}(\boldsymbol{y} | \boldsymbol{z})&\,=\,&\prod_{m=1}^{M}  p_{\textrm{{\rv{y}}}_{m} | \textrm{{\rv{z}}}_{m}}(y_{m} | z_{m})~~~\textrm{with}~~~\bm{z}\,=\,\bm{A}\bm{x}.
\end{eqnarray}
Aside from  handling nonlinear transformations, the advantage of the generalized AMP (GAMP) algorithm over its AMP predecessor lies in its ability to  accommodate statistical priors on the  sparse vector $\bm{x}$. From another perspective, the performance of both  AMP and GAMP  can be rigorously tracked by a set of scalar update equations, known as state evolution (SE) \cite{bayati2011dynamics, barbier2019optimal} in statistics or the cavity method in statistical physics \cite{mezard2009information}.
 One limitation of AMP and GAMP, however, is that they often diverge if the sensing matrix,  $\bm{A}$, is ill-conditioned and/or has a  non-zero mean. It is precisely in this context that the  vector AMP (VAMP) algorithm  has been recently introduced and rigorously  analyzed in \cite{rangan2019vector, schniter2016vector,    rangan2019vector_IT}\footnote{It is worth mentioning here that VAMP was independently derived by two research groups in \cite{rangan2019vector} and \cite{ma2017orthogonal} but under two different names, i.e., VAMP and orthogonal AMP (OAMP), respectively.}. 
 Although, there is no theoretical guarantees that VAMP will always converge, there is strong empirical evidence that it is more resilient to mean-perturbed or badly conditioned sensing matrices $\bm{A}$, provided that the latter is  right-orthogonally invariant \cite{rangan2019vector_IT}.
%

 To date, extensions of the original AMP algorithm to the bilinear recovery problem
 were already made in \cite{matsushita2013low} and \cite{lesieur2015phase}  within the context of low-rank matrix reconstruction.
 In this context, the so-called LowRAMP algorithm introduced in \cite{lesieur2015phase} extends the method in \cite{matsushita2013low} to non-Gaussian likelihoods and establishes the associated SE analysis. It does so by simplifying the  BP messages in \cite{matsushita2013low} via second-order Taylor series approximations which become more accurate in the limit of large $N$. However, the involved approximations hold in the per-measurement low-SNR regime only \big[i.e., SNR  $= \mathcal{O}(\frac{1}{N})\big]$ and, hence, LowRAMP is not suitable for other types of measurement processes such as quantization and matrix sub-sampling. Moreover, the  ``low-rank'' assumption which is critical in both \cite{matsushita2013low} and \cite{lesieur2015phase} precludes a large number of practical applications which involve  a \textit{general-rank} decomposition of structured matrices rather than a \textit{low-rank} decomposition of unstructured matrices.
 
 GAMP itself was also extended  to the bilinear case in \cite{parker2014bilinear} thereby leading to the so-called  bilinear generalized AMP\footnote{The reader is also referred to the parametric version of BiG-AMP in \cite{ parker2016parametric}.} (BiG-AMP) algorithm. Although being completely oblivious to the \textit{low-rank} assumption, BiG-AMP inherits all the GAMP-related convergence issues and was developed for separable priors on both $\boldsymbol{U}$  and $\boldsymbol{V}$ matrices.     
 %
 %
%
%
To accommodate a larger class of $\bm{V}$  matrices in (\ref{eq:non-linear-distribution0}), Sarkar \textit{et al.}  made an attempt in \cite{sarkar2019bilinear} to generalize the VAMP framework to  bilinear recovery problems  and the algorithm introduced therein was called \textit{Bilinear Adaptive VAMP} (BAd-VAMP). In essence, BAd-VAMP is a ping-pong-like approach which reconstructs both $\boldsymbol{U}$ and $\boldsymbol{V}$ matrices 
by alternating between $i)$ the expectation-maximization (EM) algorithm \cite{dempster1977maximum}  to find the maximimum-likelihood (ML) estimate of $\boldsymbol{U}$ and $ii)$ the VAMP algorithm \cite{rangan2019vector_IT} to find the minimum mean-squared error (MMSE) estimate of $\boldsymbol{V}$. Unfortunately, due to the use of the EM algorithm, BAd-VAMP does not accommodate general priors on the matrix $\boldsymbol{U}$. For instance, trying to enforce a binary or sparsity prior on $\boldsymbol{U}$ renders the E-step of the EM algorithm computationally prohibitive.  To overcome all the aforementioned  limitations, this paper introduces a new algorithm along with its state evolution analysis to solve a broader class of the bilinear recovery problems as modelled in (\ref{eq:non-linear-distribution0}).

\subsection{Contributions}
This paper builds upon the prior work in \cite{matsushita2013low} and provides a broader solution to the bilinear recovery problem under different structured matrices beyond the ``low-rank'' assumption. Our approach for bilinear recovery does not alternate between the EM and VAMP algorithms, but rather relies entirely on message passing and it is dubbed \textit{bilinear generalized VAMP} (BiG-VAMP).
The BiG-VAMP approach that we propose in this work enables the use of arbitrary priors on both $\boldsymbol{U}$ and $\boldsymbol{V}$, thereby allowing the exploitation of other matrix structures such as finite-alphabet, binarity, sparsity, constant-modulus, assignment, etc. The proposed BiG-VAMP algorithm is suitable to a broader class of bilinear recovery problems that $i)$ cover more general prior distributions on the unknown matrices (unlike BiG-AMP) and $ii)$ does not rely on the use of the EM algorithm with automated hyperparameter tuning to estimate $\boldsymbol{U}$ (unlike BAd-VAMP).  
The key differences between BiG-VAMP and LowRAMP are, however, as follows: \begin{enumerate}
    \item BiG-VAMP provides a systematic way for handling nonlinear outputs without the ``low-SNR'' assumption which does not hold,  e.g., in the matrix completion problem. Moreover, BiG-VAMP allows maximum \textit{a posteriori} bilinear reconstruction under \textit{non-differentiable} output functions such as quantization, perceptron activation, selection, and phase-retrieval, etc.    
    \item BiG-VAMP applies to a broader class  of practical applications which involve  a \textit{general-rank} decomposition of structured matrices in addition to a \textit{low-rank} decomposition of unstructured matrices.
\end{enumerate}
 Much like BiG-AMP, BiG-VAMP is also  applicable to maximum \textit{a posteriori} (MAP) and MMSE inference problems alike, as will be explained later on. 
Furthermore, it comes with theoretical performance guarantees, established in Section \ref{sec:state-evolution}, that validate its superiority against state-of-the-art BiG-AMP, BAd-VAMP, and LowRAMP algorithms. 
\subsection*{Notation}
We use Sans Serif font (e.g., {\rv{x}}) for random variables and Serif font (e.g., $x$) for its realizations. We use boldface lowercase letters for vectors (e.g., {\textbf{\rv{x}}} and $\boldsymbol{x}$) and boldface uppercase letters for  matrices (e.g., {\textbf{\rv{X}}} and $\boldsymbol{X}$). Vectors are in column-wise orientation by default. Given any matrix $\boldsymbol{X}$, we use $\boldsymbol{x}_i$  and $x_{ij}$ to denote its $i$th column and $ij$th entry, respectively. We also denote the $i$th component of a vector $\bm{x}$ as $[\bm{x}]_i$ or $x_i$.  The operator diag($\boldsymbol{X}$) stacks the diagonal elements of $\boldsymbol{X}$ in a vector while $\boldsymbol{I}$ stands for the identity matrix. The operator tr($\boldsymbol{X}$) returns the sum of the diagonal elements of $\boldsymbol{X}$. We also use $p_{\textrm{{\rv{x}}}} (x;\bm{\theta})$, $p_{\bm{\mathsf{x}}} (\bm{x};\bm{\theta})$, and $p_{\bm{\mathsf{X}}} (\bm{X};\bm{\theta})$ to denote the pdf of random variables/vectors/matrices; as being parameterized by a set of parameters $\bm{\theta}$. Moreover, 
$\mathcal{N}(\boldsymbol{x}; \widehat{\bm{x}}, \boldsymbol{R})$ stands for the multivariate Gaussian pdf of any random  vector $\bm{\mathsf{x}}$ with mean $\widehat{\bm{x}}$  and covariance matrix $\bm{R}$. We use  $\sim$ and  $\propto$ as short-hand motations for ``distributed according to'' and ``proportional to'', respectively.  We also use $\mathbb{E}[{\textbf{\rv{x}}}|d(\boldsymbol{x})]$ to denote the expectation of $\textrm{\textbf{{\rv{x}}}} \sim d(\boldsymbol{x})$ and $\delta(\boldsymbol{x})$ refers to the Dirac delta distribution. Moreover, $\langle \boldsymbol{x}\rangle$ and $\langle \boldsymbol{X}\rangle$ return the (empirical)  average values of vectors and matrices, i.e., $\langle \boldsymbol{x}\rangle \triangleq$ $\frac{1}{N} \sum_{i=1}^{N} x_{i}$ for $\boldsymbol{x} \in \mathbb{R}^{N}$ and $\langle \boldsymbol{X}\rangle \triangleq$ $\frac{1}{N M} \sum_{i=1}^{N} \sum_{j=1}^{M} x_{ij}$ for $\boldsymbol{X} \in \mathbb{R}^{N \times M}$. Finally, the symbol $\odot$ denotes the Hadamard (i.e., elementwise) product between any two matrices.
\section{Background on the low-rank matrix reconstruction}\label{sec:background-low-rank}
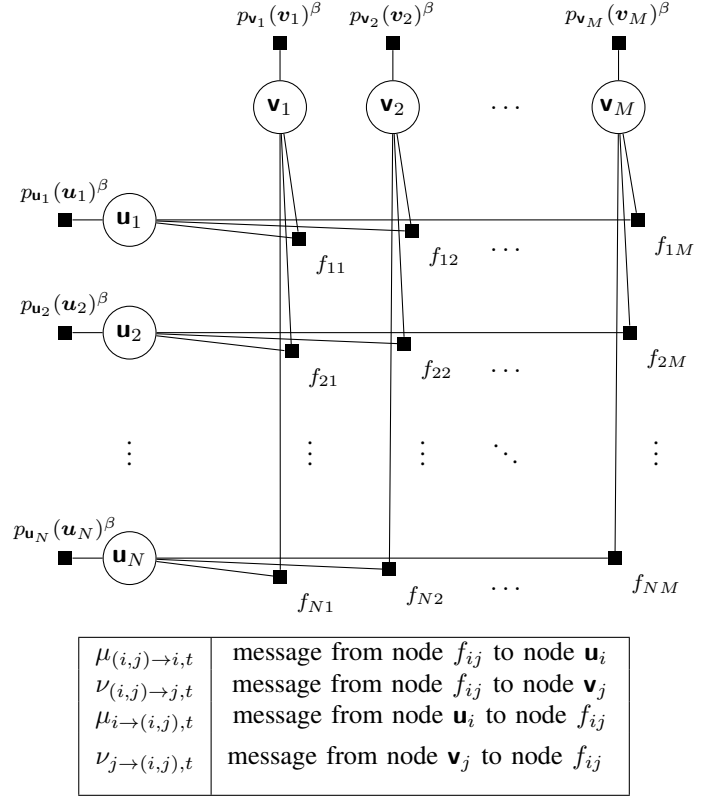
\begin{figure}[bth!]
\centering
\begin{tikzpicture}
  \node[latent, xshift=-2cm]  (v1)   {$\textrm{\textbf{{\rv{v}}}}_1$};
  \node[latent, xshift=-0.5cm]  (v2)   {$\textrm{\textbf{{\rv{v}}}}_2$};
  \node[latent, draw=none, xshift=1cm]  (vdots)   {$\ldots$};
  \node[latent, xshift=2.5cm]  (vN)   {$\textrm{\textbf{{\rv{v}}}}_M$};
  \node[latent, xshift=-4cm, yshift=-1.5cm]  (u1)   {$\textrm{\textbf{{\rv{u}}}}_1$};
  \node[latent, xshift=-4cm, yshift=-3cm]  (u2)   {$\textrm{\textbf{{\rv{u}}}}_2$};
  \node[latent, draw=none, xshift=-4cm, yshift=-4.5cm]  (udots)   {$\vdots$};
  \node[latent, xshift=-4cm, yshift=-6cm]  (uM)   {$\textrm{\textbf{{\rv{u}}}}_N$};

  \factor[right=of u1, xshift=1.4cm, yshift=-0.25cm] {f11} {below right: $f_{11}$} {u1, v1} {};
  \factor[right=of u1, xshift=2.9cm, yshift=-0.15cm] {f12} {below right:$f_{12}$} {u1, v2} {};
  \factor[right=of u1, xshift=5.9cm] {f1N} {below right:$f_{1M}$} {u1, vN} {};
  
  \factor[right=of u2, xshift=1.3cm, yshift=-0.25cm] {f21} {below right:$f_{21}$} {u2, v1} {};
  \factor[right=of u2, xshift=2.8cm, yshift=-0.15cm] {f22} {below right:$f_{22}$} {u2, v2} {};
  \factor[right=of u2, xshift=5.8cm] {f2N} {below right:$f_{2M}$} {u2, vN} {};
  
  \factor[right=of uM, xshift=1.15cm, yshift=-0.25cm] {fM1} {below right:$f_{N1}$} {uM, v1} {};
  \factor[right=of uM, xshift=2.6cm, yshift=-0.15cm] {fM2} {below right:$f_{N2}$} {uM, v2} {};
  \factor[right=of uM, xshift=5.6cm] {fMN} {below right:$f_{NM}$} {uM, vN} {};
  
  \node[latent, draw=none, xshift=-0.1cm, yshift=-4.5cm]  (fdots)   {$\vdots$};
  \node[latent, draw=none, xshift=-1.6cm, yshift=-4.5cm]  (f2dots)   {$\vdots$};
  \node[latent, draw=none, xshift=3cm, yshift=-4.5cm]  (fNdots)   {$\vdots$};
  \node[latent, draw=none, xshift=1cm, yshift=-4.5cm]  (fdots)   {$\ddots$};
  \node[latent, draw=none, xshift=1cm, yshift=-1.9cm]  (v1dots)   {$\ldots$};
  \node[latent, draw=none, xshift=1cm, yshift=-3.5cm]  (v1dots)   {$\ldots$};
  \node[latent, draw=none, xshift=1cm, yshift=-6.4cm]  (v1dots)   {$\ldots$};
  
  \factor[left=of u1] {pu1} {above:$p_{\textrm{\textbf{{\rv{u}}}}_1}(\bm{u}_1)^{\beta}$} {u1} {};
  \factor[left=of u2] {pu2} {above:$p_{\textrm{\textbf{{\rv{u}}}}_2}(\bm{u}_2)^{\beta}$} {u2} {};
  \factor[left=of uM] {puM} {above:$p_{\textrm{\textbf{{\rv{u}}}}_N}(\bm{u}_N)^{\beta}$} {uM} {};
  
  \factor[above=of v1] {pv1} {above:$p_{\textrm{\textbf{{\rv{v}}}}_1}(\bm{v}_1)^{\beta}$} {v1} {};
  \factor[above=of v2] {pv2} {above:$p_{\textrm{\textbf{{\rv{v}}}}_2}(\bm{v}_2)^{\beta}$} {v2} {};
  \factor[above=of vN] {pvN} {above:$p_{\textrm{\textbf{{\rv{v}}}}_M}(\bm{v}_M)^{\beta}$} {vN} {};
\end{tikzpicture}
\begin{center}
\begin{tabular}{ |c|c| } 
 \hline
 $\mu_{{(i,j)}\rightarrow i,t}$ & message from node $f_{ij}$ to node $\textrm{\textbf{{\rv{u}}}}_i$ \\ 
 $\nu_{{(i,j)}\rightarrow j,t}$ & message from node $f_{ij}$ to node $\textrm{\textbf{{\rv{v}}}}_j$ \\ 
 $\mu_{i\rightarrow (i,j),t}$ & message from node $\textrm{\textbf{{\rv{u}}}}_i$ to node $f_{ij}$\\
 $\nu_{j\rightarrow (i,j),t}$ & message from node $\textrm{\textbf{{\rv{v}}}}_j$ to node $f_{ij}$ \Tstrut \Bstrut\\
 \hline
\end{tabular}
\end{center}
\caption{Factor graph associated to (\ref{eq:p-uv-given-y}). The circles represent variable nodes and the squares represent factor nodes.}
\label{fig:factor-graph-Tanaka}
\end{figure}

\begin{figure}[h!]
\centering
\begin{tikzpicture}[thick,scale=1, every node/.style={scale=1.0}]
  \node[latent, xshift=-0.5cm, yshift=1cm]  (vj)   {$\textrm{\textbf{{\rv{v}}}}_j$};
  \node[latent, xshift=-3cm, yshift=-1.5cm]  (ui)   {$\textrm{\textbf{{\rv{u}}}}_i$};
  \node[latent, right=of ui, xshift=2cm, yshift=0cm, fill=light-gray]  (yij)   {$\bm{\mathsf{Y}}$};
  \node (v1) at ($(vj) + (-0.2,-1.5)$) {};
  \node (vM) at ($(vj) + (0.2,-1.5)$) {};
  \node (u11) at ($(ui) + (1.5,-0.2)$) {};
  \node (uN1) at ($(ui) + (1.5,0.2)$) {};
  %
  \factor[right=of u1, xshift=2.5cm, xscale=2.5, yscale=2.5] {fij}  {below: 
  ~~~~~~~~~~~~~~~~~~~~$\prod\limits_{i'=1}^{N}\prod\limits_{j'=1}^{M}\mathcal{N}\left(y_{i'j'};\boldsymbol{u}_{i'}^{\top} \boldsymbol{v}_{j'},\beta^{-1}\gamma_w^{-1}\right)$
  } {ui, vj, yij, v1, vM, u11, uN1}{};
  \node at ($(fij) + (-0.018,0.92)$) {\yourtightDots};
  \node at ($(fij) + (-0.94,-0.01)$) [rotate=90]{\yourtightDots};
  %
  \node (u_f_right) at ($(ui)!0.5!(fij) + (0,0.25)$) {};
  \node (u_f_left) at ($(ui)!0.5!(fij) + (-0.8,0.25)$) {};
  \draw [-latex,very thick] (u_f_right.center) -- (u_f_left.center);
  \draw[] (ui) to node[pos=0.3,above=0.85em,align=center]{
{\scriptsize \protect\tikz[inner sep=.25ex,baseline=.4ex] \protect\node[circle,draw] {1};}
  } (fij);
  %
  \node (v_f_right) at ($(vj)!0.5!(fij) + (-0.25,0)$) {}; 
  \node (v_f_left) at ($(vj)!0.5!(fij) + (-0.25,0.8)$) {};
  \draw [-latex,very thick] (v_f_right.center) -- (v_f_left.center);
  \draw[] (vj) to node[pos=0.3,right=0.1em,align=center,xshift=-0.9cm, yshift=0cm]{
{\scriptsize \protect\tikz[inner sep=.25ex,baseline=-.75ex] \protect\node[circle,draw] {2};}
  } (fij);
  \node (f_u_right) at ($(ui)!0.5!(fij) + (-0.8,-0.25)$) {};
  \node (f_u_left) at ($(ui)!0.5!(fij) + (0,-0.25)$) {};
  \draw [-latex,very thick] (f_u_right.center) -- (f_u_left.center);
  \draw[] (ui) to node[pos=0.3,below=0.85em,align=center]{
{\scriptsize  \protect\tikz[inner sep=.25ex,baseline=-1em] \protect\node[circle,draw] {3};
}
  } (fij);
  \node (f_v_right) at ($(vj)!0.5!(fij) + (0.25,0)$) {};
  \node (f_v_left) at ($(vj)!0.5!(fij) + (0.25,0.8)$) {};
  \draw [-latex,very thick] (f_v_left.center) -- (f_v_right.center);
  \draw[] (vj) to node[pos=0.3,right=0.85em,align=center]{
{\scriptsize  \protect\tikz[inner sep=.25ex,baseline=-0.4em] \protect\node[circle,draw] {4};
}
} (fij);
  \factor[left=of ui, xscale=2.5, yscale=2.5] {pui} {below:$p_{\textrm{\textbf{{\rv{u}}}}_{i}}(\bm{u}_i)^{\beta}$} {ui} {};
  \factor[above=of vj, xscale=2.5, yscale=2.5] {pvj} {right:$p_{\textrm{\textbf{{\rv{v}}}}_{j}}(\bm{v}_j)^{\beta}$} {vj} {};
  \node[latent, below=of vj, draw=none, xshift=0.01cm,yshift=-1.28cm,opacity=0, text opacity=1]  (eq)   {\textcolor{white}{\textbf{ }}};
\end{tikzpicture}
\begin{center}
\begin{tabular}{ |c|c| } 
 \hline
 \protect\tikz[inner sep=.25ex,baseline=-.75ex] \protect\node[circle,draw] {{\tiny 1}}; & $\mathcal{N}(\boldsymbol{u}_i;\mathbf{\Lambda}_{\textrm{\textbf{{\rv{u}}}}, i \rightarrow(i, j),t}^{-1}\bm{b}_{\textrm{\textbf{{\rv{u}}}}, i \rightarrow(i, j),t}, \beta^{-1}\bm{\Lambda}^{-1}_{\textrm{\textbf{{\rv{u}}}}, i \rightarrow(i, j),t})$ \Tstrut\\
 \protect\tikz[inner sep=.25ex,baseline=-.75ex] \protect\node[circle,draw] {{\tiny 2}}; & $\mathcal{N}(\boldsymbol{v}_j;\mathbf{\Lambda}_{\textrm{\textbf{{\rv{v}}}}, j \rightarrow(i, j),t}^{-1}\bm{b}_{\textrm{\textbf{{\rv{v}}}}, j \rightarrow(i, j),t}, \beta^{-1}\bm{\Lambda}^{-1}_{\textrm{\textbf{{\rv{v}}}}, j \rightarrow(i, j),t})$\Tstrut \Bstrut\\
 \hline
\end{tabular}
\end{center}
\caption{Explicit messages resulting from the CLT approximations shown here for a single cell of the entire factor graph depicted in Fig. \ref{fig:factor-graph-Tanaka}.}
\label{fig:factor-graph-y_uv}
\end{figure}
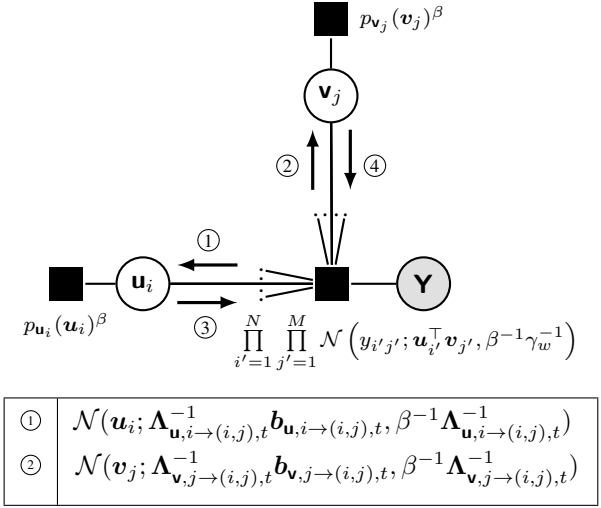

In this section, we briefly review the main results of the prior work on \textit{low-rank} matrix reconstruction in \cite{matsushita2013low} at the detail needed for a comprehensive exposition of BiG-VAMP. We emphasize, however, the fact that borrowing such results does not restrict the proposed BiG-VAMP algorithm to the bilinear
\textit{low-rank} matrix recovery as is the case in \cite{matsushita2013low}.\\
Consider the bilinear recovery  of two random independent matrices $\textrm{\textbf{{\rv{U}}}}=[\textrm{\textbf{{\rv{u}}}}_1, \dots,\textrm{\textbf{{\rv{u}}}}_N ]^{\top} \in \mathbb{R}^{N \times r}$ and $\textrm{\textbf{{\rv{V}}}}=[\textrm{\textbf{{\rv{v}}}}_1, \dots,\textrm{\textbf{{\rv{v}}}}_M ]^{\top} \in \mathbb{R}^{M \times r}$ from a linear noisy observation $\boldsymbol{Y}\,=\, \boldsymbol{U} \,\boldsymbol{V}^{\top} + \boldsymbol{W}$ $\in$ $\mathbb{R}^{N \times M}$. Given some common priors, $\mathit{p}_{\textrm{\textbf{{\rv{u}}}}}(\boldsymbol{u}_i)$ and $\mathit{p}_{\textrm{\textbf{{\rv{v}}}}}(\boldsymbol{v}_i)$, on the vectors $\boldsymbol{u}_i$ and $\boldsymbol{v}_j$, respectively, the goal of BP is to approximate their joint posterior distribution:
\begin{equation}
\begin{aligned}[b]
 &\hspace{-0.25cm}p_{\textrm{\textbf{{\rv{u}}},\textbf{{\rv{v}}}}|\bm{\mathsf{Y}}}(\boldsymbol{u}_i, \boldsymbol{v}_j | \bm{Y} ;\gamma_w^{-1}, \beta)\\
 &\hspace{0.5cm}\propto  \prod_{i=1}^{N}\prod_{j=1}^{M}p_{\textrm{{\rv{y}}}_{ij}| \textrm{\textbf{{\rv{u}}},\textbf{{\rv{v}}}}}\left(y_{ij} | \boldsymbol{u}_i, \boldsymbol{v}_j;\gamma_w^{-1}\right)^{\beta} \,p_{\textrm{\textbf{{\rv{u}}}}}(\boldsymbol{u}_i)^{\beta}\, p_{\textrm{\textbf{{\rv{v}}}}}(\boldsymbol{v}_j)^{\beta}, \label{eq:p-uv-given-y}
\end{aligned}
\end{equation}
wherein $\beta$ is a  parameter introduced here to treat the MMSE ($\beta = 1$) and MAP ($\beta=+\infty$) inference problems in a unified framework. We also assume the priors, $p_{\textrm{\textbf{{\rv{U}}}}}(\boldsymbol{U})$ and $p_{\textrm{\textbf{{\rv{V}}}}}(\boldsymbol{V})$, on the unknown matrices of $\textrm{\textbf{{\rv{U}}}}$ and $\textrm{\textbf{{\rv{V}}}}$ to be row-wise separable, i.e., $p_{\textrm{\textbf{{\rv{U}}}}}(\boldsymbol{U})=  \prod_{i=1}^{N} \mathit{p}_{\textrm{\textbf{{\rv{u}}}}}(\boldsymbol{u}_i)$ and $\mathit{p}_{\textrm{\textbf{{\rv{V}}}}}(\boldsymbol{V})=  \prod_{j=1}^{M} \mathit{p}_{\textrm{\textbf{{\rv{v}}}}}(\boldsymbol{v}_j)$. Fig. \ref{fig:factor-graph-Tanaka} depicts the factor graph associated to (\ref{eq:p-uv-given-y}) with variable nodes, $\bm{\mathsf{u}}_i$ and $\bm{\mathsf{v}}_j$, their  prior factor nodes, $\mathit{p}_{\textrm{\textbf{{\rv{u}}}}}(\boldsymbol{u}_i)^{\beta}$ and $\mathit{p}_{\textrm{\textbf{{\rv{v}}}}}(\boldsymbol{v}_j)^{\beta}$, and the labels, $f_{ij}$, which we use as a shorthand notations for the factor nodes:
\begin{eqnarray}
f(\boldsymbol{u}_i,\boldsymbol{v}_j)
&\triangleq&p_{\textrm{{\rv{y}}}_{ij}| \textrm{\textbf{{\rv{u}}},\textbf{{\rv{v}}}}}\left(y_{ij} | \boldsymbol{u}_i, \boldsymbol{v}_j,\gamma_w^{-1}\right)^{\beta},\\
\label{eq:factor_node_f_ij}&~=~&\mathcal{N}\left(y_{ij};\boldsymbol{u}_{i}^{\top} \boldsymbol{v}_{j},\beta^{-1}\gamma_w^{-1}\right).
\end{eqnarray}
By taking $\beta =1$ (i.e., minimum mean square error estimation), $p_{\textrm{\textbf{{\rv{u}}}},\textbf{{\rv{v}}}|\textrm{{\rv{y}}}_{ij}}(\boldsymbol{u}_i, \boldsymbol{v}_j | y_{ij} ; \beta)$ reduces to the true joint posterior $p_{\textrm{\textbf{{\rv{u}}}},\textbf{{\rv{v}}}|\textrm{{\rv{y}}}_{ij}}(\boldsymbol{u}_i, \boldsymbol{v}_j | y_{ij})$. In the limit $\beta \rightarrow \infty$ (i.e., maximum a posteriori estimation), however, it concentrates on the maxima of $p_{\textrm{\textbf{{\rv{u}}}},\textbf{{\rv{v}}}|\textrm{{\rv{y}}}_{ij}}(\boldsymbol{u}_i, \boldsymbol{v}_j | y_{ij})$.
Using the message derivation rules of loopy BP, the four messages defined in Fig.~\ref{fig:factor-graph-Tanaka}, , are expressed as follows (where $t$ stands for the iteration index):
\begin{eqnarray}
\label{eq:bp-messages_1}\!\!\!\!\!\!\!\! \mu_{(i, j) \rightarrow i,t}\left(\boldsymbol{u}_{i}\right) &~\propto~& \int f(\boldsymbol{u}_i,\boldsymbol{v}_j) \,\, \nu_{j \rightarrow(i, j),t-1}\left(\boldsymbol{v}_{j}\right) \mathrm{d} \boldsymbol{v}_{j},\\
\label{eq:bp-messages_2}\!\!\!\!\!\!\!\!\mu_{i \rightarrow(i, j),t+1}\left(\boldsymbol{u}_{i}\right) &~\propto~& p_{\textrm{\textbf{{\rv{u}}}}}\left(\boldsymbol{u}_{i}\right)^{\beta} \prod_{l \neq j} \mu_{(i, l) \rightarrow i,t}\left(\boldsymbol{u}_{i}\right),\\
\label{eq:bp-messages_3}\!\!\!\!\!\!\!\!\nu_{(i, j) \rightarrow j,t}\left(\boldsymbol{v}_{j}\right) &~\propto~& \int f(\boldsymbol{u}_i,\boldsymbol{v}_j) \,\, \mu_{i \rightarrow(i, j),t-1}\left(\boldsymbol{u}_{i}\right) \mathrm{d} \boldsymbol{u}_{i},\\
\label{eq:bp-messages_4}\!\!\!\!\!\!\!\!\nu_{j \rightarrow(i, j),t+1}\left(\boldsymbol{v}_{j}\right) &~\propto~& p_{\textrm{\textbf{{\rv{v}}}}}\left(\boldsymbol{v}_{j}\right)^{\beta} \prod_{k \neq i} \nu_{(k, j) \rightarrow j,t}\left(\boldsymbol{v}_{j}\right).
\end{eqnarray}
%
%
Assuming  $\textrm{\textbf{{\rv{v}}}}_{l} \sim \nu_{l \rightarrow(i, l),t}\left(\bm{v}_{l}\right)$ with mean $\boldsymbol{v}_{l \rightarrow(i, l),t}$ and covariance matrix $\beta^{-1} \bm{R}_{\bm{\mathsf{v}},l \rightarrow(i, l),t}$, it was shown in \cite{matsushita2013low} by  virtue of the central limit theorem (CLT) that the product of incoming messages, $\prod_{l \neq j} \mu_{(i, l) \rightarrow i,t}\left(\boldsymbol{u}_{i}\right)$, to any variable node $\textrm{\textbf{{\rv{u}}}}_i$ from all factor nodes $\{f_{il}\}_{l\neq j}$ can be approximated by a Gaussian random variable with mean $\boldsymbol{b}_{\textrm{\textbf{{\rv{u}}}}, i \rightarrow(i, j),t}$ and precision $\beta\mathbf{\Lambda}_{\textrm{\textbf{{\rv{u}}}}, i \rightarrow(i, j),t}$:
\begin{subequations}\label{eq:message_from_z_to_u_related_terms}
\begin{align}
\!\!\!\!\boldsymbol{b}_{\textrm{\textbf{{\rv{u}}}}, i \rightarrow(i, j),t}&\,=\,\gamma_w \sum_{l \neq j} y_{i, l} \,\widehat{\boldsymbol{v}}_{l \rightarrow(i, l),t},\\
\!\!\!\!\mathbf{\Lambda}_{\textrm{\textbf{{\rv{u}}}}, i \rightarrow(i, j),t}&\,=\,\gamma_w \sum_{l \neq j} \Big(\widehat{\boldsymbol{v}}_{l \rightarrow(i, l),t}\, \widehat{\boldsymbol{v}}_{l \rightarrow(i, l),t}^{\top} \,+\, \beta^{-1} \bm{R}_{\bm{\mathsf{v}},l \rightarrow(i, l),t} \nonumber\\
\!\!\!\!& \hspace{4.55cm}-\gamma_w~ y^2_{i, l} \,\bm{R}_{\bm{\mathsf{v}},l \rightarrow(i, l),t}\Big).
\end{align}
\end{subequations}
In other words, by dropping the normalization factor that does not depend on $\bm{u}_i$, we have: 
\begin{eqnarray}\label{eq:message_from_z_to_u}
\!\!\!\!\!\!\!\!\prod_{l \neq j} \mu_{(i,l) \rightarrow i,t}\left(\boldsymbol{u}_{i}\right)&&\nonumber\\
\!\!\!\!\!\!\!\!&&\!\!\!\!\!\!\!\!\!\!\!\!\!\!\!\!\!\!\!\!\!\!\!\!\!\!\propto\, \exp \left(-\frac{\beta}{2} \boldsymbol{u}_{i}^{\top}
\mathbf{\Lambda}_{\textrm{\textbf{{\rv{u}}}}, i \rightarrow(i, j),t}
\boldsymbol{u}_{i}+\beta \,\boldsymbol{u}_{i}^{\top} \boldsymbol{b}_{\textrm{\textbf{{\rv{u}}}}, i \rightarrow(i, j),t}\right)\!.
\end{eqnarray}
%
%
The inherent symmetry among the variable nodes $\textrm{\textbf{{\rv{u}}}}_i$ and $\textrm{\textbf{{\rv{v}}}}_j$ yields an equivalent  Gaussian approximation for $\prod_{k \neq i} \nu_{(k, j) \rightarrow j,t}\left(\boldsymbol{v}_{j}\right)$ under the density of $\textrm{\textbf{{\rv{u}}}}_{k} \sim \mu_{k \rightarrow(k, j),t}\left(\boldsymbol{u}_{k}\right)$ with mean $\boldsymbol{u}_{k \rightarrow(k, j),t}$ and covariance matrix $\beta^{-1} \bm{R}_{\bm{\mathsf{u}},k \rightarrow(k, j),t}$. That is to say:
\begin{eqnarray}\label{eq:message_from_z_to_v}
\!\!\!\!\!\!\!\!\prod_{k \neq i} \nu_{(k,j) \rightarrow j,t}\left(\boldsymbol{v}_{j}\right)\nonumber\\
\!\!\!\!\!\!\!\!&&\!\!\!\!\!\!\!\!\!\!\!\!\!\!\!\!\!\!\!\!\!\!\!\!\!\!\propto\, \exp \left(-\frac{\beta}{2} \boldsymbol{v}_{j}^{\top} \mathbf{\Lambda}_{\textrm{\textbf{{\rv{v}}}}, j \rightarrow(i, j),t} \boldsymbol{v}_{j}+\beta \,\boldsymbol{v}_{j}^{\top} \boldsymbol{b}_{\textrm{\textbf{{\rv{v}}}}, j \rightarrow(i, j),t}\right)\!,
\end{eqnarray}
with
\begin{subequations}\label{eq:message_from_z_to_v_related_terms} 
\begin{align}
\boldsymbol{b}_{\textrm{\textbf{{\rv{v}}}}, j\rightarrow(i, j),t}&\,=\,\gamma_w \sum_{k \neq i} y_{k, j} \,\widehat{\boldsymbol{u}}_{k \rightarrow(k, j),t},\\
\mathbf{\Lambda}_{\textrm{\textbf{{\rv{v}}}}, j \rightarrow(i, j),t}&\,=\,\gamma_w \sum_{k \neq i} \Big(\widehat{\boldsymbol{u}}_{k \rightarrow(k, j),t} \,\widehat{\boldsymbol{u}}_{k \rightarrow(k, j),t}^{\top} +\beta^{-1} \bm{R}_{\bm{\mathsf{u}},k \rightarrow(k, j),t} \nonumber\\
& \hspace{3.5cm}-\gamma_w \,y^2_{k, j} \,\bm{R}_{\bm{\mathsf{u}},k \rightarrow(k, j),t}\Big),
\end{align}
\end{subequations}
Pictorially, the messages given in (\ref{eq:message_from_z_to_u}) and (\ref{eq:message_from_z_to_v}) are shown in Fig.     \ref{fig:factor-graph-y_uv} as messages \protect\tikz[inner sep=.25ex,baseline=-.75ex] \protect\node[circle,draw] {\footnotesize 1}; and \protect\tikz[inner sep=.25ex,baseline=-.75ex] \protect\node[circle,draw] {\footnotesize 2}; sent by all factor nodes $\{f_{ij'}\}_{j'=1,j'\neq j}^M$ and $\{f_{i'j}\}_{i'=1,i'\neq i}^N$ to $\boldsymbol{u}_i$ and $\boldsymbol{v}_j$, respectively.
Moreover, the mean values, $\widehat{\boldsymbol{u}}_{i \rightarrow(i, j),t+1}$ and $\widehat{\boldsymbol{v}}_{j \rightarrow(i, j),t+1}$, as well as the  covariance matrices, $\beta^{-1}\bm{R}_{\bm{\mathsf{u}},i \rightarrow(i, j),t+1}$ and $\beta^{-1}\bm{R}_{\bm{\mathsf{v}},j \rightarrow(i, j),t+1}$, of messages \protect\tikz[inner sep=.25ex,baseline=-.75ex]  \protect\node[circle,draw] {\footnotesize 3}; and \protect\tikz[inner sep=.25ex,baseline=-.75ex] \protect\node[circle,draw] {\footnotesize 4};, respectively, are given by:
\begin{eqnarray}\label{eq:posterior_mean_u_new}
\!\!\!\!\widehat{\boldsymbol{u}}_{i \rightarrow(i, j),t+1}  &\,=\,& \mathbf{f}_{\mathsf{u}}(\boldsymbol{b}_{\textrm{\textbf{{\rv{u}}}}, i \rightarrow(i, j),t}, \mathbf{\Lambda}^{-1}_{\textrm{\textbf{{\rv{u}}}}, i \rightarrow(i, j),t}),\\
\!\!\!\!\label{eq:posterior_mean_v_new}\widehat{\boldsymbol{v}}_{j \rightarrow(i, j),t+1}   &\,=\,& \mathbf{f}_{\mathsf{v}}(\boldsymbol{b}_{\textrm{\textbf{{\rv{v}}}}, j \rightarrow(i, j),t}, \mathbf{\Lambda}^{-1}_{\textrm{\textbf{{\rv{v}}}}, j \rightarrow(i, j),t}),\\
\!\!\!\!\label{eq:posterior_variance_u_new}\bm{R}_{\bm{\mathsf{u}},i \rightarrow(i, j),t+1}&\,=\,& \nabla_{\boldsymbol{b}_{\textrm{\textbf{{\rv{u}}}}, i \rightarrow(i, j),t}}\mathbf{f}_{\mathsf{u}}(\boldsymbol{b}_{\textrm{\textbf{{\rv{u}}}}, i \rightarrow(i, j),t}, \mathbf{\Lambda}^{-1}_{\textrm{\textbf{{\rv{u}}}}, i \rightarrow(i, j),t})^{\mathsf{T}},\\
\!\!\!\!\label{eq:posterior_variance_v_new}\bm{R}_{\bm{\mathsf{v}},j \rightarrow(i, j),t+1}&\,=\,& \nabla_{\boldsymbol{b}_{\textrm{\textbf{{\rv{v}}}}, j \rightarrow(i, j),t}}\mathbf{f}_{\mathsf{v}}(\boldsymbol{b}_{\textrm{\textbf{{\rv{v}}}}, j \rightarrow(i, j),t}, \mathbf{\Lambda}^{-1}_{\textrm{\textbf{{\rv{v}}}}, j \rightarrow(i, j),t})^{\mathsf{T}},
\end{eqnarray}
where 
\begin{eqnarray}
\label{function_f_u} \mathbf{f}_{\mathsf{u}}(\boldsymbol{b}, \mathbf{\Lambda}^{-1}) &\,=\,&\frac{ \mathlarger{\int} \boldsymbol{u}\,p_{\textrm{\textbf{{\rv{u}}}}}\left(\boldsymbol{u}\right)^{\beta}\mathcal{N}\big(\boldsymbol{u}; \mathbf{\Lambda}^{-1}\boldsymbol{b}, \beta^{-1}\mathbf{\Lambda}^{-1}\big) d\boldsymbol{u}}{\mathlarger{\int} p_{\textrm{\textbf{{\rv{u}}}}}\left(\boldsymbol{u}\right)^{\beta}\mathcal{N}\big(\boldsymbol{u}; \mathbf{\Lambda}^{-1}\boldsymbol{b}, \beta^{-1}\mathbf{\Lambda}^{-1}\big) d\boldsymbol{u}},\\
   \label{function_f_v} \mathbf{f}_{\mathsf{v}}(\boldsymbol{b}, \mathbf{\Lambda}^{-1}) &\,=\,&\frac{ \mathlarger{\int} \boldsymbol{v}\,p_{\textrm{\textbf{{\rv{v}}}}}\left(\boldsymbol{v}\right)^{\beta}\mathcal{N}\big(\boldsymbol{v}; \mathbf{\Lambda}^{-1}\boldsymbol{b}, \beta^{-1}\mathbf{\Lambda}^{-1}\big) d\boldsymbol{v}}{\mathlarger{\int} p_{\textrm{\textbf{{\rv{v}}}}}\left(\boldsymbol{v}\right)^{\beta}\mathcal{N}\big(\boldsymbol{v}; \mathbf{\Lambda}^{-1}\boldsymbol{b}, \beta^{-1}\mathbf{\Lambda}^{-1}\big) d\boldsymbol{v}},
\end{eqnarray}
and the nabla operator, $\nabla_{\boldsymbol{x}}$,  with respect to any $n-$dimensional vector, $\bm{x}=[x_1,x_2,\cdots,x_n]^{\mathsf{T}}$, is given by:
\begin{equation}
    \nabla_{\boldsymbol{x}}~=~\left[\frac{\partial}{\partial x_1}, \frac{\partial}{\partial x_2},\cdots,\frac{\partial}{\partial x_n}\right]^{\mathsf{T}}.
\end{equation}
\par\noindent To reduce the complexity in the number of computed messages, the $(i,j)-$dependent  quantities in (\ref{eq:message_from_z_to_u_related_terms}) and (\ref{eq:message_from_z_to_v_related_terms}),  are replaced by the following $(i,j)-$independent (i.e., broadcast) ones:
\begin{subequations}\label{eq:message_from_z_to_u_related_terms_broadcast}
\begin{align} 
\!\!\!\!\!\!\boldsymbol{b}_{\textrm{\textbf{{\rv{u}}}}, i,t}&~=~\gamma_w \sum_{l} y_{i, l} \,\widehat{\boldsymbol{v}}_{l \rightarrow(i, l),t},\\
\label{eq:message_from_z_to_u_related_terms_broadcast_sub_eq_2}\!\!\!\!\!\!\mathbf{\Lambda}_{\bm{\mathsf{u}},i,t}&~=~\gamma_w \sum_{l} \Big(\widehat{\boldsymbol{v}}_{l \rightarrow(i, l),t}\, \widehat{\boldsymbol{v}}_{l \rightarrow(i, l),t}^{\top} ~+~ \left(\beta^{-1}-\gamma_w~ y^2_{i, l}\right) \bm{R}_{\bm{\mathsf{v}}_l,t}\Big).
\end{align}
\end{subequations}
\begin{subequations}\label{eq:message_from_z_to_v_related_terms_broadcast}
\begin{align} 
\!\!\!\!\!\!\boldsymbol{b}_{\textrm{\textbf{{\rv{v}}}}, j,t}&~=~\gamma_w \sum_{k} y_{j, k} \,\widehat{\boldsymbol{u}}_{k \rightarrow(k, j),t},\\
\label{eq:message_from_z_to_v_related_terms_broadcast_sub_eq_2}\!\!\!\!\!\!\mathbf{\Lambda}_{\bm{\mathsf{v}},j,t}&~=~\gamma_w \sum_{k}\! \Big(\widehat{\boldsymbol{u}}_{k \rightarrow(k, j),t}\, \widehat{\boldsymbol{u}}_{k \rightarrow(k, j),t}^{\top} +  \left(\beta^{-1}-\gamma_w y^2_{k, j}\right) \bm{R}_{\bm{\mathsf{u}}_k,t}\!\Big).
\end{align}
\end{subequations} 
after approximating the covariance matrices, $\beta^{-1}\boldsymbol{R}_{\bm{\mathsf{u}},i \rightarrow (i,j),t}$ and $\beta^{-1}\boldsymbol{R}_{\bm{\mathsf{v}},j \rightarrow (i,j),t}$, involved in (\ref{eq:message_from_z_to_u_related_terms}) and (\ref{eq:message_from_z_to_v_related_terms}) by  broadcast covariances, $\beta^{-1}\boldsymbol{R}_{\bm{\mathsf{u}}_i,t}$ and $\beta^{-1}\boldsymbol{R}_{\bm{\mathsf{v}}_j,t}$, respectively,  with a vanishing error order $O(M^{-1})$ \cite{matsushita2013low}. By recalling (\ref{eq:posterior_mean_u_new})-(\ref{eq:posterior_variance_v_new}), the underlying  broadcast means and the associated brodcacst covariances are thus givens by:
\begin{eqnarray}\label{eq:posterior_mean_u_broadcast}
\!\!\!\!\!\!\!\!\!\!\!\!\widehat{\boldsymbol{u}}_{i,t+1}  &~=~& \mathbf{f}_{\mathsf{u}}(\boldsymbol{b}_{\textrm{\textbf{{\rv{u}}}}, i,t}, \mathbf{\Lambda}^{-1}_{\textrm{\textbf{{\rv{u}}}}, i,t})\\
\label{eq:posterior_variance_u_broadcast}\boldsymbol{R}_{\bm{\mathsf{u}}_i,t+1}&\,=\,& \nabla_{\boldsymbol{b}_{\bm{\mathsf{u}},i,t}}\mathbf{f}_{\mathsf{u}}(\boldsymbol{b}_{\bm{\mathsf{u}},i,t}, \mathbf{\Lambda}^{-1}_{\textrm{\textbf{{\rv{u}}}}, i,t})^{\mathsf{T}},\\
\label{eq:posterior_mean_v_broadcast}
\!\!\!\!\!\!\!\!\!\!\!\!\widehat{\boldsymbol{v}}_{j,t+1}   &~=~& \mathbf{f}_{\mathsf{v}}(\boldsymbol{b}_{\textrm{\textbf{{\rv{v}}}}, j,t}, \mathbf{\Lambda}^{-1}_{\textrm{\textbf{{\rv{v}}}}, j,t})\\ \label{eq:posterior_variance_v_broadcast}\boldsymbol{R}_{\bm{\mathsf{v}}_j,t+1}&\,=\,& \nabla_{\boldsymbol{b}_{\bm{\mathsf{v}},j,t}}\mathbf{f}_{\mathsf{v}}(\boldsymbol{b}_{\bm{\mathsf{v}},j,t}, \mathbf{\Lambda}^{-1}_{\textrm{\textbf{{\rv{v}}}}, j,t})^{\mathsf{T}}.
\end{eqnarray}
Moreover, the $(i,j)$ posterior means, $\widehat{\boldsymbol{u}}_{i \rightarrow(i, j),t}$ and $\widehat{\boldsymbol{v}}_{j \rightarrow(i, j),t}$, are related to the their broadcast versions, $\widehat{\boldsymbol{u}}_{i,t}$ and $\widehat{\boldsymbol{v}}_{j,t}$, through small Osanger correction terms of order $O(M^{-1/2})$ as follows:
\begin{subequations}
\begin{align}
&\label{eq:posterior_mean_u} \widehat{\boldsymbol{u}}_{i \rightarrow(i, j),t} ~\approx\,~ \widehat{\boldsymbol{u}}_{i,t} ~-\underbrace{\gamma_w \,y_{ij}\, \boldsymbol{R}_{\boldsymbol{u}_i,t} \,\widehat{\boldsymbol{v}}_{j,t-1}}_{\textrm{Osanger correction term on} \, \boldsymbol{u}_{i}},
\\
&\label{eq:posterior_mean_v}\widehat{\boldsymbol{v}}_{j \rightarrow(i, j),t} ~\approx~ \widehat{\boldsymbol{v}}_{j,t} ~- \underbrace{\gamma_w \,y_{ij}\, \boldsymbol{R}_{\boldsymbol{v}_{j},t} \, \widehat{\boldsymbol{u}}_{i,t-1}}_{\textrm{Osanger correction term on}\, \boldsymbol{v}_{j}}.
\end{align}
\label{eq:posterior_means_u_v}
\end{subequations}
\par\noindent Yet, the correction terms in (\ref{eq:posterior_means_u_v}) are taken into account  during the calculation of $\boldsymbol{b}_{\textrm{\textbf{{\rv{u}}}}, i,t}$ and $\boldsymbol{b}_{\textrm{\textbf{{\rv{v}}}}, j,t}$ only. For the computation of $\mathbf{\Lambda}_{\bm{\mathsf{u}},i,t}$ and $\mathbf{\Lambda}_{\bm{\mathsf{v}},j,t}$, however, one replaces $\widehat{\boldsymbol{u}}_{i \rightarrow(i, j),t}$ by $\widehat{\boldsymbol{u}}_{i,t}$  and $\widehat{\boldsymbol{v}}_{j \rightarrow(i, j),t}$ by $\widehat{\boldsymbol{v}}_{j,t}$ after ignoring terms of vanishing order as $M$ and $N$ grow large. The algorithm in \cite{matsushita2013low} also relies on the following low-SNR approximation
\begin{equation}\label{eq:low-rank-assumption}
y_{k,j}^2 ~\approx ~\mathbb{E}[y_{k,j}^2] ~\approx  \gamma_w^{-1}.
\end{equation}
which is used in both (\ref{eq:message_from_z_to_u_related_terms_broadcast_sub_eq_2}) and (\ref{eq:message_from_z_to_v_related_terms_broadcast_sub_eq_2}). We emphasize, however, the fact that the low-SNR regime is mainly conceivable in presence of very-low-rank structures with a fully observed matrix $\bm{Y} =\bm{U}\bm{V}^{\top} +\bm{W}$.   

In conclusion, the algorithmic steps of the bilinear recovery  technique introduced in \cite{matsushita2013low} are summarized in Algorithm \ref{algo:tanaka}.  We refer the reader to the supplementary materials of \cite{matsushita2013low} for further details.
\begin{algorithm}[!t]
\notsotiny
\caption{AMP-based structured matrix reconstruction \cite{matsushita2013low}}\label{algo:tanaka}
\begin{algorithmic}[1]
\Statex $\mathbf{Require:}$ Matrix $\boldsymbol{Y}$ $\in$ $\mathbb{R}^{N \times M}$; noise precision $\gamma_w$; temperature parameter $\beta$; two denoising functions, $\mathbf{f}_{\mathsf{u}}(.)$ and $\mathbf{f}_{\mathsf{v}}(.)$, given in (\ref{function_f_u}) and (\ref{function_f_v}); number of iterations $T_{\textrm{max}}$.\vspace{0.1cm}
\State $\mathbf{Initialize}$ 
\Statex\LeftCommentNoIntent{posterior means and covariances} 
\Statex  $\boldsymbol{\widehat{U}}_{0}$, $\boldsymbol{\widehat{V}}_{0}$, $\boldsymbol{\widehat{U}}_{1}$, $\boldsymbol{\widehat{V}}_{1}$,  $\{\boldsymbol{R}_{\boldsymbol{u}_i, 1}\}_{i=1}^{N}$,
$\{\boldsymbol{R}_{\boldsymbol{v}_j, 1}\}_{j=1}^{M}$\vspace{0.1cm}
\For {$t=1,\dots, T_{\textrm{max}}$}\vspace{0.1cm}
\Statex\LeftComment{1}{Compute means and precisions related to the messages in eqs. (\ref{eq:message_from_z_to_u}) and (\ref{eq:message_from_z_to_v})}
\State $\boldsymbol{B}_{\boldsymbol{U}, t} = \gamma_w\,\Big(
\boldsymbol{Y} \,{{}\widehat{\boldsymbol{V}}_{t}} -\gamma_w \,{}\widehat{\boldsymbol{U}}_{t-1}\, \sum_{j=1}^{M}\boldsymbol{R}_{\boldsymbol{v}_j,{t}}$\Big)\label{eq:tanaka-algo-Bu}
\State $\boldsymbol{\Lambda}_{\boldsymbol{U}, t} = \gamma_{w}\Big(\boldsymbol{\widehat{V}}^{\top}_{t} \boldsymbol{\widehat{V}}_{t}  + (\frac{1}{\beta}-1)\,\sum_{j=1}^{M}\boldsymbol{R}_{\boldsymbol{v}_j,{t}}\Big)$\label{eq:tanaka-algo-Au}
\State $\boldsymbol{B}_{\boldsymbol{V}, t} = \gamma_w\,\Big(
\boldsymbol{Y}  \,{{}\widehat{\boldsymbol{U}}_{t}} -\gamma_w \,{}\widehat{\boldsymbol{V}}_{t-1}\, \sum_{i=1}^{N}\boldsymbol{R}_{\boldsymbol{u}_i,{t}}$\Big)\label{eq:tanaka-algo-Bv}
\State $\boldsymbol{\Lambda}_{\boldsymbol{V}, t} = \gamma_{w}\Big(\boldsymbol{\widehat{U}}^{\top}_{t} \boldsymbol{\widehat{U}}_{t}\,+\, (\frac{1}{\beta}-1)\,\sum_{i=1}^{N}\boldsymbol{R}_{\boldsymbol{u}_i,{t}}\Big)$\label{eq:tanaka-algo-Av}\vspace{0.15cm}
\Statex\LeftComment{1}{Update the posterior means, $\widehat{\bm{U}}=[\widehat{\bm{u}}_1,\ldots,\widehat{\bm{u}}_N]^{\mathsf{T}}$ and $\widehat{\bm{V}}=[\widehat{\bm{v}}_1,\ldots,\widehat{\bm{v}}_M]^{\mathsf{T}}$,\\of the  matrices $\boldsymbol{U}$ and $\boldsymbol{V}$ and the corresponding row-wise scaled covariance matrices}\vspace{0.01cm} 
\State $\forall i:$ ~ $\boldsymbol{\widehat{u}}_{i,t+1}= \mathbf{f}_{\mathsf{u}}(\boldsymbol{b}_{\bm{\mathsf{u}},i, t}, \boldsymbol{\Lambda}^{-1}_{\boldsymbol{U}, t})$~~~~~ with~~~~~ $\bm{B}_{\bm{U},t}=[\boldsymbol{b}_{\bm{\mathsf{u}},1, t},\ldots,\boldsymbol{b}_{\bm{\mathsf{u}},N, t}]^{\mathsf{T}}$\label{eq:tanaka-algo-posterior-mean-U} \vspace{0.12cm}
\State $\forall i:$ ~ $\boldsymbol{R}_{\boldsymbol{u}_i,{t+1}}\,=\, \nabla_{\boldsymbol{b}_{\bm{\mathsf{u}}}}\mathbf{f}_{\mathsf{u}}(\boldsymbol{b}_{\bm{\mathsf{u}},i, t}, \boldsymbol{\Lambda}^{-1}_{\boldsymbol{U}, t})^{\mathsf{T}}$
\label{eq:tanaka-algo-posterior-covarirance-U}
\vspace{0.05cm}
\State $\forall j:$ ~ $\boldsymbol{\widehat{v}}_{j,t+1}= \mathbf{f}_{\mathsf{v}}(\boldsymbol{b}_{\bm{\mathsf{v}},j, t}, \boldsymbol{\Lambda}^{-1}_{\boldsymbol{V}, t})$
~~~~~ with~~~~~ $\bm{B}_{\bm{V},t}=[\boldsymbol{b}_{\bm{\mathsf{v}},1, t},\ldots,\boldsymbol{b}_{\bm{\mathsf{v}},M,  t}]^{\mathsf{T}}$
\label{eq:tanaka-algo-posterior-mean-V}
\vspace{0.07cm}
\State \!\!$\forall j:$ ~ $\boldsymbol{R}_{\boldsymbol{v}_j,{t+1}}\,=\, \nabla_{\boldsymbol{b}_{\bm{\mathsf{v}}}}\mathbf{f}_{\mathsf{v}}(\boldsymbol{b}_{\bm{\mathsf{v}},j, t}, \boldsymbol{\Lambda}^{-1}_{\boldsymbol{V}, t})^{\mathsf{T}}$
\label{eq:tanaka-algo-posterior-covarirance-V}
\vspace{0.05cm}
\EndFor
\vspace{0.05cm}
\State \textbf{Return} $\boldsymbol{\widehat{U}}_{T_{\textrm{max}}+1}$, $\boldsymbol{\widehat{V}}_{T_{\textrm{max}}+1}$
\end{algorithmic}
\end{algorithm}
The limitation of this method, however, lies in the intractable multi-dimensional integrals in (\ref{function_f_u}) and (\ref{function_f_v}) which can be evaluated only in some specific priors, $\mathit{p}_{\textrm{\textbf{{\rv{u}}}}}(\cdot)$ and $\mathit{p}_{\textrm{\textbf{{\rv{v}}}}}(\cdot)$, such as Gaussian and community\footnote{A community prior on a random vector, $\bm{\mathsf{x}}$, in the presence of $r$ communities (i.e., clusters) is given by $p_{\textrm{\textbf{{\rv{x}}}}}(\boldsymbol{x})=\frac{1}{r} \sum_{l=1}^{r} \delta\left(\boldsymbol{x}-\boldsymbol{e}_{l}\right)$, where $\boldsymbol{e}_{l}$ is the $l^{\textrm{th}}$ canonical basis vector in $\mathbb{R}^r$.} priors. 
 It is impossible, for instance, to consider a binary prior on $\boldsymbol{\mathsf{u}}_i$ and/or $\boldsymbol{\mathsf{v}}_j$ since the underlying integrals become combinatorial sums over $2^r$ terms. In this context, the fundamental novelties brought by the proposed BiG-VAMP algorithm
 consist in its combined abilities to handle:  
\begin{itemize}
    \item A broader class  of practical applications which involve  a \textit{high-rank} decomposition of structured matrices in addition to  a \textit{low-rank} decomposition of (possibly) unstructured matrices,
    \item General priors on both $\boldsymbol{U}$ and $\boldsymbol{V}$ matrices owing to appropriate Gaussian approximation of the extrinsic information exchanged between its constituent blocks.
    \item General separable output distributions, $p_{\textrm{\textbf{{\rv{Y}}}} | \textrm{\textbf{{\rv{Z}}}}}(\boldsymbol{Y} | \boldsymbol{Z})$, in (\ref{eq:non-linear-distribution0}).
\end{itemize}
\section{The BiG-VAMP  algorithm}\label{sec:big-vamp-algorithm}
\begin{algorithm*}[!t]
\notsotiny
\caption{BiG-VAMP}\label{algo:big-vamp}
\begin{multicols}{2}
\begin{algorithmic}[1]
\Statex $\mathbf{Require:}$ Matrix $\boldsymbol{Y}$ $\in$ $\mathbb{R}^{N \times M}$; temperature parameter $\beta$; precision tolerance ($\xi=10^{-6}$);  maximum number of iterations ($T_\textrm{max}$);  two denoisers $\mathbf{g}_{\mathsf{u}}(\cdot)$ and $\mathbf{g}_{\mathsf{v}}(\cdot)$ from (\ref{new_posterior_mean_u}) and (\ref{new_posterior_mean_v});  (noise precision $\gamma_w$ only for Bi-VAMP, i.e., under bilinear observation models) 
\State $\mathbf{Initialize}$ 
\State $t\gets 1$
\Statex\LeftCommentNoIntent{posterior means, covariances and precisions} 
\Statex $\boldsymbol{\widehat{U}}^{-}_{ \mathsf{p}, 0}$, $\boldsymbol{\widehat{V}}^{-}_{ \mathsf{p}, 0}$, $\boldsymbol{\widehat{U}}_{ \mathsf{p}, 1}^{-}$, $\boldsymbol{\widehat{V}}_{  \mathsf{p}, 1}^{-}$, $\boldsymbol{\widehat{Z}}_{ \mathsf{p}, 1}^{-}$,  $\boldsymbol{R}_{\boldsymbol{U}_{ \mathsf{p}}^{-}, 1}$, $\boldsymbol{R}_{\boldsymbol{V}^{-}_{ \mathsf{p}}, 1}$,  $\gamma_{\boldsymbol{Z}^{-}_{ \mathsf{p}}, 1}$
\Statex $\boldsymbol{\widehat{U}}_{ \mathsf{p}, 1}^{+}$, $\boldsymbol{\widehat{V}}_{ \mathsf{p}, 1}^{+}$, $\boldsymbol{\widehat{Z}}_{ \mathsf{p}, 1}^{+}$, $\gamma_{\boldsymbol{U}^{-}_{ \mathsf{p}}, 1}$, $\gamma_{\boldsymbol{V}^{-}_{ \mathsf{p}}, 1}$, $\gamma_{\boldsymbol{Z}^{+}_{ \mathsf{p}}, 1}$\vspace{0.1cm}
\Statex\LeftCommentNoIntent{extrinsic means and precisions}
\Statex $\boldsymbol{\widehat{U}}^{-}_{ \mathsf{e}, 1}$, $\boldsymbol{\widehat{V}}^{-}_{ \mathsf{e}, 1}$, $\boldsymbol{\widehat{Z}}^{-}_{ \mathsf{e}, 1}$, $\gamma_{\boldsymbol{U}^{-}_{ \mathsf{e}}, 1}$, $\gamma_{\boldsymbol{V}^{-}_{ \mathsf{e}}, 1}$, $\gamma_{\boldsymbol{Z}^{-}_{ \mathsf{e}},1}$
\Statex  $\boldsymbol{\widehat{U}}^{+}_{ \mathsf{e}, 1}$, $\boldsymbol{\widehat{V}}^{+}_{ \mathsf{e}, 1}$, $\boldsymbol{\widehat{Z}}^{+}_{ \mathsf{e}, 1}$, $\gamma_{\boldsymbol{U}^{+}_{ \mathsf{e}}, 1}$, $\gamma_{\boldsymbol{V}^{+}_{ \mathsf{e}}, 1}$, $\gamma_{\boldsymbol{Z}^{+}_{ \mathsf{e}},1}$ \vspace{0.1cm}
\Statex\LeftCommentNoIntent{means and precisions in eqs. (\ref{eq:message_from_z_to_u}), (\ref{eq:message_from_z_to_u_related_terms}), (\ref{eq:message_from_z_to_v}) and (\ref{eq:message_from_z_to_v_related_terms})}
\Statex $\boldsymbol{B}_{\boldsymbol{U}, 1}$, $\boldsymbol{\Lambda}_{\boldsymbol{U}, 1}$, $\boldsymbol{B}_{\boldsymbol{V}, 1}$, $\boldsymbol{\Lambda}_{\boldsymbol{V}, 1}$\vspace{0.1cm}
\Repeat
\vspace{0.1cm}
\Statex\LeftCommentNoTriangle{1}{\underline{I. Approximate Bi-LMMSE step}}\vspace{0.05cm}
\Statex\LeftComment{1}{Compute the approximated message in (\ref{eq:posterior_mean_u})}
\State $\boldsymbol{B}_{\boldsymbol{U}, t} = \gamma_{\boldsymbol{Z}^{+}_{ \mathsf{e}}, t} \Big(
{{}\widehat{\boldsymbol{Z}}^{+}_{ \mathsf{e}, t}} \,{{}\widehat{\boldsymbol{V}}^{-}_{ \mathsf{p}, t}}\, - M\,\gamma_{\boldsymbol{Z}^{+}_{ \mathsf{e}},t}\,{}\widehat{\boldsymbol{U}}^{-}_{ \mathsf{p}, t-1} \,\boldsymbol{R}_{\boldsymbol{V}_{ \mathsf{p}}^{-}, t}  \,\langle\boldsymbol{\widehat{Z}}_{ \mathsf{e},t}^{+} \odot\boldsymbol{\widehat{Z}}_{ \mathsf{e},t}^{+}\rangle\Big)$ \label{eq:algo-Bu-non-low-rank}
\State $\boldsymbol{\Lambda}_{\boldsymbol{U}, t} = \gamma_{\boldsymbol{Z}^{+}_{ \mathsf{e}}, t}\Big(\boldsymbol{\widehat{V}}^{- \top}_{ \mathsf{p}, t} \boldsymbol{\widehat{V}}^{-}_{ \mathsf{p}, t}  + \frac{M}{\beta}\,\boldsymbol{R}_{\boldsymbol{V}_{ \mathsf{p}}^{-}, t} -\, M\,\gamma_{\boldsymbol{Z}^{+}_{ \mathsf{e}}, t} \,\boldsymbol{R}_{\boldsymbol{V}_{ \mathsf{p}}^{-}, t} \,\langle\boldsymbol{\widehat{Z}}_{ \mathsf{e},t}^{+} \odot\boldsymbol{\widehat{Z}}_{ \mathsf{e},t}^{+}\rangle\Big)$ \label{eq:algo-Au-non-low-rank}\vspace{0.05cm}
\Statex\LeftComment{1}{Compute the approximated message in (\ref{eq:posterior_mean_v})}
\State $\boldsymbol{B}_{\boldsymbol{V}, t} = \gamma_{\boldsymbol{Z}^{+}_{ \mathsf{e}}, t} \Big(
{{}\widehat{\boldsymbol{Z}}^{+}_{ \mathsf{e}, t}} \,{{}\widehat{\boldsymbol{U}}^{-}_{ \mathsf{p}, t}}\, - N\,\gamma_{\boldsymbol{Z}^{+}_{ \mathsf{e}}, t}\,{}\widehat{\boldsymbol{V}}^{-}_{ \mathsf{p}, t-1} \, \boldsymbol{R}_{\boldsymbol{U}_{ \mathsf{p}}^{-}, t}\,\langle\boldsymbol{\widehat{Z}}_{ \mathsf{e},t}^{+} \odot\boldsymbol{\widehat{Z}}_{ \mathsf{e},t}^{+}\rangle\Big)$\label{eq:algo-Bv-non-low-rank}\vspace{0.05cm}
\State $\boldsymbol{\Lambda}_{\boldsymbol{V}, t} = \gamma_{\boldsymbol{Z}^{+}_{ \mathsf{e}}, t}\Big(\boldsymbol{\widehat{U}}^{-\top}_{ \mathsf{p}, t} \, \boldsymbol{\widehat{U}}^{-}_{ \mathsf{p}, t}  + \frac{N}{\beta}\,\boldsymbol{R}_{\boldsymbol{U}_{ \mathsf{p}}^{-}, t} -\, N\,\gamma_{\boldsymbol{Z}^{+}_{ \mathsf{e}}, t} \,\boldsymbol{R}_{\boldsymbol{U}_{ \mathsf{p}}^{-}, t}\,\langle\boldsymbol{\widehat{Z}}_{ \mathsf{e},t}^{+} \odot\boldsymbol{\widehat{Z}}_{ \mathsf{e},t}^{+}\rangle\Big)$
\label{eq:algo-Av-non-low-rank} 
\Statex\LeftComment{1}{Update the posterior statistics $\boldsymbol{\widehat{U}}_{ \mathsf{p}, t}^{-}$, $\boldsymbol{R}_{\boldsymbol{U}_{ \mathsf{p}}^{-}, t}$, $\boldsymbol{\widehat{U}}_{ \mathsf{p},t}^{-}$, $\boldsymbol{R}_{\boldsymbol{V}_{ \mathsf{p}}^{-}, t}$}
\State $\boldsymbol{R}_{\boldsymbol{U}_{ \mathsf{p}}^{-}, t+1} = (\gamma_{\boldsymbol{U}^{+}_{ \mathsf{e}}, t} \, \boldsymbol{I} + \boldsymbol{\Lambda}_{\boldsymbol{U}, t})^{-1}$\label{eq:posterior-covariance-u-gaussian-approx}
\State $\boldsymbol{\widehat{U}}_{ \mathsf{p}, t+1}^{-}\;\;=(\boldsymbol{B}_{\boldsymbol{U}, t}+\gamma_{\boldsymbol{U}^{+}_{ \mathsf{e}}, t}\boldsymbol{\widehat{U}}_{ \mathsf{e},t}^{+}) \, \boldsymbol{R}_{\boldsymbol{U}_{ \mathsf{p}}^{-}, t+1}$\label{eq:posterior-mean-u-gaussian-approx}
\State $\boldsymbol{R}_{\boldsymbol{V}_{ \mathsf{p}}^{-}, t+1} = (\gamma_{\boldsymbol{V}^{+}_{ \mathsf{e}}, t} \, \boldsymbol{I} + \boldsymbol{\Lambda}_{\boldsymbol{V},t})^{-1}$
\label{eq:posterior-covariance-v-gaussian-approx}
\State $\boldsymbol{\widehat{V}}_{ \mathsf{p},t+1}^{-}\;\;=(\boldsymbol{B}_{\boldsymbol{V},t}+\gamma_{\boldsymbol{V}^{+}_{ \mathsf{e}},t}\boldsymbol{\widehat{V}}_{ \mathsf{e},t}^{+}) \,\boldsymbol{R}_{\boldsymbol{V}_{ \mathsf{p}}^{-}, t+1}$\label{eq:posterior-mean-v-gaussian-approx}\vspace{0.1cm}
\Statex\LeftCommentNoTriangle{1}{\underline{II. Denoising step}}
\Statex\LeftComment{1}{Update the extrinsic statistics $\boldsymbol{\widehat{U}}_{ \mathsf{e},t+1}^{-}$, $\gamma_{\boldsymbol{U}^{-}_{ \mathsf{e}},t+1}$, $\boldsymbol{\widehat{V}}_{ \mathsf{e},t+1}^{-}$, $\gamma_{\boldsymbol{V}^{-}_{ \mathsf{e}}, t+1}$}
\State $\gamma_{\boldsymbol{U}_{ \mathsf{p}}^{-},t+1}\,=\,\big(\frac{1}{r}\text{Tr}(\boldsymbol{R}_{\boldsymbol{U}_{ \mathsf{p}}^{-}, t+1})\big)^{-1}$ \label{eq:algo-diag-sigma-u}
\State $\gamma_{\boldsymbol{U}^{-}_{ \mathsf{e}},t+1} = \gamma_{\boldsymbol{U}_{ \mathsf{p}}^{-},t+1} -\, \gamma_{\boldsymbol{U}^{+}_{ \mathsf{e}}, t}$ \label{eq:algo-diag-sigma-u-e}
\State $\boldsymbol{\widehat{U}}_{ \mathsf{e},t+1}^{-} \;\,= \gamma^{-1}_{\boldsymbol{U}^{-}_{ \mathsf{e}},t+1}\,\big(\gamma_{\boldsymbol{U}_{ \mathsf{p}}^{-},t+1}\boldsymbol{\widehat{U}}_{ \mathsf{p},t+1}^{-} \, - \,\gamma_{\boldsymbol{U}^{+}_{ \mathsf{e}},t}\boldsymbol{\widehat{U}}_{ \mathsf{e},t}^{+} \big)$\label{eq:algo-diag-mean-u}
\State $\gamma_{\boldsymbol{V}_{ \mathsf{p}}^{-},t+1}\,=\,\big(\frac{1}{r}\text{Tr}(\boldsymbol{R}_{\boldsymbol{V}_{ \mathsf{p}}^{-}, t+1})\big)^{-1}$ \label{eq:algo-diag-sigma-v}
\State $\gamma_{\boldsymbol{V}^{-}_{ \mathsf{e}},t+1} = \gamma_{\boldsymbol{V}_{ \mathsf{p}}^{-},t+1} -\, \gamma_{\boldsymbol{V}^{+}_{ \mathsf{e}}, t}$ \label{eq:algo-diag-sigma-v-e}
\State $\boldsymbol{\widehat{V}}_{ \mathsf{e},t+1}^{-} \;\,= \gamma^{-1}_{\boldsymbol{V}^{-}_{ \mathsf{e}},t+1}\,\big(\gamma_{\boldsymbol{V}_{ \mathsf{p}}^{-},t+1}\boldsymbol{\widehat{V}}_{ \mathsf{p},t+1}^{-} \, - \,\gamma_{\boldsymbol{V}^{+}_{ \mathsf{e}},t}\boldsymbol{\widehat{V}}_{ \mathsf{e},t}^{+} \big)$\label{eq:algo-diag-mean-v}
\Statex\LeftComment{1}{Denoising the rows, $\widehat{\boldsymbol{u}}^{-}_{i,\mathsf{p},t+1}$, of   $\boldsymbol{\widehat{U}}_{\mathsf{p},t+1}^{-}$ and the columns $\widehat{\boldsymbol{v}}^{-}_{j,\mathsf{p},t+1}$ of $\boldsymbol{\widehat{V}}_{ \mathsf{p},t+1}^{-}$}\vspace{0.1cm}
\State $\forall i:$ update the $i$th row
of   $\boldsymbol{\widehat{U}}_{ \mathsf{p},t+1}^{+}$ as $\, \widehat{\boldsymbol{u}}^{+}_{i,\mathsf{p},t+1}= \mathbf{g}_{\mathsf{u}}(\widehat{\boldsymbol{u}}^{-}_{i,\mathsf{e},t+1}, \gamma^{-1}_{\boldsymbol{U}^{-}_{ \mathsf{e}},t+1}),$\label{denoiser_u}
\State $\forall j:$ update the $j$th column
of $\boldsymbol{\widehat{V}}_{ \mathsf{p},t+1}^{+}$ as $\,\widehat{\boldsymbol{v}}^{+}_{j,\mathsf{p},t+1}\,=\, \mathbf{g}_{\mathsf{v}}(\widehat{\boldsymbol{v}}^{-}_{j,\mathsf{e},t+1}, \gamma^{-1}_{\boldsymbol{V}^{-}_{ \mathsf{e}},t+1}),$ \label{denoiser_v}
\State $\gamma_{\boldsymbol{U}^{+}_{ \mathsf{p}},t+1} = \gamma_{\boldsymbol{U}^{-}_{ \mathsf{e}},t+1} \, \left(\frac{1}{N}\sum_{i=1}^N\big\langle \mathbf{g}^{\prime}_{\mathsf{u}}(\widehat{\boldsymbol{u}}^{-}_{i,\mathsf{e},t+1}, \gamma^{-1}_{\boldsymbol{U}^{-}_{ \mathsf{e}},t+1})\big\rangle\right)^{-1} $ \label{eq:algo-diag-sigma-u-alpha}
\State $\gamma_{\boldsymbol{V}^{+}_{ \mathsf{p}},t+1} = \gamma_{\boldsymbol{V}^{-}_{ \mathsf{e}},t+1} \,\left(\frac{1}{M}\sum_{j=1}^M\big\langle \mathbf{g}^{\prime}_{\mathsf{v}}(\widehat{\boldsymbol{v}}^{-}_{j,\mathsf{e},t+1}, \gamma^{-1}_{\boldsymbol{U}^{-}_{ \mathsf{e}},t+1})\big\rangle\right)^{-1} $\label{eq:algo-diag-sigma-v-alpha}\vspace{0.1cm}
\Statex\LeftComment{1}{update the extrinsic statistics $\boldsymbol{\widehat{U}}_{ \mathsf{e},t+1}^{+}$, $\gamma_{\boldsymbol{U}^{+}_{ \mathsf{e},t+1}}$, $\boldsymbol{\widehat{V}}_{ \mathsf{e},t+1}^{+}$, $\gamma_{\boldsymbol{V}^{+}_{ \mathsf{e}},t+1}$} 
\State $\gamma_{\boldsymbol{U}^{+}_{ \mathsf{e}},t+1} = \gamma_{\boldsymbol{U}^{+}_{ \mathsf{p}},t+1}- \gamma_{\boldsymbol{U}^{-}_{ \mathsf{e}},t+1}$ \label{eq:algo-gamma-Ue+}
\State $\gamma_{\mathbf{V}^{+}_{ \mathsf{e}},t+1} = \gamma_{\boldsymbol{V}^{+}_{ \mathsf{p}},t+1}- \gamma_{\boldsymbol{V}^{-}_{ \mathsf{e}},t+1}$
\State $\boldsymbol{\widehat{U}}_{ \mathsf{e},t+1}^{+} \;\,= \gamma^{-1}_{\boldsymbol{U}^{+}_{ \mathsf{e}},t+1}\,\big(\gamma_{\boldsymbol{U}^{+}_{ \mathsf{p}},t+1}\,\boldsymbol{\widehat{U}}_{ \mathsf{p},t+1}^{+} - \gamma_{\boldsymbol{U}^{-}_{ \mathsf{e}},t+1}\,\boldsymbol{\widehat{U}}_{ \mathsf{e},t+1}^{-}\big)$
\State $\boldsymbol{\widehat{V}}_{ \mathsf{e},t+1}^{+} \;\,= \gamma^{-1}_{\boldsymbol{V}^{+}_{ \mathsf{e}},t+1}\,\big(\gamma_{\boldsymbol{V}^{+}_{ \mathsf{p}},t+1}\,\boldsymbol{\widehat{V}}_{ \mathsf{p},t+1}^{+} - \gamma_{\boldsymbol{V}^{-}_{ \mathsf{e}},t+1}\boldsymbol{\widehat{V}}_{ \mathsf{e},t+1}^{-} \big)$ \label{eq:algo-mean-Ve+}\vspace{0.1cm}
\Statex\LeftCommentNoTriangle{1}{\underline{III. Generalized output step}}
\Statex\LeftComment{1}{ Compute the posterior statistics $\boldsymbol{\widehat{Z}}_{ \mathsf{p},t+1}^{-}$ and $\gamma_{\mathbf{Z}^{-}_{ \mathsf{p}},t+1}$}
\State $\boldsymbol{\widehat{Z}}_{ \mathsf{p},t+1}^{-} ~=~ \boldsymbol{\widehat{U}}_{  \mathsf{p},t+1}^{-}\,\boldsymbol{\widehat{V}}_{ \mathsf{p},t+1}^{-\top}+ \frac{\gamma_{\boldsymbol{Z}^{+}_{ \mathsf{e}},t}}{\beta}\,\boldsymbol{\widehat{Z}}_{ \mathsf{e},t}^{+}\,\text{Tr}(\boldsymbol{R}_{\boldsymbol{U}_{ \mathsf{p}}^{-},t+1}\,\boldsymbol{R}_{\boldsymbol{V}_{ \mathsf{p}}^{-},t+1}^{\top})$ \label{eq:algo-posterior-mean-z}
\State $\gamma_{\boldsymbol{Z}^{-}_{ \mathsf{p}},t+1} =\gamma_{\boldsymbol{Z}^{+}_{ \mathsf{e}},t}+MN\,\Big(\text{Tr}\big(\frac{M\,N}{\beta}\,\boldsymbol{R}_{\boldsymbol{U}_{ \mathsf{p}}^{-},t+1}\,\boldsymbol{R}_{\boldsymbol{V}_{ \mathsf{p}}^{-},t+1}^{\top}$ \label{eq:algo-posterior-sigma-z}\vspace{0.1cm}
\Statex \hspace{4.2cm}$+ \,N\, \boldsymbol{R}_{\boldsymbol{U}_{ \mathsf{p}}^{-},t+1} \, \boldsymbol{\widehat{V}}_{ \mathsf{p},t+1}^{-\top}\, \boldsymbol{\widehat{V}}^{-}_{ \mathsf{p},t+1}$\vspace{0.1cm}
\Statex  \hspace{4.6cm}$+ \,`M\, \boldsymbol{R}_{\boldsymbol{V}_{ \mathsf{p}}^{-},t+1}\, \boldsymbol{\widehat{U}}_{ \mathsf{p},t+1}^{-\top}\,  \boldsymbol{\widehat{U}}^{-}_{ \mathsf{p},t+1}\big)\Big)^{-1}$
\Statex\LeftComment{1}{ Compute the extrinsic statistics $\boldsymbol{\widehat{Z}}_{ \mathsf{e},t+1}^{-}$ and $\gamma_{\boldsymbol{Z}^{-}_{ \mathsf{e}},t+1}$}
\State $\gamma_{\boldsymbol{Z}^{-}_{ \mathsf{e}},t+1} = \gamma_{\boldsymbol{Z}^{-}_{ \mathsf{p}},t+1}- \gamma_{\boldsymbol{Z}^{+}_{ \mathsf{e}},t}$\label{extrinsic_gamma_z_minus}
\State $\boldsymbol{\widehat{Z}}_{ \mathsf{e},t+1}^{-} = \gamma^{-1}_{\boldsymbol{Z}^{-}_{ \mathsf{e}},t+1}\,\big(\boldsymbol{\widehat{Z}}_{ \mathsf{p},t}^{-}\,\gamma_{\boldsymbol{Z}^{-}_{ \mathsf{p}},t+1} - \boldsymbol{\widehat{Z}}_{ \mathsf{e},t}^{+} \,\gamma_{\boldsymbol{Z}^{+}_{ \mathsf{e}},t}\big)$
\Statex\LeftComment{1}{ Compute the posterior and extrinsic statistics $\boldsymbol{\widehat{Z}}_{ \mathsf{p},t+1}^{+}$, $\gamma_{\boldsymbol{Z}^{+}_{ \mathsf{p}},t+1}$, $\boldsymbol{\widehat{Z}}_{ \mathsf{e},t+1}^{+}$, $\gamma_{\boldsymbol{Z}^{+}_{ \mathsf{e}},t+1}$}
\State \label{update_posterior_mean_and_variance_z} Compute $\boldsymbol{\widehat{Z}}_{ \mathsf{p},t+1}^{+}$  using (\ref{eq:message_posterior_from_z_to_u_v}) and $\gamma_{\boldsymbol{Z}^{+}_{ \mathsf{p}},t+1}=\frac{1}{MN}\sum_i\sum_j \gamma_{\textrm{{\rv{z}}}_{ij,  \mathsf{p}}^{+},t+1}$ using (\ref{eq:message_posterior_from_z_to_u_v_variance})
\State $\gamma_{\boldsymbol{Z}^{+}_{ \mathsf{e}},t+1} = \gamma_{\boldsymbol{Z}^{+}_{ \mathsf{p}},t+1}- \gamma_{\boldsymbol{Z}^{-}_{ \mathsf{e}},t+1}$ \label{eq:algo-extrinsic-mu-z-minus}
\State $\boldsymbol{\widehat{Z}}_{ \mathsf{e},t+1}^{+} = \gamma_{\boldsymbol{Z}^{+}_{ \mathsf{e}},t+1}\,\big(\boldsymbol{\widehat{Z}}_{ \mathsf{p},t+1}^{+}\,\gamma_{\boldsymbol{Z}^{+}_{ \mathsf{p}},t+1} - \boldsymbol{\widehat{Z}}_{ \mathsf{e},t+1}^{-} \,\gamma_{\boldsymbol{Z}^{-}_{ \mathsf{e}},t+1}\big)$
\label{eq:algo-extrinsic-sigma-z-minus}
 \State $t \gets t + 1$\vspace{0.1cm} 
 \Until{$\Big(\big|\!\big|\boldsymbol{\widehat{U}}_{ \mathsf{p},t+1}^{+}-\boldsymbol{\widehat{U}}_{ \mathsf{p},t}^{+}\big|\!\big|^2_{\textrm{F}} + \big|\!\big|\boldsymbol{\widehat{V}}_{ \mathsf{p},t+1}^{+}-\boldsymbol{\widehat{V}}_{ \mathsf{p},t}^{+}\big|\!\big|^2_{\textrm{F}}\Big)$}
 \Statex \hspace{2cm}$\leq\xi\Big( \big|\!\big|\boldsymbol{\widehat{U}}_{ \mathsf{p},t}^{+}\big|\!\big|^2_{\textrm{F}} + \big|\!\big|\boldsymbol{\widehat{V}}_{ \mathsf{p},t}^{+}\big|\!\big|^2_{\textrm{F}}$\Big)~~\textsf{or}~~ \Big($t>T_\textrm{max}$\Big) \vspace{0.1cm}
\State \textbf{return} $\boldsymbol{\widehat{U}}_{ \mathsf{p}, T_{\textrm{max}}+1}^{+}$, $\boldsymbol{\widehat{V}}_{ \mathsf{p}, T_{\textrm{max}}+1}^{+}$
\end{algorithmic}
\end{multicols}
\end{algorithm*}

Before delving into the derivation details, we first introduce BiG-VAMP which runs iteratively according to the algorithmic steps of Algorithm~\ref{algo:big-vamp}. There, $t$ stands for the iteration index and subscripts $ \mathsf{p}$ and $ \mathsf{e}$ are used to distinguish ``posterior'' and ``extrinsic'' variables, respectively. As a visual reminder, we also use the hat symbol ``$~\,\widehat{}~\,$'' to refer to  mean values.   Moreover, Algorithm \ref{algo:big-vamp}  updates the means and precisions of all messages simultaneously (i.e., for all $i$ and $j$ at the same time).  For instance, at each iteration $t$, the entire matrix $\bm{B}_{\bm{U},t} ~\triangleq~ [\boldsymbol{b}_{\textrm{\textbf{{\rv{u}}}}, 1,t}, \boldsymbol{b}_{\textrm{\textbf{{\rv{u}}}}, 2,t},\ldots,\boldsymbol{b}_{\textrm{\textbf{{\rv{u}}}}, N,t}]^{\mathsf{T}}$ is updated where $\{\boldsymbol{b}_{\textrm{\textbf{{\rv{u}}}}, i,t}\}_{i=1}^{N}$ is the $t^{th}$ update of the message pertaining to the $\{i^{th}\}_{i=1}^{N}$   variable node $\{\textrm{\textbf{{\rv{u}}}}_i\}_{i=1}^{N}$.  For better illustration, the block diagram of Algorithm~\ref{algo:big-vamp} is depicted in Fig. \ref{fig:block-diagram} whereby we show its different constituent blocks, namely the different denoisers as they interact with the so-called bi-LMMSE module through the extrinsic information (cf. Section \ref{subsec:BiGVAMP-gauss-approx} for more details). In the sequel, we first briefly discuss the ``low rank'' assumption used in \cite{matsushita2013low} and \cite{lesieur2015phase} which is no longer needed for the derivation of all BiG-VAMP messages. Then, we describe the Gaussian approximation of the extrinsic information as a key means to handle general priors on both $\boldsymbol{U}$ and $\boldsymbol{V}$ matrices in bilinear models. Finally, we extend the results to the generalized bilinear models.

\subsection{Bilinear Vector Approximate Message Passing (Bi-VAMP) with general rank}\label{subsec:BiGVAMP-gauss-approx}
In this case, the data matrix, $\boldsymbol{Y}$, is obtained from the  bilinear observation  model in (\ref{eq:bilinear-model}) with $\phi(.)$ being the identity, i.e., $\phi(x) ~= ~x, ~\forall\, x\in \mathbb{R}$. That is to say:
\begin{equation}  
\boldsymbol{Y}~=~  \bm{U}\bm{V}^{\mathsf{T}}\,+\,\boldsymbol{W},
\end{equation}
where the noise components, $w_{ij}$, are mutually independent and Gaussian distributed with zero mean and variance $\gamma_w^{-1}$. All the algorithmic steps of Bi-VAMP  which will be explained in this section are summarized in Algorithm \ref{algo:big-vamp} after excluding the update equations pertaining to the ``generalized output step'' (i.e., lines \ref{eq:algo-posterior-mean-z}-\ref{eq:algo-extrinsic-sigma-z-minus}),  while replacing $\gamma_{\bm{Z}^+_{\mathsf{e}}}$ by $\gamma_w$ and $\widehat{\bm{Z}}^+_{\mathsf{e}}$ by $\bm{Y}$.
\par
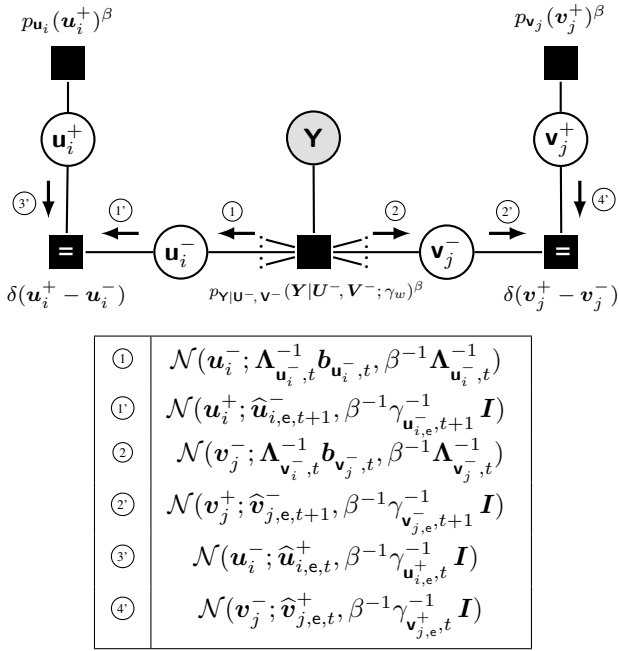
\begin{figure}
\centering
\begin{tikzpicture}[thick,scale=1.0, every node/.style={scale=1.0}]
  \node[latent]  (ui2)   {$\textrm{\textbf{{\rv{u}}}}_{i}^-$};
  \node[latent, xshift=-1.5cm, yshift=1.5cm]  (ui1)   {$\textrm{\textbf{{\rv{u}}}}_{i}^+$};
  
  \node[latent, right=of ui2, xshift=1.8cm]  (vj2)   {$\textrm{\textbf{{\rv{v}}}}_{j}^-$};
  \node[latent,  xshift=5.05cm, yshift=1.5cm]  (vj1)   {$\textrm{\textbf{{\rv{v}}}}_{j}^+$};
  
  
  
  \node[latent, above=of fij, xshift=2.25cm, yshift=1.4cm, fill=light-gray]  (yij)   {$\bm{\mathsf{Y}}$};
  
  \node (v1) at ($(vj2) + (-1,0.2)$) {};
  \node (vM) at ($(vj2) + (-1,-0.2)$) {};
  \node (u11) at ($(ui2) + (1,0.2)$) {};
  \node (uN1) at ($(ui2) + (1,-0.2)$) {};
  
  \factor[right=of ui2, xshift=0.75cm, xscale=2.5, yscale=2.5] {fij}  {below:  \scalebox{0.8}{ $ p_{\textrm{{\rv{Y}}} | \textrm{\textbf{{\rv{U}}}}^-\!,\, \textrm{\textbf{{\rv{V}}}}^-}(\bm{Y} | \boldsymbol{U}^-\!, \boldsymbol{V}^-;\gamma_w)^{\beta}$}
  } {ui2, vj2, yij, v1, vM, u11, uN1}{};
  
  \node at ($(fij) + (0.7,0)$) [rotate=90] {\yourtightDots};
  \node at ($(fij) + (-0.7,0)$) [rotate=-90] {\yourtightDots};
  
  \factor[left=of ui2, xshift=-0.5cm, xscale=2.5, yscale=2.5] {eq_ui1_ui2}  {below: $\delta(\boldsymbol{u}_{i}^+ - \boldsymbol{u}_{i}^-)$ } {ui2, ui1}{};
  
  \factor[right=of vj2, xshift=0.5cm, xscale=2.5, yscale=2.5] {eq_vj1_vj2}  {below:  $\delta(\boldsymbol{v}_{j}^+ - \boldsymbol{v}_{j}^-)$} {vj2, vj1}{};
  
  \factor[above=of ui1, xscale=2.5, yscale=2.5] {pui1} {above:$p_{\textrm{\textbf{{\rv{u}}}}_{i}}(\bm{u}_i^+)^{\beta}$} {ui1} {};
  
  \factor[above=of vj1, xscale=2.5, yscale=2.5] {pvj1} {above:$p_{\textrm{\textbf{{\rv{v}}}}_{j}}(\bm{v}_j^+)^{\beta}$} {vj1} {};

  \node (f_v2_right) at ($(vj2)!0.5!(fij) + (-0.1,0.25)$) {};
  \node (f_v2_left) at ($(vj2)!0.5!(fij) + (0.4,0.25)$) {};
  \draw [-latex,very thick] (f_v2_right.center) -- (f_v2_left.center);
  \node[right=of f_v2_left, xshift=-1.6cm, yshift=0.3cm] (id_f_v2) {\protect\tikz[inner sep=.25ex,baseline=-.75ex] \protect\node[circle,draw] {{\tiny 2}};};

  
  \node (f_u2_right) at ($(ui2)!0.5!(fij) + (0.1,0.25)$) {};
  \node (f_u2_left) at ($(ui2)!0.5!(fij) + (-0.4,0.25)$) {};
  \draw [-latex,very thick] (f_u2_right.center) -- (f_u2_left.center);
  \node[above=of f_u2_left, xshift=0.2cm, yshift=-1.225cm] (id_f_u2) {\protect\tikz[inner sep=.25ex,baseline=-.75ex] \protect\node[circle,draw] {{\tiny 1}};};
  
  
   \node (d_u1_top) at ($(ui2)!0.5!(eq_ui1_ui2) + (-0.3,0.25)$) {};
  \node (d_u1_bottom) at ($(ui2)!0.5!(eq_ui1_ui2) + (0.2,0.25)$) {};
  \draw [-latex,very thick] (d_u1_bottom.center) -- (d_u1_top.center);
  \node[pos=0.9,above=of d_u1_top, xshift=-0.8cm, yshift=-0.85cm] (id_d_u1) {\protect\tikz[inner sep=.25ex,baseline=-.75ex] \protect\node[circle,draw] {{\tiny 1'}};};
  
  \node (u1_d_top) at ($(ui1)!0.5!(eq_ui1_ui2) + (-0.25,0.2)$) {};
  \node (u1_d_bottom) at ($(ui1)!0.5!(eq_ui1_ui2) + (-0.25,-0.3)$) {};
  \draw [-latex,very thick] (u1_d_top.center) -- (u1_d_bottom.center);
  \node[left=of u1_d_bottom, xshift=1.1cm, yshift=0.2cm] (id_u1_d) {\protect\tikz[inner sep=.25ex,baseline=-.75ex] \protect\node[circle,draw] {{\tiny 3'}};};
  
  \node (d_v1_top) at ($(vj2)!0.5!(eq_vj1_vj2) + (0.3,0.25)$) {};
  \node (d_v1_bottom) at ($(vj2)!0.5!(eq_vj1_vj2) + (-0.2,0.25)$) {};
  \draw [-latex,very thick] (d_v1_bottom.center) -- (d_v1_top.center);
  \node[above=of d_v1_top, xshift=-0.25cm, yshift=-1.25cm] (id_d_v1) {\protect\tikz[inner sep=.25ex,baseline=-.75ex] \protect\node[circle,draw] {{\tiny 2'}};};
  
  \node (v1_d_top) at ($(vj1)!0.5!(eq_vj1_vj2) + (0.25,0.2)$) {};
  \node (v1_d_bottom) at ($(vj1)!0.5!(eq_vj1_vj2) + (0.25,-0.3)$) {};
  \draw [-latex,very thick] (v1_d_top.center) -- (v1_d_bottom.center);
  \node[right=of v1_d_bottom, yshift=0.25cm, xshift=-1.1cm] (id_u1_d) {\protect\tikz[inner sep=.25ex,baseline=-.75ex] \protect\node[circle,draw] {{\tiny 4'}};};
  
  \node[latent, below=of ui1, draw=none, xshift=-0.025cm,yshift=.22cm,opacity=0, text opacity=1]  (eq)   {\textcolor{white}{\textbf{=}}};
  \node[latent, below=of vj1, draw=none, xshift=0cm,yshift=.22cm,opacity=0, text opacity=1]  (eq)   {\textcolor{white}{\,\textbf{=}}};
  \node[latent, right=of ui2, xshift=0.04cm, draw=none,,opacity=0, text opacity=1]  (eq)   {\textcolor{white}{\textbf{ }}};

\end{tikzpicture}
\begin{center}
\begin{tabular}{ |c|c| } 
 \hline
 \protect\tikz[inner sep=.25ex,baseline=-.75ex] \protect\node[circle,draw] {{\tiny 1}}; & $\mathcal{N}(\boldsymbol{u}_{i}^-;  \mathbf{\Lambda}_{\textrm{\textbf{{\rv{u}}}}_{i}^-,t}^{-1}\boldsymbol{b}_{\textrm{\textbf{{\rv{u}}}}_{i}^-,t},\beta^{-1}\mathbf{\Lambda}^{-1}_{\textrm{\textbf{{\rv{u}}}}_{i}^-,t})$ \Tstrut\\ 
 \protect\tikz[inner sep=.25ex,baseline=-.75ex] \protect\node[circle,draw] {{\tiny 1'}}; & $\mathcal{N}(\boldsymbol{u}_{i}^+; {{}\widehat{\boldsymbol{u}}}^{-}_{i,\mathsf{e},t+1},\beta^{-1}\gamma^{-1}_{\textrm{\textbf{{\rv{u}}}}_{i,\mathsf{e}}^-,t+1}\, \boldsymbol{I})$ \Tstrut\\
  \protect\tikz[inner sep=.25ex,baseline=-.75ex] \protect\node[circle,draw] {{\tiny 2}}; & $\mathcal{N}(\boldsymbol{v}_{j}^-;\mathbf{\Lambda}_{\textrm{\textbf{{\rv{v}}}}_{i }^-,t}^{-1}\boldsymbol{b}_{\textrm{\textbf{{\rv{v}}}}_{ j }^-,t}, \beta^{-1}\mathbf{\Lambda}^{-1}_{\textrm{\textbf{{\rv{v}}}}_{j }^-,t})$ \\ 
 \protect\tikz[inner sep=.25ex,baseline=-.75ex] \protect\node[circle,draw] {{\tiny 2'}}; &  $\mathcal{N}(\boldsymbol{v}_{j}^+;{{}\widehat{\boldsymbol{v}}}^{-}_{j,\mathsf{e},t+1}, \beta^{-1}\gamma^{-1}_{\textrm{\textbf{{\rv{v}}}}_{j,\mathsf{e}}^-,t+1}\, \boldsymbol{I})$ \Tstrut\\
 \protect\tikz[inner sep=.25ex,baseline=-.75ex] \protect\node[circle,draw] {{\tiny 3'}}; & $\mathcal{N}(\boldsymbol{u}_{i}^-; {{}\widehat{\boldsymbol{u}}}^{+}_{i,\mathsf{e},t}, \beta^{-1}\gamma^{-1}_{\textrm{\textbf{{\rv{u}}}}_{i,\mathsf{e}}^+,t}\, \boldsymbol{I})$ \Tstrut\\
 \protect\tikz[inner sep=.25ex,baseline=-.75ex] \protect\node[circle,draw] {{\tiny 4'}}; & $\mathcal{N}(\boldsymbol{v}_{j}^-; {{}\widehat{\boldsymbol{v}}}^{+}_{j,\mathsf{e},t}, \beta^{-1}\gamma^{-1}_{\textrm{\textbf{{\rv{v}}}}_{j,\mathsf{e}}^+,t}\, \boldsymbol{I})$\Tstrut \Bstrut\\
 \hline
\end{tabular}
\end{center}
\caption{Factor graph under generalized priors on $\boldsymbol{U}$ and $\boldsymbol{V}$ along with the Gaussian approximations for the extrinsic information (as reflected by the index $\mathsf{e}$) that handles the equality constraints.}
\label{fig:factor-graph-general-priors}
\end{figure}

 To sidestep the problem of computing the intractable integrals in (\ref{function_f_u}) and (\ref{function_f_v}) during the evaluation of the mean and variance of messages \protect\tikz[inner sep=.25ex,baseline=-.75ex] \protect\node[circle,draw] {\footnotesize 3}; and \protect\tikz[inner sep=.25ex,baseline=-.75ex] \protect\node[circle,draw] {\footnotesize 4}; in Fig. \ref{fig:factor-graph-y_uv}, we proceed as follows. We rewrite the original posterior factorization in (\ref{eq:p-uv-given-y}) by splitting $\textrm{\textbf{{\rv{u}}}}_i$ (resp. $\textrm{\textbf{{\rv{v}}}}_j$) into two identical variables with equality constraints in between, i.e., $\textrm{\textbf{{\rv{u}}}}_{i}^+ = \textrm{\textbf{{\rv{u}}}}_{i}^-$ (resp. $\textrm{\textbf{{\rv{v}}}}_{j}^+ = \textrm{\textbf{{\rv{v}}}}_{j}^-$), thereby yielding the following equivalent factorization:
\begin{eqnarray}
\label{eq:p-uv-given-y-with-equality-constraints}
p_{\textrm{\textbf{{\rv{u}}}}_{i}^+, \textrm{\textbf{{\rv{u}}}}_{i}^-, \textrm{\textbf{{\rv{v}}}}_{j}^+, \textrm{\textbf{{\rv{v}}}}_{j}^- |  \bm{\mathsf{Y}}}\left(\boldsymbol{u}_{i}^+, \boldsymbol{u}_{i}^-, \boldsymbol{v}_{j}^+, \boldsymbol{v}_{j}^- | \bm{Y} ; \gamma_w^{-1}, \beta\right)&&\nonumber\\
&&\!\!\!\!\!\!\!\!\!\!\!\!\!\!\!\!\!\!\!\!\!\!\!\!\!\!\!\!\!\!\!\!\!\!\!\!\!\!\!\!\!\!\!\!\!\!\!\!\!\!\!\!\!\!\!\!\!\!\!\!\!\!\!\!\!\!\!\!\!\!\!\!\!\!\!\!\!\propto \,
 \prod_{i=1}^{N}\prod_{j=1}^{M}p_{\textrm{{\rv{y}}}_{ij} | \textrm{\textbf{{\rv{u}}}}_{i}^-, \textrm{\textbf{{\rv{v}}}}_{j}^-}\left(y_{ij} | \boldsymbol{u}_{i}^-, \boldsymbol{v}_{j}^-;\gamma_w^{-1}\right)^{\beta}\nonumber\\
&&\!\!\!\!\!\!\!\!\!\!\!\!\!\!\!\!\!\!\!\!\!\!\!\!\!\! \times\,\delta(\boldsymbol{u}_{i}^--\boldsymbol{u}_{i}^+)p_{\textrm{\textbf{{\rv{u}}}}}(\boldsymbol{u}_{i}^+)^{\beta}\nonumber\\
 && \!\!\!\!\!\!\!\!\!\!\!\!\!\!\times\,\delta(\boldsymbol{v}_{j}^--\boldsymbol{v}_{j}^+)  
 \, p_{\textrm{\textbf{{\rv{v}}}}}(\boldsymbol{v}_{j}^+)^{\beta}\!.\nonumber\\
\end{eqnarray}
The new factorization in (\ref{eq:p-uv-given-y-with-equality-constraints}) transforms the original factor graph in Fig. \ref{fig:factor-graph-y_uv} into the new one depicted in Fig.~\ref{fig:factor-graph-general-priors}. To handle the newly introduced equality constraints, the new variables $\textrm{\textbf{{\rv{u}}}}_{i}^+,~ \textrm{\textbf{{\rv{u}}}}_{i}^-$, $\textrm{\textbf{{\rv{v}}}}_{j}^+$, and $ \textrm{\textbf{{\rv{v}}}}_{j}^-$ are rather regarded as processing nodes which exchange scalar  messages/beliefs in the form of component-wise (i.e., decoupled) Gaussian densities. The beliefs provided by $\bm{\mathsf{u}}_i^-$ and $\bm{\mathsf{v}}_i^-$ on $\bm{\mathsf{u}}_i^+$ and $\bm{\mathsf{v}}_i^+$ (and vice versa) are known as the \textit{extrinsic information} and are modelled by the  Gaussian messages \protect\tikz[inner sep=.25ex,baseline=-.75ex] \protect\node[circle,draw] {\footnotesize 1'};, \protect\tikz[inner sep=.25ex,baseline=-.75ex] \protect\node[circle,draw] {\footnotesize 2'};, \protect\tikz[inner sep=.25ex,baseline=-.75ex] \protect\node[circle,draw] {\footnotesize 3'}; and \protect\tikz[inner sep=.25ex,baseline=-.75ex] \protect\node[circle,draw] {\footnotesize 4'};  in Fig.~\ref{fig:factor-graph-general-priors} whose parameters will be calculated later in this section.
\par The decoupling in the messages on each side of the equality nodes results in two simple types of denoising functions, namely $\mathbf{f}_{\mathsf{u}}(\cdot,\cdot)$ \big[resp. $\mathbf{f}_{\mathsf{v}}(\cdot,\cdot)$\big] and $\mathbf{g}_{\mathsf{u}}(\cdot,\cdot)$ \big[resp. $\mathbf{g}_{\mathsf{v}}(\cdot,\cdot)$\big] to recover $\boldsymbol{u}_{i}^-$ (resp. $\boldsymbol{v}_{j}^-$) and  $\boldsymbol{u}_{i}^+$ (resp. $\boldsymbol{v}_{j}^+$). Such decoupled message passing is made possible by ignoring the off-diagonal elements of the error covariance matrices calculated on the side of $\bm{\mathsf{u}}_{i}^-$ and $\bm{\mathsf{v}}_{j}^-$ nodes. Although the integrals of the denoising functions,  $\mathbf{f}_{\mathsf{u}}(\cdot,\cdot)$ and  $\mathbf{f}_{\mathsf{v}}(\cdot,\cdot)$, in (\ref{function_f_u}) and (\ref{function_f_v}) are still required for Bi-VAMP, they are now evaluated in closed form after replacing the actual priors, $\mathit{p}_{\textrm{\textbf{{\rv{u}}}}}(\cdot)$ and $\mathit{p}_{\textrm{\textbf{{\rv{v}}}}}(\cdot)$, with  the extrinsic Gaussian messages \protect\tikz[inner sep=.25ex,baseline=-.75ex] \protect\node[circle,draw] {\footnotesize 3'}; and \protect\tikz[inner sep=.25ex,baseline=-.75ex] \protect\node[circle,draw] {\footnotesize 4'};, respectively.
Messages \protect\tikz[inner sep=.25ex,baseline=-.75ex] \protect\node[circle,draw] {\footnotesize 1}; and \protect\tikz[inner sep=.25ex,baseline=-.75ex] \protect\node[circle,draw] {\footnotesize 2}; are updated in lines  \ref{eq:algo-Bu-non-low-rank}--\ref{eq:algo-Av-non-low-rank} of Algorithm \ref{algo:big-vamp} in a matrix form (i.e., $\boldsymbol{B}_{\boldsymbol{U},t}$, $\boldsymbol{B}_{\boldsymbol{V},t}$, $\boldsymbol{\Lambda}_{\boldsymbol{U},t}$, and $\boldsymbol{\Lambda}_{\boldsymbol{V},t}$) as discussed in Section \ref{sec:background-low-rank}.  
%
%
%
%
Note, however, that Bi-VAMP avoids the approximation in (\ref{eq:low-rank-assumption}) which is valid for the very-low-rank structure  only. To that end, Bi-VAMP incorporates explicitly the contribution of all $y_{k,j}^2$ in (\ref{eq:message_from_z_to_u_related_terms_broadcast_sub_eq_2}) and (\ref{eq:message_from_z_to_v_related_terms_broadcast_sub_eq_2}) during the computation of the broadcast precision matrices $\boldsymbol{\Lambda}_{\boldsymbol{U},t}$ and $\boldsymbol{\Lambda}_{\boldsymbol{V},t}$, respectively. 
\par The posterior estimates of $\boldsymbol{u}_{i}^-$ and $\boldsymbol{v}_{j}^-$ and their error covariance matrices are updated in closed forms  as shown in lines \ref{eq:posterior-covariance-u-gaussian-approx}--\ref{eq:posterior-mean-v-gaussian-approx} of Algorithm~\ref{algo:big-vamp}. The resulting regularized matrix inverse structure therein suggests that the updates in lines \ref{eq:algo-Bu-non-low-rank}--\ref{eq:posterior-mean-v-gaussian-approx} are in essence performing approximate bi-LMMSE recovery of $\bm{U}$ and $\bm{V}$ under Gaussian prior information. Based on the i.i.d. assumption, we  further reduce the  posterior covariance matrices  to common scalar variances obtained by simply averaging their diagonal entries  (see lines \ref{eq:algo-diag-sigma-u} and \ref{eq:algo-diag-sigma-v} in Algorithm \ref{algo:big-vamp}) while ignoring the off-diagonal part.
%
%
%
%
%
As a result, the posterior covariance matrices are given by the  scaled identities, $\gamma^{-1}_{\boldsymbol{U}^{-}_{ \mathsf{p}},t+1}\,\boldsymbol{I}$ and $\gamma^{-1}_{\boldsymbol{V}^{-}_{ \mathsf{p}},t+1}\,\boldsymbol{I}$.
 Such approximation of  messages by their means and scalar variances is a common practice in the message passing paradigm, also known as expectation propagation (EP) principle\footnote{In error control codes literature, the concept of EP is also know as the turbo principle.}. 
After computing the posterior estimates and the  associated common precision (i.e., $\gamma_{\boldsymbol{U}^{-}_{ \mathsf{p}},t+1}$ and $\gamma_{\boldsymbol{V}^{-}_{ \mathsf{p}},t+1}$), each processing node  $\bm{\mathsf{u}}^-_i$  (resp. $\bm{\mathsf{v}}^-_j$)  subtracts  the contribution of its incoming extrinsic message \protect\tikz[inner sep=.25ex,baseline=-.75ex] \protect\node[circle,draw] {\footnotesize 3'}; \big(resp. \protect\tikz[inner sep=.25ex,baseline=-.75ex] \protect\node[circle,draw] {\footnotesize 4'};\big) before returning the extrinsic messages  \protect\tikz[inner sep=.25ex,baseline=-.75ex] \protect\node[circle,draw] {\footnotesize 1'};  \big(resp. \protect\tikz[inner sep=.25ex,baseline=-.75ex] \protect\node[circle,draw] {\footnotesize 2'};\big) to the other side of the equality node. Formally speaking, this amounts to approximating the posterior messages by  Gaussian distributions with the same means and variances \big(i.e., $\mathcal{N}(\bm{u}_i^+; \widehat{\bm{u}}^{-}_{i,\mathsf{p},t+1},\beta^{-1}\gamma_{\bm{\bm{U}}^-_{\mathsf{p}},t+1}^{-1}\bm{I})$ and $\mathcal{N}(\bm{v}_j^+; \widehat{\bm{v}}^{-}_{j,\mathsf{p},t+1},\beta^{-1}\gamma_{\bm{\bm{V}}^-_{\mathsf{p}},t+1}^{-1}\bm{I})$\big) and extracting the extrinsic messages  as follows:
\begin{eqnarray}
\mathcal{N}(\bm{u}_i^+;  \widehat{\bm{u}}^{-}_{i,\mathsf{e},t+1},\beta^{-1}\gamma_{\bm{U}^-_{\mathsf{e}},t+1}^{-1}\bm{I})&\propto&\frac{\mathcal{N}(\bm{u}_i^+; \widehat{\bm{u}}^{-}_{i,\mathsf{p},t+1},\frac{1}{\beta}\,\gamma_{\bm{U}^-_{\mathsf{p}},t+1}^{-1}\bm{I})}{\mathcal{N}(\bm{u}_i^+; \bm{\widehat{u}}^{+}_{i,\mathsf{e},t},\frac{1}{\beta}\gamma_{\bm{U}^+_{\mathsf{e}},t}^{-1}\bm{I})}, \nonumber\\ 
\mathcal{N}(\bm{v}_j^+;  \widehat{\bm{v}}^{-}_{j,\mathsf{e},t+1},\beta^{-1}\gamma_{\bm{V}^-_{\mathsf{e}},t+1}^{-1}\bm{I})&\propto&\frac{\mathcal{N}(\bm{v}_j^+; \widehat{\bm{v}}^{-}_{j,\mathsf{p},t+1},\frac{1}{\beta}\gamma_{\bm{V}^-_{\mathsf{p}},t+1}^{-1}\bm{I})}{\mathcal{N}(\bm{v}_j^+; \bm{\widehat{v}}^{+}_{j,\mathsf{e},t},\frac{1}{\beta}\gamma_{\bm{V}^+_{\mathsf{e}},t}^{-1}\bm{I})}. \nonumber
\end{eqnarray} 
By doing so, the extrinsic/posterior means and precisions are related as follows:
\begin{eqnarray}
\!\!\!\!\!\!\!\!\label{extrinsic_precision_u-}\gamma_{\bm{U}^-_{\mathsf{e}},t+1} &~=~& \gamma_{\bm{U}^-_{\mathsf{p}},t+1}\,-\,\gamma_{\bm{U}^+_{\mathsf{e}},t},\\
\!\!\!\!\!\!\!\!\label{extrinsic_precision_v-}\gamma_{\bm{V}^-_{\mathsf{e}},t+1} &~=~& \gamma_{\bm{V}^-_{\mathsf{p}},t+1}\,-\,\gamma_{\bm{V}^+_{\mathsf{e}},t},\\
\!\!\!\!\!\!\!\!\label{extrinsic_mean_u-}\widehat{\bm{u}}^{-}_{i,\mathsf{e},t+1}&~=~& \gamma_{\bm{U}^-_{\mathsf{e}},t+1}^{-1}\big(\gamma_{\bm{U}^-_{\mathsf{p}},t+1}\widehat{\bm{u}}^{-}_{i,\mathsf{p},t+1}\,+\,\gamma_{\bm{U}^+_{\mathsf{e}},t}\widehat{\bm{u}}^{+}_{i,\mathsf{e},t}\big),\\
\!\!\!\!\!\!\!\!\label{extrinsic_mean_v-}\widehat{\bm{v}}^{-}_{j,\mathsf{e},t+1}&~=~& \gamma_{\bm{V}^-_{\mathsf{e}},t+1}^{-1}\big(\gamma_{\bm{V}^-_{\mathsf{p}},t+1}\widehat{\bm{v}}^{-}_{j,\mathsf{p},t+1}\,+\,\gamma_{\bm{V}^+_{\mathsf{e}},t}\widehat{\bm{v}}^{+}_{j,\mathsf{e},t}\big).
\end{eqnarray}
These extrinsic precisions and means are updated in lines  \ref{eq:algo-diag-sigma-u-e},  \ref{eq:algo-diag-mean-u}, \ref{eq:algo-diag-sigma-v-e}, and \ref{eq:algo-diag-mean-v} of Algorithm \ref{algo:big-vamp}.
 Given this extrinsic information,
 the denoising functions, $\mathbf{g}_{\mathsf{u}}(\cdot,\cdot)$, and $\mathbf{g}_{\mathsf{v}}(\cdot,\cdot)$, used to estimate  $\boldsymbol{u}_{i}^+$ and $\boldsymbol{v}_{j}^+$, along with their respective divergences, $\mathbf{g_{\mathsf{u}}^{\prime}}(\cdot,\cdot)$ and $\mathbf{g_{\mathsf{v}}^{\prime}}(\cdot,\cdot)$ in lines \ref{denoiser_u}--\ref{eq:algo-diag-sigma-v-alpha} of Algorithm~\ref{algo:big-vamp}
 are given by:
\begin{figure*}[ht]
\centering
\begin{tikzpicture}[thick,scale=0.85, every node/.style={transform shape}]
  \node[block, fill=gray!15] (p_u) {\Large Denoiser \\\vspace {0.2 cm} $p_\textrm{\textbf{{\rv{U}}}}(\bm{U})$};
  \node[block, fill=blue!15, right= 5cm of p_u,text width=3.5cm] (z_uv) {\Large Bi-LMMSE\\\vspace {0.3 cm}$\delta\big(\mathbf{\textrm{\textbf{{\rv{Z}}}}-\textrm{\textbf{{\rv{U}}}}\textrm{\textbf{{\rv{V}}}}^{\top}}\big)$};
  \node[block, fill=green!15, right= 5cm of z_uv,minimum width=3.5cm, text width=2cm] (phi_z){\Large Denoiser\vspace{0.2cm} $p_{\textrm{\textbf{{\rv{Y}}}} | \textrm{\textbf{{\rv{Z}}}}}(\boldsymbol{Y} | \boldsymbol{Z})$};
  \node[blockV, fill=gray!15, above=5cm of z_uv] (p_v) {\Large Denoiser\\\vspace {0.2 cm} $p_\textrm{\textbf{{\rv{V}}}}(\bm{V})$};
  \node[blockExt,right=of p_u, xshift=0.4cm, yshift=-1.5cm] (ext_pu_to_z_uv) {$\mathrm{\textbf{ext}}$};
  \node[blockExt,left=of z_uv, xshift=-0.4cm, yshift=1.5cm] (ext_z_uv_to_p_u) {$\mathrm{\textbf{ext}}$};
  \node[blockExt,right=of z_uv, xshift=0.4cm, yshift=-1.5cm] (ext_z_uv_to_y) {$\mathrm{\textbf{ext}}$};
  \node[blockExt,left=of phi_z, xshift=-0.4cm, yshift=1.5cm] (ext_y_to_z_uv) {$\mathrm{\textbf{ext}}$};
  \node[blockExt,above=of z_uv, xshift=0.6cm, yshift=0.4cm] (ext_z_uv_to_v) {$\mathrm{\textbf{ext}}$};
  \node[blockExt,below=of p_v, xshift=-0.6cm, yshift=-0.4cm] (ext_v_to_z_uv) {$\mathrm{\textbf{ext}}$};
  \draw [-latex,very thick] ([yshift=-4.25em]p_u.east) -- 
  node [midway,below=0em,align=center ] { $\boldsymbol{\widehat{U}}_{ \textsf{p}}^{+}$}
  node [midway,below=1.6em,align=center ] {$\gamma_{\boldsymbol{U}^{+}_{ \textsf{p}}}$}
  (ext_pu_to_z_uv.west);
  \draw [-latex,very thick] (ext_pu_to_z_uv) --
  node [midway,below=0em,align=center ] { $\boldsymbol{\widehat{U}}_{ \textsf{e}}^{+}$}
  node [midway,below=1.6em,align=center ] {$\gamma_{\boldsymbol{U}^{+}_{ \textsf{e}}}$}
  ([yshift=-4.25em]z_uv.west)
  node [pos=0.25](ext_between_pu_z_uv){};
  \draw [-latex,very thick] ([yshift=-4.25em]z_uv.east) --
  node [midway,below=0em,align=center ] { $\boldsymbol{\widehat{Z}}_{ \textsf{p}}^{-}$}
  node [midway,below=1.6em,align=center ] {$\gamma_{\boldsymbol{Z}^{-}_{ \textsf{p}}}$}
  (ext_z_uv_to_y.west);
  \draw [-latex,very thick] (ext_z_uv_to_y) --
  node [midway,below=0em,align=center ] { $\boldsymbol{\widehat{Z}}_{ \textsf{e}}^{-}$}
  node [midway,below=1.7em,align=center ] {$\gamma_{\boldsymbol{Z}^{-}_{ \textsf{e}}}$}
  ([yshift=-4.25em]phi_z.west)
  node [pos=0.25](ext_between_z_uv_y){};
  \draw [-latex,very thick] ([yshift=4.25em]phi_z.west) --
  node [midway,above=0em,align=center ] { $\boldsymbol{\widehat{Z}}_{ \textsf{p}}^{+}$}
  node [midway,above=1.6em,align=center ] {$\gamma_{\boldsymbol{Z}^{+}_{ \textsf{p}}}$}
  (ext_y_to_z_uv.east);
  \draw [-latex,very thick] (ext_y_to_z_uv) --
  node [midway,above=0em,align=center ] { $\boldsymbol{\widehat{Z}}_{ \textsf{e}}^{+}$}
  node [midway,above=1.7em,align=center ] {$\gamma_{\boldsymbol{Z}^{+}_{ \textsf{e}}}$}
  ([yshift=4.25em]z_uv.east)
  node [pos=0.25](ext_between_y_z_uv){};
  \draw [-latex,very thick] ([yshift=4.25em]z_uv.west) --
  node [midway,above=0em,align=center ] { $\boldsymbol{\widehat{U}}_{ \textsf{p}}^{-}$}
  node [midway,above=1.7em,align=center ] {$\boldsymbol{R}_{\boldsymbol{U}_{ \textsf{p}}^{-}}$}
  (ext_z_uv_to_p_u.east);
  \draw [-latex,very thick] (ext_z_uv_to_p_u) --
  node [midway,above=0em,align=center ] { $\boldsymbol{\widehat{U}}_{ \textsf{e}}^{-}$}
  node [midway,above=1.7em,align=center ] {$\gamma_{\boldsymbol{U}^{-}_{ \textsf{e}}}$}
  ([yshift=4.25em]p_u.east)
  node [pos=0.25](ext_between_z_uv_pu){};
  \draw [-latex,very thick] ([xshift=1.75em]z_uv.north) --
  node [pos=0.6,right=0em,align=center ] { $\boldsymbol{\widehat{V}}_{ \textsf{p}}^{-}$}
  node [pos=0.2,right=0em,align=center ] {$\boldsymbol{R}_{\boldsymbol{V}_{ \textsf{p}}^{-}}$}
  (ext_z_uv_to_v.south);
  \draw [-latex,very thick] (ext_z_uv_to_v.north) --
  node [pos=0.6,right=0em,align=center ] { $\boldsymbol{\widehat{V}}_{ \textsf{e}}^{-}$}
  node [pos=0.4,right=0em,align=center ] {$\gamma_{\boldsymbol{V}^{-}_{ \textsf{e}}}$}
  ([xshift=1.75em]p_v.south)
  node [pos=0.25](ext_between_z_uv_pv){};
  \draw [-latex,very thick] ([xshift=-1.75em]p_v.south) --
  node [pos=0.6,left=0em,align=center ] { $\boldsymbol{\widehat{V}}_{ \textsf{p}}^{+}$}
  node [pos=0.25,left=0em,align=center ] {$\gamma_{\boldsymbol{V}^{+}_{ \textsf{p}}}$}
  (ext_v_to_z_uv.north);
  \draw [-latex,very thick] (ext_v_to_z_uv.south) --
  node [pos=0.6,left=0em,align=center ] { $\boldsymbol{\widehat{V}}_{ \textsf{e}}^{+}$}
  node [pos=0.4,left=0em,align=center ] {$\gamma_{\boldsymbol{V}^{+}_{ \textsf{e}}}$}
  ([xshift=-1.75em]z_uv.north)
  node [pos=0.25](ext_between_pv_z_uv){};
  
  \draw [-latex,very thick] (ext_between_pu_z_uv.center) --
  (ext_z_uv_to_p_u.south);
  \draw [-latex,very thick] (ext_between_z_uv_y.center) --
  (ext_y_to_z_uv.south);
  \draw [-latex,very thick] (ext_between_y_z_uv.center) --
  (ext_z_uv_to_y.north);
  \draw [-latex,very thick] (ext_between_z_uv_pu.center) --
  (ext_pu_to_z_uv.north);
  \draw [-latex,very thick] (ext_between_z_uv_pv.center) --
  (ext_v_to_z_uv.east);
  \draw [-latex,very thick] (ext_between_pv_z_uv.center) --
  (ext_z_uv_to_v.west);
\end{tikzpicture}
\caption{Block diagram of BiG-VAMP  with its four modules: two denoising  modules (MMSE or MAP) incorporating the prior information, $p_\textrm{\textbf{{\rv{U}}}}(\cdot)$ and $p_\textrm{\textbf{{\rv{V}}}}(\cdot)$, the approximate bi-LMMSE module,  
 and one output denoising module (MMSE or MAP) handling the observation model  $p_{\textrm{\textbf{{\rv{Y}}}} | \textrm{\textbf{{\rv{Z}}}}}(\boldsymbol{Y} | \boldsymbol{Z})$. The four modules exchange extrinsic information/messages through the \protect\tikz[inner sep=.25ex,baseline=-.75ex] \protect\node[rectangle,draw,thick,minimum width=0.45cm,minimum height=0.45cm] {\footnotesize \textbf{ext}}; blocks.}.
\label{fig:block-diagram}
\end{figure*}
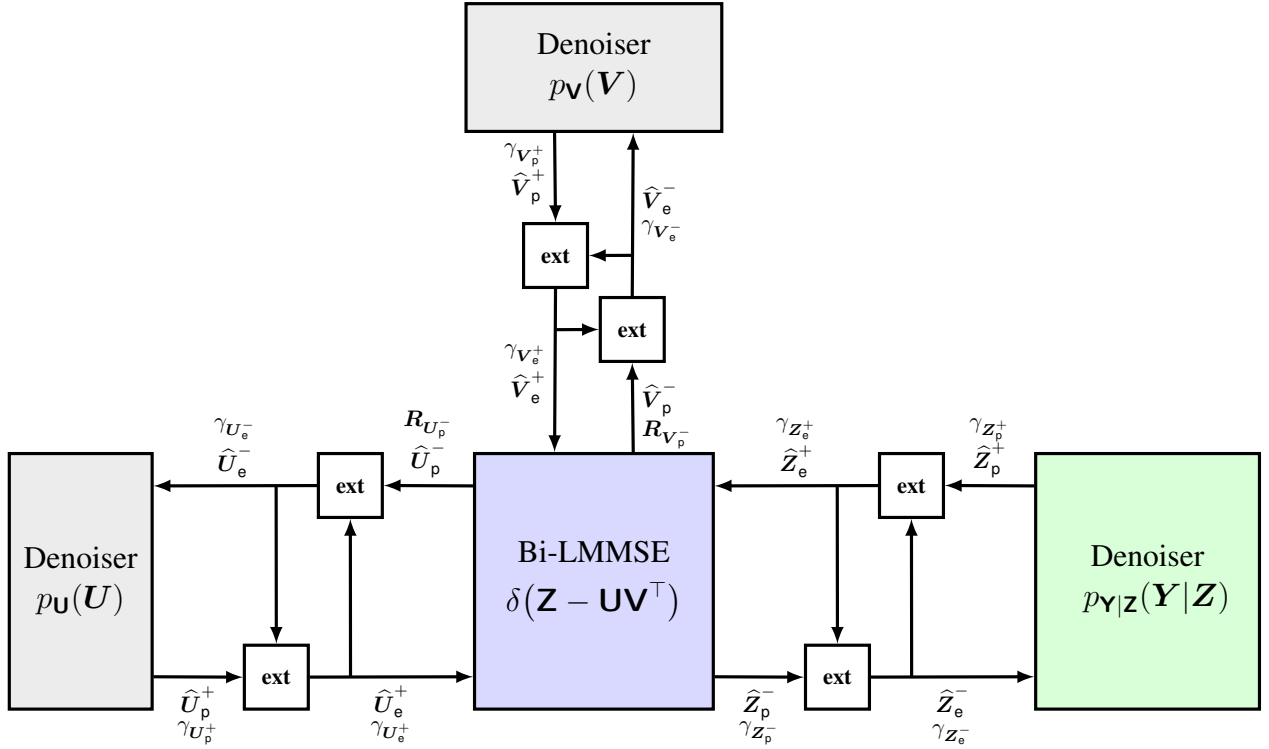
\begin{eqnarray}
\!\!\!\!\!\!\!\!\!\!\!\!\label{new_posterior_mean_u}\mathbf{g}_{\mathsf{u}}(\widehat{\bm{u}} , \gamma^{-1}_{\boldsymbol{U}}) &~=~& \frac{ \mathlarger{\int} \boldsymbol{u} \,p_{\textrm{\textbf{{\rv{u}}}}}\left(\boldsymbol{u}\right)^{\beta}\mathcal{N}(\boldsymbol{u}; \widehat{\bm{u}} ,\beta^{-1}\gamma^{-1}_{\boldsymbol{U}}\, \boldsymbol{I})\,  d\boldsymbol{u}}{\mathlarger{\int}  \,p_{\textrm{\textbf{{\rv{u}}}}}\left(\boldsymbol{u}\right)^{\beta}\mathcal{N}(\boldsymbol{u}; \widehat{\bm{u}} ,\beta^{-1}\gamma^{-1}_{\boldsymbol{U}}\, \boldsymbol{I})\,  d\boldsymbol{u}},\\
\!\!\!\!\!\!\!\!\!\!\!\!\label{new_posterior_mean_v}\mathbf{g}_{\mathsf{v}}( \widehat{\bm{v}} , \gamma^{-1}_{\boldsymbol{V}}) &~=~& \frac{ \mathlarger{\int} \boldsymbol{v} \,p_{\textrm{\textbf{{\rv{v}}}}}\left(\boldsymbol{v}\right)^{\beta}\mathcal{N}(\boldsymbol{v}; \widehat{\bm{v}} ,\beta^{-1}\gamma^{-1}_{\boldsymbol{V}}\, \boldsymbol{I})\,  d\boldsymbol{v}}{\mathlarger{\int}  \,p_{\textrm{\textbf{{\rv{v}}}}}\left(\boldsymbol{v}\right)^{\beta}\mathcal{N}(\boldsymbol{v}; \widehat{\bm{v}} ,\beta^{-1}\gamma^{-1}_{\boldsymbol{V}}\, \boldsymbol{I})\,  d\boldsymbol{v}},\\
\!\!\!\!\!\!\!\!\!\!\!\!\label{new_posterior_variance_u}\big[\mathbf{g}_{\mathsf{u}}^{\prime}(\widehat{\bm{u}} , \gamma^{-1}_{\boldsymbol{U}})\big]_{\ell} &~ =~& \frac{\partial\big[\mathbf{g}_{\mathsf{u}}( \widehat{\bm{u}}, \gamma^{-1}_{\boldsymbol{U}})\big]_{\ell}}{\partial\big[ \widehat{\bm{u}}\big]_{\ell}},~\ell=1,\ldots,r,\\
\!\!\!\!\!\!\!\!\!\!\!\!\label{new_posterior_variance_v}\big[\mathbf{g}_{\mathsf{v}}^{\prime}(\widehat{\bm{v}} , \gamma^{-1}_{\boldsymbol{V}})\big]_{\ell} & ~=~& \frac{\partial\big[\mathbf{g}_{\mathsf{v}}( \widehat{\bm{v}} , \gamma^{-1}_{\boldsymbol{V}})\big]_{\ell}}{\partial\big[ \widehat{\bm{v}}\big]_{\ell}},~~\ell=1,\ldots, r.
\end{eqnarray}
In essence,
$ \widehat{\bm{u}}^{+}_{i,\mathsf{p},t+1}=\mathbf{g}_{\mathsf{u}}( \widehat{\bm{u}}^{-}_{i,\mathsf{e},t+1} , \gamma^{-1}_{\boldsymbol{U}^{-}_{ \mathsf{e}},t+1})$ and $\widehat{\bm{v}}^{+}_{j,\mathsf{p},t+1} = \mathbf{g}_{\mathsf{v}}( \widehat{\bm{v}}^{-}_{j,\mathsf{e},t+1} , \gamma^{-1}_{\boldsymbol{V}^{-}_{ \mathsf{e}},t+1})$ are the posterior means of $\boldsymbol{\mathsf{u}}_{i}^+$ and $\boldsymbol{\mathsf{v}}_{j}^+$, respectively. In addition, their common posterior precisions updated in lines \ref{eq:algo-diag-sigma-u-alpha} and \ref{eq:algo-diag-sigma-v-alpha} of Algorithm \ref{algo:big-vamp} are given by: 
\begin{eqnarray}
\gamma_{\boldsymbol{U}^{+}_{ \mathsf{p}},t+1} & \,=\,& \gamma_{\boldsymbol{U}^{-}_{ \mathsf{e}},t+1} \, \left(\frac{1}{N}\sum_{i=1}^N\big\langle \mathbf{g}^{\prime}_{\mathsf{u}}(\widehat{\boldsymbol{u}}^{-}_{i,\mathsf{e},t+1}, \gamma^{-1}_{\boldsymbol{U}^{-}_{ \mathsf{e}},t+1})\big\rangle\right)^{-1}\hspace{-0.4cm},\\
\gamma_{\boldsymbol{V}^{+}_{ \mathsf{p}},t+1} &\,=\,& \gamma_{\boldsymbol{V}^{-}_{ \mathsf{e}},t+1} \,\left(\frac{1}{M}\sum_{j=1}^M\big\langle \mathbf{g}^{\prime}_{\mathsf{v}}(\widehat{\boldsymbol{v}}^{-}_{j,\mathsf{e},t+1}, \gamma^{-1}_{\boldsymbol{U}^{-}_{ \mathsf{e}},t+1})\big\rangle\right)^{-1}.
\end{eqnarray}
 Notice that unlike the multi-dimensional integrals in (\ref{function_f_u}) and (\ref{function_f_v}) which restrict the existing low-rank matrix recovery algorithms in \cite{matsushita2013low} and \cite{lesieur2015phase} to the case of Gaussian and community priors, all the one-dimensional integrals
 involved in (\ref{new_posterior_mean_u}) and (\ref{new_posterior_mean_v}) 
can be found analytically for almost all statistical priors of practical interest.   As one example, the intractable posterior mean of $\textrm{\textbf{{\rv{u}}}}_{i}$ in (\ref{function_f_u}) with a binary prior becomes straightforwardly
equal to $\mathbf{g}_{\mathsf{u}}( \widehat{\bm{u}}^{-}_{i,\mathsf{e},t+1} , \gamma^{-1}_{\boldsymbol{U}^{-}_{ \mathsf{e}},t+1})\,=\,\textrm{tanh}(\gamma_{\boldsymbol{U}^{-}_{ \mathsf{e}},t+1} \widehat{\bm{u}}^{-}_{i,\mathsf{e},t+1} )$ instead of summing over $2^r$ terms.  Finally, the extrinsic precisions  and means  for the messages  \protect\tikz[inner sep=.25ex,baseline=-.75ex] \protect\node[circle,draw] {\footnotesize 3'}; and \protect\tikz[inner sep=.25ex,baseline=-.75ex] \protect\node[circle,draw] {\footnotesize 4'}; are updated analogously to (\ref{extrinsic_precision_u-})-(\ref{extrinsic_mean_v-}) in lines \ref{eq:algo-gamma-Ue+}--\ref{eq:algo-mean-Ve+} of Algorithm~\ref{algo:big-vamp}.  
\subsection{From Bi-VAMP to BiG-VAMP}
In this section, we extend the Bi-VAMP algorithm introduced in Section
\ref{subsec:BiGVAMP-gauss-approx} to the generalized bilinear model given in (\ref{eq:non-linear-distribution0}) so as to complete the derivation of BiG-VAMP. To that end, we introduce the intermediate random matrix, $\textrm{\textbf{{\rv{Z}}}}\triangleq \textrm{\textbf{{\rv{U}}}}\,\textrm{\textbf{{\rv{V}}}}^{\top}$, whose $ij$th entry is given by  $\textrm{{\rv{z}}}_{ij}=\textrm{\textbf{{\rv{u}}}}_i^{\top} \textrm{\textbf{{\rv{v}}}}_j$.
 \begin{figure}[!h]
\centering
\begin{tikzpicture}[thick,scale=1.0, every node/.style={scale=1.0}]
  
  \node[latent]  (ui2)   {$\textrm{\textbf{{\rv{u}}}}_{i}^-$};
  \node[latent, xshift=-1cm, yshift=-1.5cm]  (ui1)   {$\textrm{\textbf{{\rv{u}}}}_{i}^+$};
  
  \node (u1extra) at ($(ui2) + (0.2,0.9)$) {};
  \node (uNextra) at ($(ui2) + (-0.2,0.9)$) {};
  \node at ($(ui2) + (0,0.85)$) {\mytightDots};
  \draw  (ui2.north) -- (u1extra);
  \draw  (ui2.north) -- (uNextra);
  
  \node[latent, right=of ui2, xshift=1.8cm]  (vj2)   {$\textrm{\textbf{{\rv{v}}}}_{j}^-$};
  \node[latent,  xshift=4.55cm, yshift=-1.5cm]  (vj1)   {$\textrm{\textbf{{\rv{v}}}}_{j}^+$};
  
  \node (v1extra) at ($(vj2) + (0.2,0.9)$) {};
  \node (vMextra) at ($(vj2) + (-0.2,0.9)$) {};
  \node at ($(vj2) + (0,0.85)$) {\mytightDots};
  \draw  (vj2.north) -- (v1extra);
  \draw  (vj2.north) -- (vMextra);
  
  \node[latent, right= of ui2, xshift=0cm, yshift=-3.7cm]  (zijp)   {$\textrm{{\rv{z}}}_{ij}^{+}$};
  
  \node[latent, right= of ui2, xshift=0cm, yshift=-1.5cm]  (zijm)   {$\textrm{{\rv{z}}}_{ij}^{-}$};
  
  \factor[right=of ui2, xshift=0.75cm, xscale=2.5, yscale=2.5] {fij}  {above:  $\delta(z_{ij}^--\boldsymbol{u}_{i}^{-\top}\,\boldsymbol{v}_{j}^-)$} {ui2, vj2, zijm}{};
  
  \factor[left=of ui2, xscale=2.5, yscale=2.5] {eq_ui1_ui2}  {} {ui2, ui1}{};
  
  \draw[] (eq_ui1_ui2) to node[pos=0.5,right=0.8em,align=center, xshift=-2cm, yshift=0.6cm]{$\footnotesize{\delta(\boldsymbol{u}_{i}^+ - \boldsymbol{u}_{i}^-)}$} (eq_ui1_ui2);
  
  \factor[right=of vj2, xscale=2.5, yscale=2.5] {eq_vj1_vj2}  {} {vj2, vj1}{};
  
  \draw[] (eq_vj1_vj2) to node[pos=0.5,right=0.8em,align=center, xshift=-0.75cm, yshift=0.6cm]{$\footnotesize{\delta(\boldsymbol{v}_{j}^+ - \boldsymbol{v}_{j}^-)}$} (eq_vj1_vj2);
  
  \factor[below=of zijm, xscale=2.5, yscale=2.5] {eq_zijp_zijm}  {left:  $\delta(z_{ij}^+ - z_{ij}^-)$} {zijp, zijm}{};
  
  \node[latent, left=of zijp, yshift=-1cm, fill=light-gray]  (yij)   {$y_{ij}$};
  \factor[below=of zijp, xscale=2.5, yscale=2.5, yshift=0cm] {yz-ij}  {right: $p_{\textrm{{\rv{y}}}_{ij} | \textrm{{\rv{z}}}_{ij}^{+}}(y_{ij}|z_{ij}^{+})^{\beta}$} {zijp, yij}{};
  
  \factor[below=of ui1, xscale=2.5, yscale=2.5] {pui1} {left:$p_{\textrm{\textbf{{\rv{u}}}}_{i}^+}(\bm{u}_i)^{\beta}$} {ui1} {};
  
  \factor[below=of vj1, xscale=2.5, yscale=2.5] {pvj1} {right:$p_{\textrm{\textbf{{\rv{v}}}}_{j}^+}(\bm{v}_j)^{\beta}$} {vj1} {};

  \node (v2_f_right) at ($(vj2)!0.5!(fij) + (0.2,-0.3)$) {};
  \node (v2_f_left) at ($(vj2)!0.5!(fij) + (-0.3,-0.3)$) {};
  \draw [-latex,very thick] (v2_f_right.center) -- (v2_f_left.center);
  \node[right=of v2_f_right, xshift=-1.6cm, yshift=-0.3cm] (id_v2_f) {\protect\tikz[inner sep=.25ex,baseline=-.75ex] \protect\node[circle,draw] {{\tiny 4}};};
  
  \node (u2_f_right) at ($(ui2)!0.5!(fij) + (0.4,-0.3)$) {};
  \node (u2_f_left) at ($(ui2)!0.5!(fij) + (-0.1,-0.3)$) {};
  \draw [-latex,very thick] (u2_f_left.center) -- (u2_f_right.center);
  \node[left=of u2_f_left, xshift=1.6cm, yshift=-0.3cm] (id_u2_f) {\protect\tikz[inner sep=.25ex,baseline=-.75ex] \protect\node[circle,draw] {{\tiny 3}};};
  
  \node (z_f_top) at ($(zijp)!0.5!(eq_zijp_zijm) + (0.3,0.1)$) {};
  \node (z_f_bottom) at ($(zijp)!0.5!(eq_zijp_zijm) + (0.3,-0.05)$) {};
  \draw [-latex,very thick] (z_f_bottom.south) -- (z_f_top.north);
  \draw[] (zijp) to node[pos=0.5,right=0.8em,align=center]{{\scriptsize \protect\tikz[inner sep=.25ex,baseline=-1.9ex] \protect\node[circle,draw] {\tiny 6'}; }  } (eq_zijp_zijm);

  \node (f_z_top) at ($(zijm)!0.5!(fij) + (-0.3,-1)$) {};
  \node (f_z_bottom) at ($(zijm)!0.5!(fij) + (-0.3,-1.65)$) {};
  \draw [-latex,very thick] (f_z_top.south) -- (f_z_bottom.north);
  \draw[] (eq_zijp_zijm) to node[pos=0.4,left=0.8em,align=center]{{\scriptsize \protect\tikz[inner sep=.25ex,baseline=.4ex] \protect\node[circle,draw] {\tiny 5'};}} (zijm);

  \node[latent, below=of zijm, draw=none, xshift=-0.015cm,yshift=.7cm,opacity=0, text opacity=1]  (eq)   {\textcolor{white}{\textbf{=}}};
  
  \node[latent, below=of zijp, draw=none, xshift=-0.015cm,yshift=.35cm,opacity=0, text opacity=1]  (eq)   {\textcolor{white}{\textbf{=}}};
  
  \node[latent, above=of zijm, draw=none, xshift=0.03cm,yshift=-0.25cm,opacity=0, text opacity=1]  (eq)   {\textcolor{white}{\textbf{}}};
  
  \node[latent, above=of vj1, draw=none, xshift=0cm,yshift=-0.23cm,opacity=0, text opacity=1]  (eq)   {\textcolor{white}{\textbf{=}}};
  
  \node[latent, above=of ui1, draw=none, xshift=-0.03cm,yshift=-0.23cm,opacity=0, text opacity=1]  (eq)   {\textcolor{white}{\textbf{=}}};
  
\end{tikzpicture}
\begin{center}
\begin{tabular}{ |c|c| } 
 \hline
 \protect\tikz[inner sep=.25ex,baseline=-.75ex] \protect\node[circle,draw] {{\tiny 5'}}; & $\mathcal{N}(z_{ij}^{-}; \widehat{z}^{-}_{ij,  \mathsf{e},t+1}, \beta^{-1}\gamma^{-1}_{\bm{Z}_{ \mathsf{e}}^{\,-},t+1})$ \Tstrut\\ 
 \protect\tikz[inner sep=.25ex,baseline=-.75ex] \protect\node[circle,draw] {{\tiny 6'}}; & $\mathcal{N}(z_{ij}^{+}; \widehat{z}^{+}_{ij,  \mathsf{e},t}, \beta^{-1}\gamma^{-1}_{\bm{Z}_{\mathsf{e}}^+,t})$ \Bstrut\\
 \hline
\end{tabular}
\end{center}
\caption{Factor graph for the generalized bilinear signal recovery problem}.
\label{fig:factor-graph-y_uv-non-linear}
\end{figure}
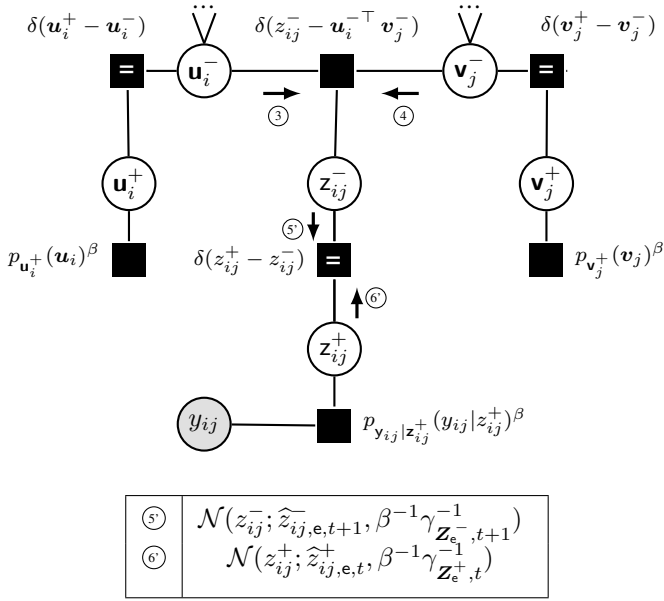
\noindent We again resort to the expectation propagation (aka turbo) principle to approximate the posterior messages \protect\tikz[inner sep=.25ex,baseline=-.75ex] \protect\node[circle,draw] {\footnotesize 5'}; and \protect\tikz[inner sep=.25ex,baseline=-.75ex] \protect\node[circle,draw] {\footnotesize 6'}; in Fig. \ref{fig:factor-graph-y_uv-non-linear} by  Gaussian distributions, $\mathcal{N}(z_{ij}^{-}; \widehat{z}^{-}_{ij,  \mathsf{e},t+1}, \beta^{-1}\gamma^{-1}_{\bm{Z}_{\mathsf{e}}^{\,-},t+1})$ and $\mathcal{N}(z_{ij}^{+}; \widehat{z}^{+}_{ij,  \mathsf{e},t}, \beta^{-1}\gamma^{-1}_{\bm{Z}_{\mathsf{e}}^+,t})$, respectively, whose means and variances are calculated in the sequel. To start with, by defining $\textrm{{\rv{z}}}_{ij}^-\,\triangleq\, \textrm{\textbf{{\rv{u}}}}_i^{-\top} \textrm{\textbf{{\rv{v}}}}_j^-$, the posterior mean and variance, $\widehat{z}^{+}_{ij,  \mathsf{p}}$ and $\gamma^{-1}_{\textrm{{\rv{z}}}_{ij,  \mathsf{p}}^+}$, of $\mathsf{z}_{ij}^+\,\triangleq\, \textrm{\textbf{{\rv{u}}}}_i^{+\top} \textrm{\textbf{{\rv{v}}}}_j^+$
under the scalar likelihood, $p_{\textrm{{\rv{y}}}_{ij}|\textrm{{\rv{z}}}_{ij}^+}(y_{ij}|z_{ij}^+)$,  are obtained
as follows:
\begin{eqnarray}
 \label{eq:message_posterior_from_z_to_u_v}\widehat{z}^{\,+}_{ij,  \mathsf{p},t+1}&~=~&   g_{\mathsf{z}}(y_{ij},\widehat{z}^{\,-}_{ij,  \mathsf{e},t+1}, \gamma^{-1}_{\bm{Z}_{\mathsf{e}}^-,t+1}),\\
 \gamma_{\textrm{{\rv{z}}}_{ij,  \mathsf{p}}^{\,+},t+1}^{-1}&=& \label{eq:message_posterior_from_z_to_u_v_variance}\gamma_{\bm{Z}_{\mathsf{e}}^{-},t+1}^{-1} \,\frac{\partial g_{\mathsf{z}}\big(y_{ij},\widehat{z}^{\,-}_{ij,  \mathsf{e},t+1}, \gamma^{-1}_{\bm{Z}_{\mathsf{e}}^-,t+1}\big)}{\partial \widehat{z}^{\,-}_{ij,  \mathsf{e},t+1}},
\end{eqnarray}
with $g_{\mathsf{z}}(y,\widehat{z},\gamma_z^{-1})$ being the following scalar denoising function: 
\begin{equation}
    g_{\mathsf{z}}(y,\widehat{z},\gamma_z^{-1})~=~ \frac{\int_{-\infty}^{+\infty} z\;\mathcal{N}(z; \widehat{z}, \beta^{-1}\gamma^{-1}_{z}) \;p_{\textrm{{\rv{y}}}|\textrm{{\rv{z}}}}(y|z)^{\beta}\; dz}{\int_{-\infty}^{+\infty} \;\mathcal{N}(z; \widehat{z}, \beta^{-1}\gamma^{-1}_{z}) \;p_{\textrm{{\rv{y}}}|\textrm{{\rv{z}}}}(y|z)^{\beta}\; dz}.
\end{equation}
%
%
Now, we detail the derivation of the scalar extrinsic message \protect\tikz[inner sep=.25ex,baseline=-.75ex] \protect\node[circle,draw] {\footnotesize 5'}; which is approximated by the Gaussian density\footnote{Note that by virtue of the CLT $\textrm{{\rv{z}}}_{ij}^-\triangleq \textrm{\textbf{{\rv{u}}}}_i^{-\top} \textrm{\textbf{{\rv{v}}}}_j^-$ (up to an appropriate scaling) converges to a Gaussian random variable in the large system limit. Hence, finding its first- and second-order statistics is enough to completely specify its distribution. Approximating its extrinsic information by a Gaussian density is thus equivalent to performing exact message passing.} $\mathcal{N}(z_{ij}^+; \widehat{z}^{-}_{ij,  \mathsf{e},t+1}, \beta^{-1}\gamma^{-1}_{\bm{Z}_{\mathsf{e}}^{\,-},t+1})$ with a common variance for all nodes. We start with the derivation of the individual variance, $\beta^{-1}\gamma^{-1}_{z_{ij,\mathsf{e}}^-,t+1}$, of $\textrm{{\rv{z}}}^{-}_{ij}$:
\begin{eqnarray}\label{eq:non-linear-prior-precision-0}
\beta^{-1}\gamma^{-1}_{\textrm{{\rv{z}}}_{ij, \mathsf{e}}^{\,-},t+1} &~=~ \mathbb{E}\Big[\big(\textrm{\textbf{{\rv{u}}}}_i^{-\top} \textrm{\textbf{{\rv{v}}}}_j^- - \mathbb{E}[\textrm{\textbf{{\rv{u}}}}_i^{-\top} \textrm{\textbf{{\rv{v}}}}_j^-]\big)^2 \Big].
\end{eqnarray}
In (\ref{eq:non-linear-prior-precision-0}), the expectation is taken with respect to the densities on $\bm{\mathsf{u}}_i^-$ and $\bm{\mathsf{v}}_j^-$ while assuming them to be independent.Those densities are given by messages \protect\tikz[inner sep=.25ex,baseline=-.75ex] \protect\node[circle,draw] {\footnotesize 3}; and \protect\tikz[inner sep=.25ex,baseline=-.75ex] \protect\node[circle,draw] {\footnotesize 4};, at iteration at iteration $t$,
whose first- and second-order statistics were already evaluated in (\ref{eq:posterior_mean_u_new})-(\ref{eq:posterior_variance_v_new}). Therefore, it follows from (\ref{eq:non-linear-prior-precision-0})  that:
\begin{equation}\label{eq:non-linear-prior-precision-1} 
\Scale[1]{
\!\!\!\!\begin{aligned}[b]
&\beta^{-1}\gamma^{-1}_{\textrm{{\rv{z}}}_{ij, \mathsf{e}}^{\,-},t+1} \\
&\,=\, \mathbb{E}\big[(\textrm{\textbf{{\rv{u}}}}_i^{-\top} \textrm{\textbf{{\rv{v}}}}^-_j)^2\big]\,-\, \big(\widehat{\boldsymbol{u}}_{i\rightarrow (i,j),t+1}^{\top} \widehat{\boldsymbol{v}}_{j\rightarrow (i,j),t+1}\big)^{2},\\
&\,=\,\mathbb{E}\Big[\textrm{Tr}\big(\textrm{\textbf{{\rv{u}}}}_i^-\,\textrm{\textbf{{\rv{u}}}}_i^{-\top} \textrm{\textbf{{\rv{v}}}}^-_j\,\textrm{\textbf{{\rv{v}}}}^{-\top}_j\big)\Big] \,-\, \big(\widehat{\boldsymbol{u}}_{i\rightarrow (i,j),t+1}^{\top} \widehat{\boldsymbol{v}}_{j\rightarrow (i,j),t+1}\big)^{2},\\
&\,=\,\textrm{Tr}\Big(\mathbb{E}\big[\textrm{\textbf{{\rv{u}}}}^-_i\,\textrm{\textbf{{\rv{u}}}}_i^{-\top}\big]\, \mathbb{E}\big[ \textrm{\textbf{{\rv{v}}}}^-_j\,\textrm{\textbf{{\rv{v}}}}_j^{-\top}\big]\Big) \,-\, \big(\widehat{\boldsymbol{u}}_{i\rightarrow (i,j),t+1}^{\top} \widehat{\boldsymbol{v}}_{j\rightarrow (i,j),t+1}\big)^{2}.
\end{aligned}
}
\end{equation}
Using the covriance identity, $\textrm{cov}(\textrm{\textbf{{\rv{x}}}}, \textrm{\textbf{{\rv{x}}}}) = \mathbb{E}[\textrm{\textbf{{\rv{x}}}}\,\textrm{\textbf{{\rv{x}}}}^{\top}] - \mathbb{E}[\textrm{\textbf{{\rv{x}}}} ]\,\mathbb{E}[\textrm{\textbf{{\rv{x}}}}]^{\top}$, for any random vector $\bm{\mathsf{x}}$,  (\ref{eq:non-linear-prior-precision-1}) is equivalent to:
\begin{eqnarray}\label{eq:non-linear-prior-precision-2}
\!\!\!\!\!\!\!\!\!\!\beta^{-1}\,\gamma^{-1}_{z_{ij,\mathsf{e}}^-,t+1}
&& \nonumber\\
\!\!\!\!\!\!\!\!\!\!&&\!\!\!\!\!\!\!\!\!\!\!\!\!\!\!\!\!\!\!\!\!\!\!\!\!\!\!\!\!=\,\textrm{Tr}\Big(\Big[\textrm{cov}(\textrm{\textbf{{\rv{u}}}}^-_i, \textrm{\textbf{{\rv{u}}}}^-_i) 
+ \widehat{\boldsymbol{u}}_{i\rightarrow (i,j),t+1} \, \widehat{\boldsymbol{u}}_{i\rightarrow (i,j),t+1}^{\top}\Big]\Big.\nonumber\\
\!\!\!\!\!\!\!\!\!\!&&\Big.\!\!\!\!\!\times\Big[\textrm{cov}(\textrm{\textbf{{\rv{v}}}}^-_j, \textrm{\textbf{{\rv{v}}}}^-_j) 
+ \widehat{\boldsymbol{v}}_{j\rightarrow (i,j),t+1} \, \widehat{\boldsymbol{v}}_{j\rightarrow (i,j),t+1}^{\top}\Big] \Big)\nonumber\\
\!\!\!\!\!\!\!\!\!\!&&\quad\quad~- \left(\widehat{\boldsymbol{u}}_{i\rightarrow (i,j),t+1}^{\top} \widehat{\boldsymbol{v}}_{j\rightarrow (i,j),t+1}\right)^{2}\!\!.
\end{eqnarray}
By recalling the fact that $\textrm{cov}(\textrm{\textbf{{\rv{u}}}}^-_i, \textrm{\textbf{{\rv{u}}}}^-_i) \,\triangleq \, \beta^{-1}\boldsymbol{R}_{\bm{u}_i,t+1}= \beta^{-1}\boldsymbol{R}_{\boldsymbol{U}_{ \mathsf{p}}^-,t+1}$ and  $\textrm{cov}(\textrm{\textbf{{\rv{v}}}}^-_j, \textrm{\textbf{{\rv{v}}}}^-_j)  \,\triangleq\,\beta^{-1}\boldsymbol{R}_{\bm{v}_j,t+1}= \beta^{-1}\boldsymbol{R}_{\boldsymbol{V}_{ \mathsf{p}}^-,t+1}$, with common $\boldsymbol{R}_{\boldsymbol{U}_{ \mathsf{p}}^-,t+1}$ and $\boldsymbol{R}_{\boldsymbol{V}_{ \mathsf{p}}^-,t+1}$ from line~\ref{eq:posterior-covariance-u-gaussian-approx} and \ref{eq:posterior-covariance-v-gaussian-approx} in Algorithm~\ref{algo:big-vamp}, 
it follows that:
\begin{eqnarray}\label{eq:non-linear-prior-precision-3}
\!\!\!\!\!\!\!\!\!\!\!\!\,\gamma^{-1}_{z_{ij,\mathsf{e}}^-,t+1}
&=&\textrm{Tr}\Big(\beta^{-1}\boldsymbol{R}_{\boldsymbol{U}_{ \mathsf{p}}^-,t+1}\boldsymbol{R}_{\boldsymbol{V}_{ \mathsf{p}}^-,t+1}\nonumber\\
&&
\quad\quad\,+\, \boldsymbol{R}_{\boldsymbol{V}_{ \mathsf{p}}^-,t+1}\widehat{\boldsymbol{u}}_{i\rightarrow (i,j),t+1} \, \widehat{\boldsymbol{u}}_{i\rightarrow (i,j),t+1}^{\top}\Big.\nonumber\\
\!\!\!\!\!\!\!\!\!\!\!\!&&\Big.~~\quad\quad\quad
+ \,\boldsymbol{R}_{\boldsymbol{U}_{ \mathsf{p}}^-,t+1}\widehat{\boldsymbol{v}}_{j\rightarrow (i,j),t+1} \, \widehat{\boldsymbol{v}}_{j\rightarrow (i,j),t+1}^{\top} \Big).
\end{eqnarray}
In (\ref{eq:non-linear-prior-precision-3}), we further replace $\widehat{\boldsymbol{u}}_{i\rightarrow (i,j),t+1}$ and $\widehat{\boldsymbol{v}}_{j\rightarrow (i,j),t+1}$ by their broadcast versions $\widehat{\boldsymbol{u}}^-_{i,\mathsf{p},t+1}$ and $\widehat{\boldsymbol{v}}^-_{j,\mathsf{p},t+1}$, respectively, while incurring  a negligible error  after ignoring terms of vanishing order as $M$ and $N$ grow large:   
\begin{eqnarray}\label{eq:non-linear-prior-precision-4}
\!\!\!\!\!\!\!\!\!\!\!\!\,\gamma^{-1}_{z_{ij,\mathsf{e}}^-,t+1}&\,=\,&
\textrm{Tr}\Big[\beta^{-1}\boldsymbol{R}_{\boldsymbol{U}_{ \mathsf{p}}^-,t+1}\boldsymbol{R}_{\boldsymbol{V}_{ \mathsf{p}}^-,t+1}\nonumber\\
&&\quad\quad\quad
\,+\, \boldsymbol{R}_{\boldsymbol{V}_{ \mathsf{p}}^-,t+1}\widehat{\boldsymbol{u}}^-_{i,\mathsf{p},t+1} \, \widehat{\boldsymbol{u}}_{i,\mathsf{p},t+1}^{-\top}\,\nonumber\\&\,\,&\quad\quad\quad\quad\quad\quad
+ \,\boldsymbol{R}_{\boldsymbol{U}_{ \mathsf{p}}^-,t+1}\widehat{\boldsymbol{v}}^-_{j,\mathsf{p},t+1} \, \widehat{\boldsymbol{v}}_{j,\mathsf{p},t+1}^{-\top} \Big].
\end{eqnarray}
As usually done in approximate message passing practices, we combine the individual variances, $\gamma^{-1}_{z_{ij,\mathsf{e}}^-,t+1}$, into one common variance, $\gamma_{\bm{Z}^-_{\mathsf{e}},t+1}^{-1}$, for all nodes:
\begin{equation}
\begin{aligned}
   \gamma_{\bm{Z}^-_{\mathsf{e}},t+1}^{-1} &~=~\frac{1}{MN} \sum_{i=1}^{N}\sum_{j=1}^{M} \gamma^{-1}_{z_{ij,\mathsf{e}}^-,t+1}\\
   &~\approx~ \textrm{Tr}\left(\beta^{-1}\boldsymbol{R}_{\boldsymbol{U}_{ \mathsf{p}}^-,t+1}\boldsymbol{R}_{\boldsymbol{V}_{ \mathsf{p}}^-,t+1}\right.\\
   &~~\quad\quad\quad\left.\,+\, \boldsymbol{R}_{\boldsymbol{V}_{ \mathsf{p}}^-,t+1}\left[\textstyle\frac{1}{N}\sum_{i=1}^N \widehat{\boldsymbol{u}}_{i,\mathsf{p},t+1} \, \widehat{\boldsymbol{u}}_{i,\mathsf{p},t+1}^{\top}\right]\right.\nonumber\\
\!\!\!\!\!\!\!\!\!\!\!\!&\left.\quad\quad\quad\quad\quad
+ \,\boldsymbol{R}_{\boldsymbol{U}_{ \mathsf{p}}^-,t+1}\left[\textstyle\frac{1}{M} \sum_{j=1}^M \widehat{\boldsymbol{v}}_{j,\mathsf{p},t+1} \,  \widehat{\boldsymbol{v}}_{j,\mathsf{p},t+1}^{\top}\right] \right),
\end{aligned}
\end{equation}
where terms of vanishing order are ignored as $M$ and $N$ grow large. This is used to find the posterior precision, $\gamma_{\bm{Z}_{\mathsf{p}}^-}$, as follows:
\begin{equation}\label{eq:non-linear-posterior-precision}
\gamma_{\bm{Z}_{\mathsf{p}}^-,t+1} ~=~ \gamma_{\bm{Z}_{\mathsf{e}}^+,t}~+~\gamma_{\bm{Z}^-_{\mathsf{e}},t+1}.
\end{equation}
Moreover,  the posterior mean, $\widehat{z}^{\,-}_{ij,  \mathsf{p},t+1}$ is evaluated as follows: 

\begin{equation} \label{eq:non-linear-posterior-mean}
\begin{aligned}
&\widehat{z}^{\,-}_{ij,  \mathsf{p},t+1}  \\
&~~=\, \gamma_{\bm{Z}_{\mathsf{p}}^-,t+1}^{-1}\,\Big(\gamma_{\bm{Z}_{\mathsf{e}}^+,t}\widehat{z}^{\,+}_{ij, \mathsf{e},t} + \gamma_{\bm{z}_{ij,\mathsf{e}}^-,t+1}\,\widehat{\boldsymbol{u}}_{i \rightarrow(i, j),t+1}^{\top}  \widehat{\boldsymbol{v}}_{j \rightarrow(i, j),t+1} \Big).
\end{aligned}
\end{equation}
Recall here the Osanger correction terms in (\ref{eq:posterior_mean_u}) and (\ref{eq:posterior_mean_v}) which lead to the following approximations of  $\widehat{\boldsymbol{u}}_{i \rightarrow(i, j),t+1}$ and $\widehat{\boldsymbol{v}}_{j \rightarrow(i, j),t+1}$:
\begin{subequations}\label{eq:approx-ui-vj}
\begin{align}
&\widehat{\boldsymbol{u}}_{i \rightarrow (i,j),t+1}~ \approx~ \widehat{\boldsymbol{u}}^-_{i,\mathsf{p},t+1}-\gamma_{\bm{Z}_{\mathsf{e}}^+,t} \,\widehat{z}_{i j, \mathsf{e},t}^{\,+}\,\boldsymbol{R}_{\boldsymbol{U}_{ \mathsf{p}}^-,t+1}\, \widehat{\boldsymbol{v}}^-_{j,\mathsf{p},t}\label{eq:approx-ui},\\
& \widehat{\boldsymbol{v}}_{j \rightarrow (i,j),t+1} ~\approx~ \widehat{\boldsymbol{v}}^-_{j,\mathsf{p},t+1}-\gamma_{\bm{Z}_{\mathsf{e}}^+,t} \,\widehat{z}_{i j, \mathsf{e},t}^{\,+} \,\boldsymbol{R}_{\boldsymbol{V}_{ \mathsf{p}}^-,t+1}\, \widehat{\boldsymbol{u}}^-_{i,\mathsf{p},t}.\label{eq:approx-vj}
\end{align}
\end{subequations}
These approximated messages are injected back into (\ref{eq:non-linear-posterior-mean}) thereby yielding the approximate posterior mean in (\ref{eq:non-linear-posterior-mean-approx}) on the top of the next page.
\begin{figure*}[!t]
\begin{subequations}
\label{eq:non-linear-posterior-mean-approx}
\begin{align}
\widehat{z}^{\,-}_{ij,  \mathsf{p},t+1}  &~\approx~ \gamma_{\bm{Z}_{\mathsf{p}}^-,t+1}^{-1}\,\bigg[{\gamma_{\bm{Z}_{\mathsf{e}}^+,t}\widehat{z}^{\,+}_{ij, \mathsf{e},t}} ~+~ 
\gamma_{z_{ij,\mathsf{e}}^-,t+1}
\Big(\widehat{\boldsymbol{u}}^-_{i,\mathsf{p},t+1}-\gamma_{\bm{Z}_{\mathsf{e}}^+,t} \,\widehat{z}_{i j, \mathsf{e},t}^{\,+}\, \boldsymbol{R}_{\boldsymbol{U}_{ \mathsf{p}}^-,t+1}\, \widehat{\boldsymbol{v}}^-_{j,\mathsf{p},t}\Big)^{\top} \Big( \widehat{\boldsymbol{v}}^-_{j,\mathsf{p},t+1}-\gamma_{\bm{Z}_{\mathsf{e}}^+,t} \,\widehat{z}_{i j, \mathsf{e},t}^{\,+} \,\boldsymbol{R}_{\boldsymbol{V}_{ \mathsf{p}}^-,t+1}\, \widehat{\boldsymbol{u}}^-_{i,\mathsf{p},t}\Big) \bigg],\\
&~=~ \gamma_{\bm{Z}_{\mathsf{p}}^-,t+1}^{-1}\,\bigg[{\gamma_{\bm{Z}_{\mathsf{e}}^+,t}\widehat{z}^{\,+}_{ij, \mathsf{e},t}} + 
\gamma_{z_{ij,\mathsf{e}}^-,t+1}
\Big( \widehat{\boldsymbol{u}}_{i,\mathsf{p},t+1}^{-\top}\,\widehat{\boldsymbol{v}}^-_{j,\mathsf{p},t+1}\,-\,\gamma_{\bm{Z}_{\mathsf{e}}^+,t}\, \widehat{z}_{i j, \mathsf{e},t}^{\,+} \,\widehat{\boldsymbol{u}}_{i,\mathsf{p},t+1}^{-\top} \,\boldsymbol{R}_{\boldsymbol{V}_{ \mathsf{p}}^-,t+1}\, \widehat{\boldsymbol{u}}^-_{i,\mathsf{p},t}\nonumber
\\&\hspace{4.4cm}\,-\,\gamma_{\bm{Z}_{\mathsf{e}}^+,t}\, \widehat{z}_{i j, \mathsf{e},t}^{\,+} \,\widehat{\boldsymbol{v}}_{j,\mathsf{p},t}^{-\top} \,\boldsymbol{R}_{\boldsymbol{U}_{ \mathsf{p}}^-,t+1}\, \widehat{\boldsymbol{v}}^-_{j,\mathsf{p},t+1}\,+\,(\gamma_{\bm{Z}_{\mathsf{e}}^+,t}\, \widehat{z}_{i j, \mathsf{e},t}^{\,+})^2 \,\widehat{\boldsymbol{v}}_{j,\mathsf{p},t}^{-\top}\boldsymbol{R}_{\boldsymbol{U}_{ \mathsf{p}}^-,t+1}\, \,\boldsymbol{R}_{\boldsymbol{V}_{ \mathsf{p}}^-,t+1}\, \widehat{\boldsymbol{u}}_{i,\mathsf{p},t}^-\Big)\bigg],\\
&~\approx~ \gamma_{\bm{Z}_{\mathsf{p}}^-,t+1}^{-1}\,\bigg[{\gamma_{\bm{Z}_{\mathsf{e}}^+,t}\widehat{z}^{\,+}_{ij, \mathsf{e},t}} + 
\gamma_{z_{ij,\mathsf{e}}^-,t+1}
\Big( \widehat{\boldsymbol{u}}_{i,\mathsf{p},t+1}^{-\top}\,\widehat{\boldsymbol{v}}^-_{j,\mathsf{p},t+1}\,-\,\gamma_{\bm{Z}_{\mathsf{e}}^+,t}\, \widehat{z}_{i j, \mathsf{e},t}^{\,+} \,\textrm{Tr}\big(\boldsymbol{R}_{\boldsymbol{V}_{ \mathsf{p}}^-,t+1}\, \widehat{\boldsymbol{u}}^-_{i,\mathsf{p},t}\widehat{\boldsymbol{u}}_{i,\mathsf{p},t+1}^{-\top}\big)\nonumber\\
&\hspace{11cm}\,-\,\gamma_{\bm{Z}_{\mathsf{e}}^+,t}\, \widehat{z}_{i j, \mathsf{e},t}^{\,+} \,\textrm{Tr}\big(\boldsymbol{R}_{\boldsymbol{U}_{ \mathsf{p}}^-,t+1}\, \widehat{\boldsymbol{v}}^-_{j,\mathsf{p},t+1}\widehat{\boldsymbol{v}}_{j,\mathsf{p},t}^{-\top}\big)\Big)\bigg],\label{eq:approx-cov-cov}\\
&~\approx~ \gamma_{\bm{Z}_{\mathsf{p}}^-,t+1}^{-1}\,\bigg[{\gamma_{\bm{Z}_{\mathsf{e}}^+,t}\widehat{z}^{\,+}_{ij, \mathsf{e},t}} + 
\gamma_{z_{ij,\mathsf{e}}^-,t+1}
\Big( \widehat{\boldsymbol{u}}_{i,\mathsf{p},t+1}^{-\top}\,\widehat{\boldsymbol{v}}^-_{j,\mathsf{p},t+1}\,-\,\gamma_{\bm{Z}_{\mathsf{e}}^+,t}\, \widehat{z}_{i j, \mathsf{e},t}^{\,+} \,\textrm{Tr}\big(\boldsymbol{R}_{\boldsymbol{V}_{ \mathsf{p}}^-,t+1}\, \widehat{\boldsymbol{u}}^-_{i,\mathsf{p},t+1}\widehat{\boldsymbol{u}}_{i,\mathsf{p},t+1}^{-\top}\big)\nonumber\\
&\hspace{10.5cm}\,-\,\gamma_{\bm{Z}_{\mathsf{e}}^+,t}\, \widehat{z}_{i j, \mathsf{e},t}^{\,+} \,\textrm{Tr}\big(\boldsymbol{R}_{\boldsymbol{U}_{ \mathsf{p}}^-,t+1}\, \widehat{\boldsymbol{v}}^-_{j,\mathsf{p},t+1}\widehat{\boldsymbol{v}}_{j,\mathsf{p},t+1}^{-\top}\big)\Big)\bigg],\label{eq:approx-cov-cov_t_t-1}
\\
&~=~ \gamma_{\bm{Z}_{\mathsf{p}}^-,t+1}^{-1}\,\bigg[{\gamma_{\bm{Z}_{\mathsf{e}}^+,t}\widehat{z}^{\,+}_{ij, \mathsf{e},t}} ~+~ 
\gamma_{z_{ij,\mathsf{e}}^-,t+1}
\Big( \widehat{\boldsymbol{u}}_{i,\mathsf{p},t+1}^{-\top}\,\widehat{\boldsymbol{v}}^-_{j,\mathsf{p},t+1}\,-\,\gamma_{\bm{Z}_{\mathsf{e}}^+,t}\,\gamma_{z_{ij,\mathsf{e}}^-,t+1}^{-1} \widehat{z}_{i j, \mathsf{e},t}^{\,+} \Big)\nonumber\\
&\hspace{9.5cm}~+~ \beta^{-1}\gamma_{\bm{Z}_{\mathsf{e}}^+,t}\gamma_{z_{ij,\mathsf{e}}^-,t+1}\, \widehat{z}_{i j, \mathsf{e},t}^{\,+} \,\textrm{Tr}\big(\,\boldsymbol{R}_{\boldsymbol{U}_{ \mathsf{p}}^-,t+1}\,\boldsymbol{R}_{\boldsymbol{V}_{ \mathsf{p}}^-,t+1}\big)\bigg].\label{eq:non-linear-posterior-mean-approx-final}
\end{align}
\end{subequations}
\rule{\textwidth}{0.4 pt}
\end{figure*}
The approximation in (\ref{eq:approx-cov-cov}) is a result of  dropping the last term which has a vanishing order as $M \rightarrow \infty$. The approximation in (\ref{eq:approx-cov-cov_t_t-1}) follows from the observation that one makes an error of vanishing order (as both $M$ and $N$ grow large) due to the fact that  $\widehat{\boldsymbol{u}}^-_{i,\mathsf{p},t}\widehat{\boldsymbol{u}}_{i,\mathsf{p},t+1}^{-\top}\,\approx\,\widehat{\boldsymbol{u}}^-_{i,\mathsf{p},t+1}\widehat{\boldsymbol{u}}_{i,\mathsf{p},t+1}^{-\top}$ and $\widehat{\boldsymbol{v}}^-_{j,\mathsf{p},t+1}\widehat{\boldsymbol{v}}_{j,\mathsf{p},t}^{-\top}\,\approx\,\widehat{\boldsymbol{v}}^-_{j,\mathsf{p},t+1}\widehat{\boldsymbol{v}}_{j,\mathsf{p},t+1}^{-\top}$. Then using (\ref{eq:non-linear-prior-precision-4}) in (\ref{eq:approx-cov-cov_t_t-1}) leads to (\ref{eq:non-linear-posterior-mean-approx-final}) in which we further replace $\gamma_{z_{ij,\mathsf{e}}^-,t+1}$ by $\gamma_{\bm{Z}_{\mathsf{e}}^-,t+1}$ thereby yielding:
\begin{equation} \label{eq:non-linear-posterior-mean-approx2}
\Scale[1]{
\begin{aligned}[b]
&\widehat{z}^{\,-}_{ij,  \mathsf{p},t+1}~=~\frac{\gamma_{\bm{Z}_{\mathsf{e}}^-,t+1}}{\gamma_{\bm{Z}_{\mathsf{p}}^-,t+1} } \Big(\widehat{\boldsymbol{u}}_{i,\mathsf{p},t+1}^{-\top} \widehat{\boldsymbol{v}}^-_{j,\mathsf{p},t+1} \\
&\hspace{2.8cm}+\beta^{-1}\gamma_{\bm{Z}_{ \mathsf{e}}^+,t}\widehat{z}^{\,+}_{ij, \mathsf{e},t}\, \textrm{Tr}\big(\boldsymbol{R}_{\boldsymbol{U}_{ \mathsf{p}}^-,t+1}\,\boldsymbol{R}_{\boldsymbol{V}_{ \mathsf{p}}^-,t+1} \big) \Big).
\end{aligned}
}
\end{equation}
After plugging the expression of $\gamma_{\bm{Z}_{\mathsf{e}}^-,t+1}$ as obtained from (\ref{eq:non-linear-posterior-precision}),  that is $\gamma_{\bm{Z}_{\mathsf{e}}^-,t+1} = \gamma_{\bm{Z}_{\mathsf{p}}^-,t+1}-\gamma_{\bm{Z}_{\mathsf{e}}^+,t}$, (\ref{eq:non-linear-posterior-mean-approx2}) becomes:
\begin{equation}
\Scale[1]{
\begin{aligned}[b]
\widehat{z}^{\,-}_{ij,  \mathsf{p},t+1} &~=~ \frac{\gamma_{\bm{Z}_{\mathsf{p}}^-,t+1}- \gamma_{\bm{Z}_{\mathsf{e}}^+,t}}{\gamma_{\bm{Z}_{\mathsf{p}}^-,t+1}} \Big(\widehat{\boldsymbol{u}}_{i,\mathsf{p},t+1}^{-\top} \widehat{\boldsymbol{v}}^-_{j,\mathsf{p},t+1} \\
&\hspace{1.7cm}\,+\beta^{-1} \gamma_{\bm{Z}_{\mathsf{e}}^+,t}\widehat{z}^{\,+}_{ij, \mathsf{e},t}\, \textrm{Tr}\big( \boldsymbol{R}_{\boldsymbol{U}_{ \mathsf{p}}^-,t+1}\,\boldsymbol{R}_{\boldsymbol{V}_{ \mathsf{p}}^-,t+1} \big) \Big).
\end{aligned}}
\end{equation}
Finally, we assume that $\gamma_{\bm{Z}_{\mathsf{p}}^-,t+1} \gg \gamma_{\bm{Z}_{\mathsf{e}}^+,t}$ which follows from the observation that the information on $z_{ij}=\bm{u}_i^{\top}\bm{v}_j$ that is brought by the strong structure on both $\bm{U}$ and $\bm{V}$ overwhelms the information brought by the observation $\bm{Y}$. This leads to:
\begin{equation}
\Scale[1]{
\begin{aligned}[b]
\widehat{z}^{\,-}_{ij,  \mathsf{p},t+1}&\,~\approx~\,  \widehat{\boldsymbol{u}}_{i,\mathsf{p},t+1}^{-\top} \widehat{\boldsymbol{v}}^-_{j,\mathsf{p},t+1}\\
&\hspace{1cm}+\beta^{-1}\gamma_{\bm{Z}_{\mathsf{e}}^+,t}\widehat{z}^{+}_{ij, \mathsf{e},t}\, \textrm{Tr}\big(\boldsymbol{R}_{\boldsymbol{U}_{ \mathsf{p}}^-,t+1}\,\boldsymbol{R}_{\boldsymbol{V}_{ \mathsf{p}}^-,t+1}\big) .
\end{aligned}}
\label{eq:non-linear-posterior-mean-final}
\end{equation}
In summary, (\ref{eq:non-linear-posterior-mean-final}) and (\ref{eq:non-linear-posterior-precision}) are the element-wise expressions of 
(\ref{eq:posterior_z_mean_batch}) and (\ref{eq:posterior_z_precision_batch}), respectively:
\begin{subequations}\label{eq:posterior_z}
\small{
\begin{align}
\boldsymbol{\widehat{Z}}_{ \mathsf{p},t+1}^{-} &~=~ \boldsymbol{\widehat{U}}_{ \mathsf{e},t+1}^{-}\,\boldsymbol{\widehat{V}}_{ \mathsf{e},t+1}^{-\top}+ \frac{\gamma_{\bm{Z}^{+}_{ \mathsf{e}},t}}{\beta}\,\boldsymbol{\widehat{Z}}_{ \mathsf{e},t}^{+}\,\text{Tr}\left(\boldsymbol{R}_{\boldsymbol{U}_{ \mathsf{p}}^-,t+1}\,\boldsymbol{R}_{\boldsymbol{V}_{ \mathsf{p}}^-,t+1}\right),\label{eq:posterior_z_mean_batch}\\
\gamma_{\bm{Z}^{-}_{ \mathsf{p}},t+1} &~=~\gamma_{\bm{Z}^{+}_{ \mathsf{e}},t}+MN\,\text{Tr}\bigg(\frac{M\,N}{\beta}\,\boldsymbol{R}_{\boldsymbol{U}_{ \mathsf{p}}^-,t+1}\,\boldsymbol{R}_{\boldsymbol{V}_{ \mathsf{p}}^-,t+1} \nonumber\\
&\hspace{3.1cm}+ N\, \boldsymbol{R}_{\boldsymbol{U}_{ \mathsf{p}}^-,t+1} \nonumber\, \boldsymbol{\widehat{V}}_{ \mathsf{p},t+1}^{-\top}\, \boldsymbol{\widehat{V}}_{ \mathsf{p},t+1}^{-}\\
&\hspace{3.5cm}+ M\, \boldsymbol{R}_{\boldsymbol{V}_{ \mathsf{p}}^{-},t+1}\, \boldsymbol{\widehat{U}}_{ \mathsf{p},t+1}^{-\top}\, \boldsymbol{\widehat{U}}^{-}_{ \mathsf{p},t+1}\bigg)^{-1}\hspace{-0.3cm}, \label{eq:posterior_z_precision_batch}
\end{align}
}%
\end{subequations}
\noindent which correspond to lines \ref{eq:algo-posterior-mean-z}--\ref{eq:algo-posterior-sigma-z} in Algorithm~\ref{algo:big-vamp}.
The  extrinsic values, $\widehat{z}^{-}_{ij,  \mathsf{e},t+1}$ and $\gamma^{-1}_{\bm{Z}_{\mathsf{e}}^-,t+1}$, for message \protect\tikz[inner sep=.25ex,baseline=-.75ex] \protect\node[circle,draw] {\footnotesize 5'}; in Fig. \ref{fig:factor-graph-y_uv-non-linear} are then easily evaluated. Finally, the extrinsic mean and variance, $\widehat{z}^{+}_{ij,  \mathsf{e},t+1}$ and $\gamma^{-1}_{\bm{Z}_{\mathsf{e}}^+,t+1}$, of message \protect\tikz[inner sep=.25ex,baseline=-.75ex] \protect\node[circle,draw] {\footnotesize 6'}; can be calculated from the posterior mean, $\widehat{z}^{+}_{ij,  \mathsf{p},t+1}$ and common precision, $\gamma_{\bm{Z}_{\mathsf{p}}^+,t+1}$, as specified in lines \ref{eq:algo-extrinsic-mu-z-minus}--\ref{eq:algo-extrinsic-sigma-z-minus} of Algorithm~\ref{algo:big-vamp}.
\section{State Evolution}\label{sec:state-evolution}
Our main goal is to understand the behavior of the proposed BiG-VAMP algorithm in the asymptotic regime for a certain class of $\boldsymbol{U}$ and $\boldsymbol{V}$ matrices, i.e.,  when they both  have zero mean i.i.d. priors. For simplicity, we will focus on MMSE estimation, i.e.,  $\beta=1$. In our case, the asymptotic regime refers to the case where $ r, N, M  \rightarrow +\infty$ with $\frac{r}{N} \rightarrow \beta_{u} = \mathcal{O}(1)$ and  $\frac{r}{M} \rightarrow \beta_{v} = \mathcal{O}(1)$ for some fixed ratios $\beta_{u}\leq 1$ and $\beta_{v}\leq 1$. \\
In approximate message passing practices, a state evolution ansatz is based on the following concentration of measure for the precision variables in the asymptotic regime:
\begin{subequations}\label{eq:Linear-SE-results 1}
\begin{align}
\lim _{M,N \rightarrow \infty}\left(\gamma_{\boldsymbol{U}_{ \mathsf{p}}^{+},t}, \gamma_{\boldsymbol{U}_{ \mathsf{e}}^{+},t}\right)=\left(\bar\gamma_{\boldsymbol{U}_{ \mathsf{p}}^{+},t}, \bar\gamma_{\boldsymbol{U}_{ \mathsf{e}}^{+},t}\right), \\
\lim _{M,N  \rightarrow \infty}\left(\gamma_{\boldsymbol{U}_{ \mathsf{p}}^{-},t}, \gamma_{\boldsymbol{U}_{ \mathsf{e}}^{-},t}\right)=\left(\bar\gamma_{\boldsymbol{U}_{ \mathsf{p}}^{-},t}, \bar\gamma_{\boldsymbol{U}_{ \mathsf{e}}^{-},t}\right), \\
\lim _{M,N  \rightarrow \infty}\left(\gamma_{\boldsymbol{V}_{ \mathsf{p}}^{+},t}, \gamma_{\boldsymbol{V}_{ \mathsf{e}}^{+},t}\right)=\left(\bar\gamma_{\boldsymbol{V}_{ \mathsf{p}}^{+},t}, \bar\gamma_{\boldsymbol{V}_{ \mathsf{e}}^{+},t}\right), \\
\lim _{M,N  \rightarrow \infty}\left(\gamma_{\boldsymbol{V}_{ \mathsf{p}}^{-},t}, \gamma_{\boldsymbol{V}_{ \mathsf{e}}^{-},t}\right)=\left(\bar\gamma_{\boldsymbol{V}_{ \mathsf{p}}^{-},t}, \bar\gamma_{\boldsymbol{V}_{ \mathsf{e}}^{-},t}\right),\\
\lim _{M,N \rightarrow \infty}\left(\gamma_{\boldsymbol{Z}_{ \mathsf{p}}^{+},t}, \gamma_{\boldsymbol{Z}_{ \mathsf{e}}^{+},t}\right)=\left(\bar\gamma_{\boldsymbol{Z}_{ \mathsf{p}}^{+},t}, \bar\gamma_{\boldsymbol{Z}_{ \mathsf{e}}^{+},t}\right), \\
\lim _{M,N \rightarrow \infty}\left(\gamma_{\boldsymbol{Z}_{ \mathsf{p}}^{-},t}, \gamma_{\boldsymbol{Z}_{ \mathsf{e}}^{-},t}\right)=\left(\bar\gamma_{\boldsymbol{Z}_{ \mathsf{p}}^{-},t}, \bar\gamma_{\boldsymbol{Z}_{ \mathsf{e}}^{-},t}\right).
\end{align}
\end{subequations}
\subsection{State Evolution of Bi-VAMP}\label{subsec:linear-case-model}
Recall here that Bi-VAMP applies to the bilinear observation model in which the data matrix, $\boldsymbol{Y}$, is obtained according to (\ref{eq:bilinear-model}) while taking  $\phi(.)$ to be the identity, i.e., $\phi(x) ~= ~x, ~\forall\, x\in \mathbb{R}$. That is to say:
\begin{equation}
\boldsymbol{Y}~=~  \bm{U}\bm{V}^{\mathsf{T}}\,+\,\boldsymbol{W}.
\end{equation}
$$
$$
%
%
%
in which $\boldsymbol{W}$ $\in$ $\mathbb{R}^{N \times M}$ is the additive white Gaussian noise matrix whose entries are assumed to be mutually independent with mean zero and variance $\sqrt{MN}\gamma_w^{-1}$, i.e., $w_{i,j}~\,\sim\,\mathcal{N}(w_{ij}; 0, \sqrt{MN}\gamma_w^{-1})$. Note here that by letting $M$ and $N$ grow unboundedly  one must scale the noise variance by $\sqrt{MN}$ to maintain the same signal-to-noise ratio irrespectively of the values of $M$ and $N$. We emphasize the fact that this scaling is,  however, required for the purpose of SE analysis only. For ease of notation, we will also drop the iteration index, $t$, and reintroduce it in the final state evolution recursion.
  Recall here that the algorithmic steps of Bi-VAMP are summarized in Algorithm \ref{algo:big-vamp} excluding the update equations pertaining to the ``generalized output step'' (i.e., lines \ref{eq:algo-posterior-mean-z} to \ref{eq:algo-extrinsic-sigma-z-minus}) while replacing $\widehat{\bm{Z}}^+_{\mathsf{e}}$ by $\bm{Y}$ and $\gamma_{\bm{Z}^+_{\mathsf{e}}}$ by $\gamma_w/\sqrt{MN}$ (after taking into account the aforementioned appropriate scaling by $\sqrt{MN}$). Consequently, from the update equations in lines \ref{eq:algo-Au-non-low-rank},  \ref{eq:algo-Av-non-low-rank}, \ref{eq:posterior-covariance-u-gaussian-approx}, and \ref{eq:posterior-covariance-v-gaussian-approx} of Algorithm \ref{algo:big-vamp}, it follows that in the large system limit the component-wise  MSEs of the bi-LMMSE denoisers  are given by (\ref{eq:u2-final-error-function}) and (\ref{eq:v2-final-error-function}) displayed on the top of this  page.
\begin{figure*}[!t]
\begin{equation}\label{eq:u2-final-error-function}
\begin{aligned}
\mathcal{E}_{u^-}&~=\lim _{r \rightarrow \infty} \frac{1}{r} \operatorname{Tr}\left(\boldsymbol{R}_{\boldsymbol{U}^{-}_{ \mathsf{p}}}\right)\,= \lim _{M,N \rightarrow \infty} \frac{1}{r} \operatorname{Tr}\left(\left[\gamma_{\boldsymbol{U}^{+}_{ \mathsf{e}}} \, \boldsymbol{I} +  \frac{\gamma_{w}}{\sqrt{MN}}\left({\widehat{\boldsymbol{V}}_{ \mathsf{p}}^{-\top}}\widehat{\boldsymbol{V}}_{ \mathsf{p}}^{-}\,+\, \frac{M}{\beta}\,\boldsymbol{R}_{\boldsymbol{V}_{ \mathsf{p}}^{-}} -\,\frac{M\gamma_{w}}{\sqrt{MN}} \,\boldsymbol{R}_{\boldsymbol{V}_{ \mathsf{p}}^{-}} \,\langle\boldsymbol{Y} \odot\boldsymbol{Y}\rangle\right)\right]^{-1}\right),\!\!\!\!\!\!
\end{aligned}
\end{equation}
\begin{equation}\label{eq:v2-final-error-function}
\begin{aligned}
\mathcal{E}_{v^-}&~=\lim _{r \rightarrow \infty} \frac{1}{r} \operatorname{Tr}\left(\boldsymbol{R}_{\boldsymbol{V}^{-}_{ \mathsf{p}}}\right)\,=\lim _{M,N \rightarrow \infty} \frac{1}{r} \operatorname{Tr}\left(\left[\gamma_{\boldsymbol{V}^{+}_{ \mathsf{e}}} \, \boldsymbol{I} + \frac{\gamma_{w}}{\sqrt{MN}}\left({\widehat{\boldsymbol{U}}_{ \mathsf{p}}^{-\top}}\widehat{\boldsymbol{U}}_{ \mathsf{p}}^{-}\,+\, \frac{N}{\beta}\,\boldsymbol{R}_{\boldsymbol{U}_{ \mathsf{p}}^{-}} -\, \frac{N\gamma_{w}}{\sqrt{MN}} \,\boldsymbol{R}_{\boldsymbol{U}_{ \mathsf{p}}^{-}} \,\langle\boldsymbol{Y} \odot\boldsymbol{Y}\rangle\right)\right]^{-1}\right).\!\!\!\!\!\
\end{aligned}
\end{equation} 
\rule{\textwidth}{0.4 pt}
\end{figure*}
Moreover, we assume that for large enough $M$ and $N$ the  element-wise errors in the matrix updates, $\widehat{\bm{U}}_{ \mathsf{p}}^-$ and  $\widehat{\bm{V}}_{ \mathsf{p}}^-$, are i.i.d. with zero mean  thereby leading to:
\begin{subequations}
\begin{align}
 \boldsymbol{R}_{\boldsymbol{U}_{ \mathsf{p}}^{-}}~\approx~\bar\gamma_{\boldsymbol{U}_{ \mathsf{p}}^{-}}^{-1}\boldsymbol{I}_{r},\label{eq:limit 11-13-19-1}\\
 \boldsymbol{R}_{\boldsymbol{V}_{ \mathsf{p}}^{-}}~\approx~\bar\gamma_{\boldsymbol{V}_{ \mathsf{p}}^{-}}^{-1}\boldsymbol{I}_{r},\label{eq:limit 11-13-19-1_V}
\end{align}
\end{subequations}
For large enough $M$ and $N$, we also make use of the following approximation:
\begin{equation}\label{eq:limit 11-13-19-1_Z}
\langle\boldsymbol{Y} \odot\boldsymbol{Y}\rangle~\approx~\bar{\sigma}_u^2\bar{\sigma}_v^2r\,+\,\sqrt{MN}\gamma_w^{-1},
\end{equation}
which follows from the observation that
$\sigma_u^2\triangleq\langle \bm{U}\odot\bm{U} \rangle\,\approx\,\bar{\sigma}_u^2\,\triangleq\,\mathbb{E}[\mathsf{u}^2_{i,\ell}|p_{\mathsf{u}}(u)]$ and $\sigma_v^2\triangleq\langle  \bm{V}\odot\bm{V}\rangle\,\approx\,\bar{\sigma}_v^2\,\triangleq\,\mathbb{E}[\mathsf{v}^2_{j\ell}|p_{\mathsf{v}}(v)]$ $\forall i,j,\ell$.   Here, $p_{\mathsf{u}}(u)$ [resp., $p_{\mathsf{v}}(v)$] is a common prior on the entries of the matrix $\bm{\mathsf{U}}$ [resp., $\bm{\mathsf{V}}$]. By the same virtue, we also approximate  $\gamma_{\boldsymbol{U}^{+}_{ \mathsf{e}}}$ and $\gamma_{\boldsymbol{V}^{+}_{ \mathsf{e}}}$ by $\bar{\gamma}_{\boldsymbol{U}^{+}_{ \mathsf{e}}}$ and  $\bar{\gamma}_{\boldsymbol{V}^{+}_{ \mathsf{e}}}$, respectively.
After using these approximations 
in (\ref{eq:u2-final-error-function}) and (\ref{eq:v2-final-error-function}), it follows that:
\begin{equation}\label{eq:u2-final-error_function_simplified}
\begin{aligned}
\!\!\!\!\mathcal{E}_{u^-}(\widetilde\gamma_{\boldsymbol{U}_{ \mathsf{e}}^{+}})&~=\lim _{r \rightarrow \infty} \frac{1}{r} \operatorname{Tr}\left(\left[\widetilde\gamma_{\boldsymbol{U}_{ \mathsf{e}}^{+}} \, \boldsymbol{I}_r \,+\,  \frac{\gamma_{w}}{\sqrt{MN}}\widehat{\boldsymbol{V}}_{ \mathsf{p}}^{-\top}\widehat{\boldsymbol{V}}_{ \mathsf{p}}^{-}\right)^{-1}\right),\!\!\!\!\!\!
\end{aligned} 
\end{equation}
\begin{equation}\label{eq:v2-final-error-function_function_simplified}
\!\!\!\!\begin{aligned}
\mathcal{E}_{v^-}(\widetilde\gamma_{\boldsymbol{V}_{ \mathsf{e}}^{+}})&~=\lim _{r \rightarrow \infty} \frac{1}{r} \operatorname{Tr}\left(\left[\widetilde\gamma_{\boldsymbol{V}_{ \mathsf{e}}^{+}} \, \boldsymbol{I}_r \,+\,  \frac{\gamma_{w}}{\sqrt{MN}}\widehat{\boldsymbol{U}}_{ \mathsf{p}}^{-\top}\widehat{\boldsymbol{U}}_{ \mathsf{p}}^{-}\right]^{-1}\right),\!\!\!\!\!\!
\end{aligned}
\end{equation} 
where
\begin{eqnarray}
\widetilde\gamma_{\boldsymbol{U}_{ \mathsf{e}}^{+}}&~=~&\bar\gamma_{\boldsymbol{U}^{+}_{ \mathsf{e}}}+\frac{1}{\beta}\sqrt{\frac{\beta_u}{\beta_v}}\gamma_w\bar\gamma_{\boldsymbol{V}_{ \mathsf{p}}^{-}}^{-1}\nonumber\\
&&~~~~~-\sqrt{\frac{\beta_u}{\beta_v}}\gamma_w\bar\gamma_{\boldsymbol{V}_{ \mathsf{p}}^{-}}^{-1}\left(\sqrt{\beta_u\beta_v}\bar\sigma_u^2\bar\sigma_v^2\gamma_w\,+\,1\right)\hspace{-0.05cm},\\
\widetilde\gamma_{\boldsymbol{V}_{ \mathsf{e}}^{+}}&~=~&\bar\gamma_{\boldsymbol{V}^{+}_{ \mathsf{e}}}+\frac{1}{\beta}\sqrt{\frac{\beta_v}{\beta_u}}\gamma_w\bar\gamma_{\boldsymbol{U}_{ \mathsf{p}}^{-}}^{-1}\nonumber\\
&&~~~~~-\sqrt{\frac{\beta_v}{\beta_u}}\gamma_w\bar\gamma_{\boldsymbol{U}_{ \mathsf{p}}^{-}}^{-1}\left(\sqrt{\beta_u\beta_v}\bar\sigma_u^2\bar\sigma_v^2\gamma_w\,+\,1\right)\hspace{-0.05cm},
\end{eqnarray}
To find the limits in (\ref{eq:u2-final-error_function_simplified}) and (\ref{eq:v2-final-error-function_function_simplified}), we define the following two matrices:
\begin{subequations}
\label{eq:H-1-H2}
\begin{align}
    \boldsymbol{H}_{u}~=~\frac{1}{\sqrt{(\bar\sigma_u^2-\bar{\gamma}_{\boldsymbol{U}^{-}_{ \mathsf{p}}}^{-1})N}}\,\widehat{\boldsymbol{U}}_{ \mathsf{p}}^{-}.\\
\boldsymbol{H}_{v}~=~\frac{1}{\sqrt{(\bar\sigma_v^2-\bar{\gamma}_{\boldsymbol{V}^{-}_{ \mathsf{p}}}^{-1})M}}\,\widehat{\boldsymbol{V}}_{ \mathsf{p}}^{-}.
\end{align}
\end{subequations}
Under the matched conditions\footnote{That is to say the true variance of the  MMSE estimation error in the entries of $\widehat{\boldsymbol{V}}_{ \mathsf{p}}^{-}$ (resp., $\widehat{\boldsymbol{U}}_{ \mathsf{p}}^{-}$) is equal the one predicted by the algorithm, i.e.,   $\bar{\gamma}_{\boldsymbol{V}^{-}_{ \mathsf{p}}}^{-1}$ (resp., $\bar{\gamma}_{\boldsymbol{U}^{-}_{ \mathsf{p}}}^{-1}$). The matched condition assumption is common in previous works on state evolution analysis and it holds true if the algorithm at hand is optimum.}, the entries of $\bm{H}_u$ (resp., $\bm{H}_v$) are i.i.d. with zero mean  and variance  $\frac{1}{N}$ (resp., $\frac{1}{M}$).
For ease of notation, we also define the two quantities:
\begin{subequations}
\label{eq: parameters}
\begin{align}
    \alpha_{u}~\triangleq~\sqrt{\frac{\beta_v}{\beta_u}}\frac{\gamma_{w}(\bar\sigma_u^2-\bar{\gamma}_{\boldsymbol{U}^{-}_{ \mathsf{p}}}^{-1})}{\widetilde{\gamma}_{\boldsymbol{V}^{+}_{ \mathsf{e}}}},\\
 \alpha_{v}~\triangleq~\sqrt{\frac{\beta_u}{\beta_v}}\frac{\gamma_{w}(\bar\sigma_v^2-\bar{\gamma}_{\boldsymbol{V}^{-}_{ \mathsf{p}}}^{-1})}{\widetilde{\gamma}_{\boldsymbol{U}^{+}_{ \mathsf{e}}}}.
\end{align}
\end{subequations}
\\Now, using a well-known result in random matrix theory  (cf. eq. (1.16) in \cite{tulino2004random}) --- we show that: 
\begin{equation}\label{eq:limit-Tolino-1}
\begin{aligned}
\mathcal{E}_{u^-}(\widetilde\gamma_{\boldsymbol{U}_{ \mathsf{e}}^{+}}) &~=~\lim _{r \rightarrow \infty} \frac{1}{r} \operatorname{Tr}\left(\Big[\widetilde{\gamma}_{\boldsymbol{U}^{+}_{ \mathsf{e}}} \, \boldsymbol{I} + \frac{\gamma_{w}}{\sqrt{MN}}{\widehat{\boldsymbol{V}}_{ \mathsf{p}}^{-\top}}\widehat{\boldsymbol{V}}_{ \mathsf{p}}^{-}\Big]^{-1}\right),\\
&~=~\lim _{r \rightarrow \infty} \frac{\widetilde{\gamma}_{\boldsymbol{U}^{+}_{ \mathsf{e}}}^{-1}}{r}\operatorname{Tr}\Bigg(\bigg[\boldsymbol{I} + \sqrt{\frac{\beta_u}{\beta_v}}\gamma_{w}\widetilde{\gamma}_{\boldsymbol{U}^{+}_{ \mathsf{e}}}^{-1}(\bar\sigma_v^2-\bar{\gamma}_{\boldsymbol{V}^{-}_{ \mathsf{p}}}^{-1})\bigg.\\ & \hspace{5.05cm}\bigg.\boldsymbol{H}_{v}^{\top}\boldsymbol{H}_{v}\bigg]^{-1}\Bigg),\\
&~=~ \widetilde{\gamma}_{\boldsymbol{U}^{+}_{ \mathsf{e}}}^{-1}\left(1-\frac{\mathcal{F}( \alpha_{v},\beta_{v})}{4 \beta_{v}  \alpha_{v}}\right),
\end{aligned} 
\end{equation}
wherein the function  $\mathcal{F}(., .)$ is defined as:
\begin{equation}\label{eq:tolino-function}
\mathcal{F}(x, z)~=~\left(\sqrt{x(1+\sqrt{z})^{2}+1}-\sqrt{x(1-\sqrt{z})^{2}+1}\ \right)^{2}\!,
\end{equation}
Similarly we show that: 
\begin{equation}\label{eq:limit-Tolino-2}
\begin{aligned} 
\mathcal{E}_{v^-}(\widetilde\gamma_{\boldsymbol{V}_{ \mathsf{e}}^{+}}) &~=~\lim _{r \rightarrow \infty} \frac{1}{r} \operatorname{Tr}\left(\left[\widetilde\gamma_{\boldsymbol{V}^{+}_{ \mathsf{e}}} \, \boldsymbol{I} + \frac{\gamma_{w}}{\sqrt{MN}}{\widehat{\boldsymbol{U}}_{ \mathsf{p}}^{-\top}}\widehat{\boldsymbol{U}}_{ \mathsf{p}}^{-}\right]^{-1}\right),\\
&~=~\lim _{r \rightarrow \infty} \frac{\widetilde\gamma_{\boldsymbol{V}^{+}_{ \mathsf{e}}}^{-1}}{r}\operatorname{Tr}\Bigg(\bigg[\boldsymbol{I} +\sqrt{\frac{\beta_v}{\beta_u}}\gamma_{w}\widetilde\gamma_{\boldsymbol{V}^{+}_{ \mathsf{e}}}^{-1}(\bar{\sigma}_u^2-\gamma_{\boldsymbol{U}^{-}_{ \mathsf{p}}}^{-1})\bigg.\\ & \hspace{5.05cm}\bigg.\boldsymbol{H}_{u}^{\top}\boldsymbol{H}_{u}\bigg]^{-1}\Bigg),\\
&~=~ \widetilde{\gamma}_{\boldsymbol{V}^{+}_{ \mathsf{e}}}^{-1}\left(1-\frac{\mathcal{F}( \alpha_{u},\beta_{u})}{4 \beta_{u}  \alpha_{u}}\right).
\end{aligned}
\end{equation}
Recall here that as $ r, N, M  \rightarrow +\infty$ we have $\frac{r}{N} \rightarrow \beta_{u}$ and  $\frac{r}{M} \rightarrow \beta_{v}$.  
 Now, the output variances of the MMSE denoisers of $\bm{U}$ and $\bm{V}$ matrices are obtained from lines \ref{eq:algo-diag-sigma-u-alpha} and \ref{eq:algo-diag-sigma-v-alpha} of Algorithm \ref{algo:big-vamp} as follows:
 \begin{eqnarray}
 \label{eq:SE-diag-sigma-u-alpha}
 \mathcal{E}_{u^+}(\gamma_{\boldsymbol{U}_{ \mathsf{e}}^{-}})~\triangleq~\frac{1}{\gamma_{\boldsymbol{U}^{+}_{ \mathsf{p}}}} &~=~& \frac{1}{N\gamma_{\boldsymbol{U}^{-}_{ \mathsf{e}}}} \,\sum_{i=1}^{N} \langle \mathbf{g}^{\prime}_{\mathsf{u}}(\widehat{\boldsymbol{u}}^{-}_{i}, \gamma^{-1}_{\boldsymbol{U}^{-}_{ \mathsf{e}}})\rangle,\\
 \label{eq:SE-diag-sigma-v-alpha}
 \mathcal{E}_{v^+}(\gamma_{\boldsymbol{V}_{ \mathsf{e}}^{-}})~\triangleq~\frac{1}{\gamma_{\boldsymbol{V}^{+}_{ \mathsf{p}}}} &~=~& \frac{1}{M\gamma_{\boldsymbol{V}^{-}_{ \mathsf{e}}}} \,\sum_{j=1}^{M}\langle \mathbf{g}^{\prime}_{\mathsf{v}}(\widehat{\boldsymbol{v}}^{-}_{j}, \gamma^{-1}_{\boldsymbol{U}^{-}_{ \mathsf{e}}})\rangle.
 \end{eqnarray}
 In the large system limits, the empirical averages involved in (\ref{eq:SE-diag-sigma-u-alpha}) and (\ref{eq:SE-diag-sigma-v-alpha}) can be approximated by the following statistical averages in which we use the fact that $\gamma_{\boldsymbol{U}^{+}_{ \mathsf{p}}}\longrightarrow\bar{\gamma}_{\boldsymbol{U}^{+}_{ \mathsf{p}}}$ and $\gamma_{\boldsymbol{V}^{+}_{ \mathsf{p}}}\longrightarrow\bar{\gamma}_{\boldsymbol{V}^{+}_{ \mathsf{p}}}$ as $M$ and $N$ grow large:
\begin{equation}\label{eq:u1-sensitivity-function}
\begin{aligned}
\mathcal{E}_{u^+}(\bar\gamma_{\boldsymbol{U}_{ \mathsf{e}}^{-}})~\triangleq~\frac{1}{\bar\gamma_{\boldsymbol{U}_{ \mathsf{e}}^{-}}}\mathbb{E}\left[g^{\prime}_{\textrm{{\rv{u}}}}(\widehat{u}_{\mathsf{e}}^{\,-},\bar\gamma_{\boldsymbol{U}_{ \mathsf{e}}^{-}}^{-1})\middle\vert\mathit{p}_{\textrm{\rv{u}}}(u)p_{\widehat{\mathsf{u}}_{\mathsf{e}}^{\,-}|\mathsf{u}}(\widehat{u}_{\mathsf{e}}^{\,-}|u)\right]
\end{aligned}
\end{equation}
\begin{equation}\label{eq:v1-sensitivity-function} 
\begin{aligned}
\mathcal{E}_{v^+}(\bar\gamma_{\boldsymbol{V}_{ \mathsf{e}}^{-}})~\triangleq~\frac{1}{\bar\gamma_{\boldsymbol{V}_{ \mathsf{e}}^{-}}}\mathbb{E}\left[g^{\prime}_{\textrm{{\rv{v}}}}(\widehat{v}_{\mathsf{e}}^{\,-},\bar\gamma_{\boldsymbol{V}_{ \mathsf{e}}^{-}}^{-1})\middle\vert\mathit{p}_{\textrm{\rv{v}}}(v)p_{\widehat{\mathsf{v}}_{\mathsf{e}}^{\,-}|\mathsf{v}}(\widehat{v}_{\mathsf{e}}^{\,-}|v)\right]
\end{aligned}
\end{equation}
In (\ref{eq:u1-sensitivity-function}) and (\ref{eq:v1-sensitivity-function}), $p_{\widehat{\mathsf{u}}_{\mathsf{e}}^{-}|\mathsf{u}}(\widehat{u}_{\mathsf{e}}^{-}|u)$ and $p_{\widehat{\mathsf{v}}_{\mathsf{e}}^{-}|\mathsf{v}}(\widehat{v}_{\mathsf{e}}^{-}|v)$  correspond to  the scalar models  $\widehat{\mathsf{u}}_{\mathsf{e}}^{-}\,=\,\mathsf{u}\,+\,\mathsf{w}_u$ and $\widehat{\mathsf{v}}_{\mathsf{e}}^{-}\,=\,\mathsf{v}\,+\,\mathsf{w}_v$ where under the matched conditions, we have $\mathsf{w}_u\sim\mathcal{N}(w_u;0,\bar\gamma_{\boldsymbol{U}_{ \mathsf{e}}^{-}}^{-1})$ and $\mathsf{w}_v\sim\mathcal{N}(w_v;0,\bar\gamma_{\boldsymbol{V}_{ \mathsf{e}}^{-}}^{-1})$.
Under all the aforementioned assumptions and  matched conditions, we can now describe  in Algorithm~\ref{algo:Linear big-vamp SE} our main result, which is the SE recursion equations for Bi-VAMP.
\begin{algorithm}[h!]
\small
\caption{Bi-VAMP State Evolution}\label{algo:Linear big-vamp SE}
\begin{algorithmic}[1]
\Statex $\mathbf{Require:}$ Noise precision $\gamma_w$; set $\beta=1$; $\beta_u$ and $\beta_v$; number of $~~~~~~~~~~~~~~~~ \textrm{iterations}~ T_\textrm{max}$. 
\vspace{0.1cm}
\Statex $\mathbf{Initialization:}$  extrinsic precisions  $\bar\gamma_{\boldsymbol{U}_{ \mathsf{e}}^{+},1}$, $\bar\gamma_{\boldsymbol{V}_{ \mathsf{e}}^{+},1}$, $\bar\gamma_{\boldsymbol{U}_{ \mathsf{p}}^{-},1}$, $\bar\gamma_{\boldsymbol{V}_{ \mathsf{p}}^{-},1}$
\For {$t=1,\dots, T_{\textrm{max}}$}
\Statex \LeftComment{1}{\scriptsize compute the effective inverse noise variance for the Bi-LMMSE block} 
\State 
$\widetilde\gamma_{\boldsymbol{U}_{ \mathsf{e}}^{+},t+1}~=~\bar\gamma_{\boldsymbol{U}^{+}_{ \mathsf{e}},t}
-\sqrt{\frac{\beta_u}{\beta_v}}\gamma_w\bar\gamma_{\boldsymbol{V}_{ \mathsf{p}}^{-},t}^{-1}\left(\sqrt{\beta_u\beta_v}\bar\sigma_u^2\bar\sigma_v^2\gamma_w\,+\,1\right)$
\Statex\hspace{3.5cm}$+\frac{1}{\beta}\sqrt{\frac{\beta_u}{\beta_v}}\gamma_w\bar\gamma_{\boldsymbol{V}_{ \mathsf{p}}^{-},t}^{-1}$\vspace{0.1cm}
\State 
$\widetilde\gamma_{\boldsymbol{V}_{ \mathsf{e}}^{+},t+1}~=~\bar\gamma_{\boldsymbol{V}^{+}_{ \mathsf{e}},t}
-\sqrt{\frac{\beta_v}{\beta_u}}\gamma_w\bar\gamma_{\boldsymbol{U}_{ \mathsf{p}}^{-},t}^{-1}\left(\sqrt{\beta_u\beta_v}\bar\sigma_u^2\bar\sigma_v^2\gamma_w\,+\,1\right)$
\Statex \hspace{3.5cm}$+\frac{1}{\beta}\sqrt{\frac{\beta_v}{\beta_u}}\gamma_w\bar\gamma_{\boldsymbol{U}_{ \mathsf{p}}^{-},t}^{-1}$\vspace{0.1cm}
\Statex \LeftComment{1}{\scriptsize compute the analytical posterior and extrinsic precision of  $\textrm{\textbf{{\rv{u}}}}^-$}
\State $\bar\gamma_{\boldsymbol{U}_{ \mathsf{p}}^{-},t+1} = \frac{1}{\mathcal{E}_{u^-}\big(\widetilde\gamma_{\boldsymbol{U}_{ \mathsf{e}}^{+},t+1}\big)}$  
\State $\bar\gamma_{\boldsymbol{U}_{ \mathsf{e}}^{-} ,t+1}~=~\bar\gamma_{\boldsymbol{U}_{ \mathsf{p}}^{-},t+1}~-~\bar\gamma_{\boldsymbol{U}_{ \mathsf{e}}^{+},t}$ \vspace{0.1cm}
\Statex \LeftComment{1}{\scriptsize compute the analytical posterior and extrinsic precision of $\textrm{\textbf{{\rv{v}}}}^-$}
\State $\bar\gamma_{\boldsymbol{V}_{ \mathsf{p}}^{-},t+1} = \frac{1}{\mathcal{E}_{v^-}\big(\widetilde\gamma_{\boldsymbol{V}_{ \mathsf{e}}^{+},t+1}\big)}$ 
\State $\bar\gamma_{\boldsymbol{V}_{ \mathsf{e}}^{-},t+1}~=~\bar\gamma_{\boldsymbol{V}_{ \mathsf{p}}^{-},t+1}~-~\bar\gamma_{\boldsymbol{V}_{ \mathsf{e}}^{+},t}$\vspace{0.1cm}
\Statex \LeftComment{1}{\scriptsize compute the analytical posterior and extrinsic precision of  $\textrm{\textbf{{\rv{u}}}}^+$}
\State $\bar\gamma_{ \boldsymbol{U}_{ \mathsf{p}}^{+},t+1}=\frac{1}{\mathcal{E}_{u^+}(\bar\gamma_{\boldsymbol{U}_{ \mathsf{e}}^{-},t+1})}$
\State $\bar\gamma_{\boldsymbol{U}_{ \mathsf{e}}^{+},t+1}=\bar\gamma_{\boldsymbol{U}_{ \mathsf{p}}^{+},t+1}-\bar\gamma_{\boldsymbol{U}_{ \mathsf{e},t+1}^{-}}$\vspace{0.1cm}
\Statex \LeftComment{1}{\scriptsize compute the analytical posterior and extrinsic precision of  $\textrm{\textbf{{\rv{v}}}}^+$}
\State $\bar\gamma_{ \boldsymbol{V}_{ \mathsf{p}}^{+},t+1}=\frac{1}{\mathcal{E}_{v^+}(\bar\gamma_{\boldsymbol{V}_{ \mathsf{e}}^{-},t+1})}$
\State $\bar\gamma_{\boldsymbol{V}_{ \mathsf{e}}^{+},t+1}=\bar\gamma_{\boldsymbol{V}_{ \mathsf{p}}^{+},t+1}-\bar\gamma_{\boldsymbol{V}_{ \mathsf{e}}^{-},t+1}$\vspace{0.1cm}
\EndFor
\State \textbf{Return}
$\bar\gamma_{ \boldsymbol{U}_{ \mathsf{p}}^{+},T_{\textrm{max}}+1}, \bar\gamma_{\boldsymbol{V}_{ \mathsf{p}}^{+},T_{\textrm{max}}+1}$
\end{algorithmic}
\end{algorithm} 
\\
Note here that at convergence,  one must have equality between the posterior variances  of the same variable:
\begin{subequations}
\label{eq:Linear-SE-results 2}
\begin{align}
    \mathcal{E}_{u^+}(\bar\gamma_{\boldsymbol{U}_{ \mathsf{e}}^{-},\infty})~\triangleq~\bar\gamma_{\boldsymbol{U}_{ \mathsf{p}}^{+},\infty}^{-1}~= ~\ \mathcal{E}_{u^-}(\widetilde\gamma_{\boldsymbol{U}_{ \mathsf{e}}^{+},\infty})~\triangleq~\bar\gamma_{\boldsymbol{U}_{ \mathsf{p}}^{-},\infty}^{-1} \\
\mathcal{E}_{v^+}(\bar\gamma_{\boldsymbol{V}_{ \mathsf{e}}^{-},\infty})~\triangleq~\bar\gamma_{\boldsymbol{V}_{ \mathsf{p}}^{+},\infty}^{-1}~=~ \ \mathcal{E}_{v^-}(\widetilde\gamma_{\boldsymbol{V}_{ \mathsf{e}}^{+},\infty})~\triangleq~\bar\gamma_{\boldsymbol{V}_{ \mathsf{p}}^{-},\infty}^{-1}
\end{align}
\end{subequations}
\subsection{Extension to the Generalized Bilinear Model: State Evolution of BiG-VAMP}
In this case, the observation matrix, $\boldsymbol{Y}$, is obtained from the generalized bilinear model in (\ref{eq:non-linear-distribution0}). To account for the residual error of the output denoiser (cf. Fig. \ref{fig:block-diagram}), all the previous SE equations summarized in Algorithm \ref{algo:Linear big-vamp SE} remain the same except for replacing the noise variance, $\gamma_w^{-1}$, by the extrinsic error variance, $\gamma^{-1}_{\bm{Z}_{\mathsf{e}}^+}$, whose SE update equation will be characterized in the following. Again, to maintain the same energy per each entry in the matrix $ \bm{\mathsf{U}}\bm{\mathsf{V}}^{\mathsf{T}}$, as we grow $M$ and $N$, we redefine $\bm{\mathsf{Z}}$ as $\bm{\mathsf{Z}} \triangleq\frac{1}{\sqrt[4]{MN}} \bm{\mathsf{U}}\bm{\mathsf{V}}^{\mathsf{T}}$.

 Recall form line \ref{eq:algo-posterior-sigma-z} in 
 Algorithm \ref{algo:big-vamp} that  the posterior variance, $\mathcal{E}_{z^{-}}(\gamma_{\boldsymbol{Z}_{ \mathsf{e}}^{+}})\triangleq \gamma_{\boldsymbol{Z}_{ \mathsf{p}}^{-}}^{-1}$,   of each $z_{ij}$ estimate provided by the Bi-LMMSE module is given by (after taking into account the effect of the above scaling by $\frac{1}{\sqrt[4]{MN}}$): 
\begin{eqnarray}\label{eq:z2-error-function}
\mathcal{E}_{z^{-}}(\gamma_{\boldsymbol{Z}_{ \mathsf{e}}^{+}})&\,=\,&\left(\gamma_{\boldsymbol{Z}_{ \mathsf{e}}^{+}}\,+\,\sqrt{MN}\left[\operatorname{Tr}\left(\frac{1}{\beta}\,\boldsymbol{R}_{\boldsymbol{U}_{ \mathsf{p}}^{-}} \boldsymbol{R}_{\boldsymbol{V}_{ \mathsf{p}}^{-}}^{\top}+
\right.\right.\right.\nonumber\\
&&\left.\left.\left. ~~~~\frac{1}{M}\, \boldsymbol{R}_{\boldsymbol{U}_{ \mathsf{p}}^{-}} \boldsymbol{\widehat{V}}_{ \mathsf{p}}^{-\top} \boldsymbol{\widehat{V}}^{-}_{ \mathsf{p}}+ \frac{1}{N}\, \boldsymbol{R}_{\boldsymbol{V}_{ \mathsf{p}}^{-}}  \boldsymbol{\widehat{U}}_{ \mathsf{p}}^{-\top}\boldsymbol{\widehat{U}}^{-}_{ \mathsf{p}}\right)\right]^{-1}\right)^{-1}\!\!\!\!\!\!.\nonumber\\
\end{eqnarray}
For large enough $M$ and $N$, by plugging (\ref{eq:limit 11-13-19-1}), (\ref{eq:limit 11-13-19-1_V}), and (\ref{eq:H-1-H2}) in (\ref{eq:z2-error-function}), it follows that: 
\begin{equation}\label{eq:z2-final-error-function}
\begin{aligned}
\mathcal{E}_{z^-}(\bar\gamma_{\boldsymbol{Z}_{ \mathsf{e}}^{+}})&\,=\,\bar\gamma_{\boldsymbol{Z}_{ \mathsf{p}}^{-}}^{-1} \\
&\,=\,\left(\bar\gamma_{\boldsymbol{Z}_{ \mathsf{e}}^{+}}+\frac{1}{\sqrt{\beta_u\beta_v}}\left[\frac{1}{\beta}\ \bar\gamma_{\boldsymbol{U}_{ \mathsf{p}}^{-}}^{-1}\bar\gamma_{\boldsymbol{V}_{ \mathsf{p}}^{-}}^{-1}+\ \frac{(\bar{\sigma}_v^2-\bar\gamma_{\boldsymbol{V}^{-}_{ \mathsf{p}}}^{-1})}{\bar\gamma_{\boldsymbol{U}_{ \mathsf{p}}^{-}}}\right.\right.\\ & \quad \quad\quad\quad\quad\quad\quad\quad\quad\quad\quad\quad\quad\left.\left. +\ \frac{(\bar{\sigma}_u^2-\bar\gamma_{\boldsymbol{U}^{-}_{ \mathsf{p}}}^{-1})}{\bar\gamma_{\boldsymbol{V}_{ \mathsf{p}}^{-}}}\right]^{-1}\right)^{-1}\!\!\!\!\!\!.
\end{aligned}
\end{equation}
 Now, the common component-wise variance of the output denoiser for $\bm{Z}$  is obtained from line     \ref{update_posterior_mean_and_variance_z} of Algorithm \ref{algo:big-vamp} as follows:
 \begin{eqnarray}
 \label{eq:SE-diag-sigma-z-alpha}
 \!\!\!\!\!\!\!\!\!\!\!\!\mathcal{E}_{z^+}(\gamma_{\boldsymbol{Z}_{ \mathsf{e}}^{-}})\,\triangleq\,\frac{1}{\gamma_{\boldsymbol{Z}^{+}_{ \mathsf{p}}}} &\,=\,& \frac{1}{MN\gamma_{\boldsymbol{Z}^{-}_{ \mathsf{e}}}} \,\sum_{i=1}^{N}\sum_{j=1}^{M}  g^{\prime}_{\mathsf{z}}(y_{ij},\widehat{z}^{\,-}_{ij,\mathsf{e}}, \gamma^{-1}_{\boldsymbol{Z}^{-}_{ \mathsf{e}}}).
 \end{eqnarray}
 In the large system limits, the empirical average involved in (\ref{eq:SE-diag-sigma-z-alpha})  can be approximated by the following statistical average in which we use the fact that $\gamma_{\boldsymbol{Z}^{+}_{ \mathsf{p}}}\longrightarrow\bar{\gamma}_{\boldsymbol{Z}^{+}_{ \mathsf{p}}}$ as $M$ and $N$ grow large:
\begin{equation}\label{eq:z-sensitivity-function}
\begin{aligned}
\mathcal{E}_{z^+}(\bar\gamma_{\boldsymbol{Z}_{ \mathsf{e}}^{-}})\triangleq\frac{1}{\bar\gamma_{\boldsymbol{Z}_{ \mathsf{e}}^{-}}}\mathbb{E}\left[g^{\prime}_{\textrm{{\rv{z}}}}(y,\widehat{z}_{\mathsf{e}}^{\,-},\bar\gamma_{\boldsymbol{Z}_{ \mathsf{e}}^{-}}^{-1})\middle\vert\mathit{p}_{\textrm{\rv{z}}}(z)\mathit{p}_{\textrm{\rv{y}}|\textrm{\rv{z}}}(y|z)p_{\widehat{\mathsf{z}}_{\mathsf{e}}^{\,-}|\mathsf{z}}(\widehat{z}_{\mathsf{e}}^{-}|z)\right].
\end{aligned}
\end{equation}
In (\ref{eq:z-sensitivity-function}), $p_{\widehat{\mathsf{z}}_{\mathsf{e}}^{-}|\mathsf{z}}(\widehat{z}_{\mathsf{e}}^{-}|z)$ corresponds to  the scalar model  $\widehat{\mathsf{z}}_{\mathsf{e}}^{\,-}\,=\,\mathsf{z}\,+\,\mathsf{w}_z$  where under the matched conditions, we have $\mathsf{w}_z\sim\mathcal{N}(w_z;0,\bar\gamma_{\boldsymbol{Z}_{ \mathsf{e}}^{-}}^{-1})$. Moreover, since we have $\mathsf{z}=\frac{1}{\sqrt[4]{MN}}\bm{\mathsf{u}}^{\mathsf{T}}\bm{\mathsf{v}}$, then owing to the central limit theorem we have  $p_{\mathsf{z}}(z)=\mathcal{N}(z;0,\sqrt{\beta_u\beta_v}\bar{\sigma}^2_u\bar{\sigma}^2_v)$. 
Under all the assumptions and  matched conditions that we stated above, we can now describe  in Algorithm~\ref{algo:NL-Linear big-vamp SE} our main result, which is the SE recursion  for BiG-VAMP.
%
%
%
\begin{algorithm}[h!]
\small
\caption{BiG-VAMP State Evolution}\label{algo:NL-Linear big-vamp SE}
\begin{algorithmic}[1]
\Statex $\mathbf{Require:}$  set $\beta=1$; $\beta_u$ and $\beta_v$; number of iterations $T_\textrm{max}$. 
\vspace{0.1cm}
\Statex $\mathbf{Initialization:}$  extrinsic precisions  $\bar\gamma_{\boldsymbol{Z}_{ \mathsf{e}}^{+},1}$, $\bar\gamma_{\boldsymbol{U}_{ \mathsf{e}}^{+},1}$, $\bar\gamma_{\boldsymbol{V}_{ \mathsf{e}}^{+},1}$, $\bar\gamma_{\boldsymbol{U}_{ \mathsf{p}}^{-},1}$, $~~~~~~~~~~~~~~~~~~~~~~~~~~~~~~~~~~~~~~~~~~~~~~~~\bar\gamma_{\boldsymbol{V}_{ \mathsf{p}}^{-},1}$\vspace{0.1cm}
\For {$t=1,\dots, T_{\textrm{max}}$}
\Statex \LeftComment{1}{\scriptsize compute the effective inverse noise variance for the Bi-LMMSE block} 
\State 
$\widetilde\gamma_{\boldsymbol{U}_{ \mathsf{e}}^{+},t+1}~=~\bar\gamma_{\boldsymbol{U}^{+}_{ \mathsf{e}},t} + \frac{1}{\beta}\sqrt{\frac{\beta_u}{\beta_v}}\bar\gamma_{\boldsymbol{Z}_{ \mathsf{e}}^{+},t}\bar\gamma_{\boldsymbol{V}_{ \mathsf{p}}^{-},t}^{-1}$
\Statex \hspace{2.6cm}$-\sqrt{\frac{\beta_u}{\beta_v}}\bar\gamma_{\boldsymbol{Z}_{ \mathsf{e}}^{+},t}\bar\gamma_{\boldsymbol{V}_{ \mathsf{p}}^{-},t}^{-1}\left(\sqrt{\beta_u\beta_v}\bar\sigma_u^2\bar\sigma_v^2\bar\gamma_{\boldsymbol{Z}_{ \mathsf{e}}^{+},t}\,+\,1\right)$
\State 
$\widetilde\gamma_{\boldsymbol{V}_{ \mathsf{e}}^{+},t+1}~=~\bar\gamma_{\boldsymbol{V}^{+}_{ \mathsf{e}},t}+\frac{1}{\beta}\sqrt{\frac{\beta_v}{\beta_u}}\bar\gamma_{\boldsymbol{Z}_{ \mathsf{e}}^{+},t}\bar\gamma_{\boldsymbol{U}_{ \mathsf{p}}^{-},t}^{-1}$
\Statex \hspace{2.6cm} $ - \sqrt{\frac{\beta_v}{\beta_u}}\bar\gamma_{\boldsymbol{Z}_{ \mathsf{e}}^{+},t}\bar\gamma_{\boldsymbol{U}_{ \mathsf{p}}^{-},t}^{-1}\left(\sqrt{\beta_u\beta_v}\bar\sigma_u^2\bar\sigma_v^2\bar\gamma_{\boldsymbol{Z}_{ \mathsf{e}}^{+},t}\,+\,1\right)$
\Statex \LeftComment{1}{\scriptsize compute the analytical posterior and extrinsic precision of $\textrm{\textbf{{\rv{u}}}}^-$}
\State $\bar\gamma_{\boldsymbol{U}_{ \mathsf{p}}^{-},t+1} = \frac{1}{\mathcal{E}_{u^-}\big(\widetilde\gamma_{\boldsymbol{U}_{ \mathsf{e}}^{+},t+1}\big)}$  
\State $\bar\gamma_{\boldsymbol{U}_{ \mathsf{e}}^{-} ,t+1}~=~\bar\gamma_{\boldsymbol{U}_{ \mathsf{p}}^{-},t+1}~-~\bar\gamma_{\boldsymbol{U}_{ \mathsf{e}}^{+},t}$ \vspace{0.1cm}
\Statex \LeftComment{1}{\scriptsize compute the analytical posterior and extrinsic precision of $\textrm{\textbf{{\rv{v}}}}^-$}
\State $\bar\gamma_{\boldsymbol{V}_{ \mathsf{p}}^{-},t+1} = \frac{1}{\mathcal{E}_{v^-}\big(\widetilde\gamma_{\boldsymbol{V}_{ \mathsf{e}}^{+},t+1}\big)}$ 
\State $\bar\gamma_{\boldsymbol{V}_{ \mathsf{e}}^{-},t+1}~=~\bar\gamma_{\boldsymbol{V}_{ \mathsf{p}}^{-},t+1}~-~\bar\gamma_{\boldsymbol{V}_{ \mathsf{e}}^{+},t}$\vspace{0.1cm}
\Statex \LeftComment{1}{\scriptsize compute the analytical posterior and extrinsic precision  of $\textrm{{\rv{z}}}^-$}
\State $\bar\gamma_{\boldsymbol{Z}_{ \mathsf{p}}^{-},t+1} = \frac{1}{\mathcal{E}_{z^-}\big(\widetilde\gamma_{\boldsymbol{Z}_{ \mathsf{e}}^{+},t}\big)}$ 
\State $\bar\gamma_{\boldsymbol{Z}_{ \mathsf{e}}^{-},t+1}~=~\bar\gamma_{\boldsymbol{Z}_{ \mathsf{p}}^{-},t+1}~-~\bar\gamma_{\boldsymbol{Z}_{ \mathsf{e}}^{+},t}$\vspace{0.1cm} 
\Statex \LeftComment{1}{\scriptsize compute the analytical posterior and extrinsic precision of  $\textrm{\textbf{{\rv{u}}}}^+$}
\State $\bar\gamma_{ \boldsymbol{U}_{ \mathsf{p}}^{+},t+1}=\frac{1}{\mathcal{E}_{u^+}(\bar\gamma_{\boldsymbol{U}_{ \mathsf{e}}^{-},t+1})}$
\State $\bar\gamma_{\boldsymbol{U}_{ \mathsf{e}}^{+},t+1}=\bar\gamma_{\boldsymbol{U}_{ \mathsf{p}}^{+},t+1}-\bar\gamma_{\boldsymbol{U}_{ \mathsf{e},t+1}^{-}}$\vspace{0.1cm}
\Statex \LeftComment{1}{\scriptsize compute the analytical posterior and extrinsic precision of $\textrm{\textbf{{\rv{v}}}}^+$}
\State $\bar\gamma_{ \boldsymbol{V}_{ \mathsf{p}}^{+},t+1}=\frac{1}{\mathcal{E}_{v^+}(\bar\gamma_{\boldsymbol{V}_{ \mathsf{e}}^{-},t+1})}$
\State $\bar\gamma_{\boldsymbol{V}_{ \mathsf{e}}^{+},t+1}=\bar\gamma_{\boldsymbol{V}_{ \mathsf{p}}^{+},t+1}-\bar\gamma_{\boldsymbol{V}_{ \mathsf{e}}^{-},t+1}$\vspace{0.1cm}
\Statex \LeftComment{1}{\scriptsize compute the analytical posterior and extrinsic precision of $\textrm{\rv{z}}^+$}
\State $\bar\gamma_{ \boldsymbol{Z}_{ \mathsf{p}}^{+},t+1}=\frac{1}{\mathcal{E}_{z^+}(\bar\gamma_{\boldsymbol{Z}_{ \mathsf{e}}^{-},t+1})}$
\State $\bar\gamma_{\boldsymbol{Z}_{ \mathsf{e}}^{+},t+1}=\bar\gamma_{\boldsymbol{Z}_{ \mathsf{p}}^{+},t+1}-\bar\gamma_{\boldsymbol{Z}_{ \mathsf{e}}^{-},t+1}$\vspace{0.1cm}
\EndFor
\State \textbf{Return}
$\bar\gamma_{ \boldsymbol{U}_{ \mathsf{p}}^{+},T_{\textrm{max}}+1}, \bar\gamma_{\boldsymbol{V}_{ \mathsf{p}}^{+},T_{\textrm{max}}+1}, \bar\gamma_{ \boldsymbol{Z}_{ \mathsf{p}}^{+},T_{\textrm{max}}+1}$
\end{algorithmic}
\end{algorithm} 
Note  that at convergence, on top of the  equalities in (\ref{eq:Linear-SE-results 2}), we also have:
\begin{equation}\label{asymptotic_Z}
\mathcal{E}_{z^+}(\bar\gamma_{\boldsymbol{Z}_{ \mathsf{e}}^{-},\infty})~\triangleq~\bar\gamma_{\boldsymbol{Z}_{ \mathsf{p}}^{+},\infty}^{-1}~= ~\ \mathcal{E}_{z^-}(\widetilde\gamma_{\boldsymbol{Z}_{ \mathsf{e}}^{+},\infty})~\triangleq~\bar\gamma_{\boldsymbol{Z}_{ \mathsf{p}}^{-},\infty}^{-1}. \\
\end{equation}
\section{Simulation Results}\label{sec:simulation-results}
In this section, we  assess the performance behavior of the proposed BiG-Vamp algorithm and benchmark it against BiG-AMP \cite{parker2014bilinear}, BAd-VAMP \cite{sarkar2019bilinear} and LowRAMP \cite{lesieur2015phase} algorithms for different applications, namely: 
\begin{itemize}
    \item matrix factorization,
    \item dictionary learning,
    \item matrix completion.
\end{itemize}
 In all simulations, we set $T_{\textrm{max}}=1000$ and the precision tolerance to $\xi=10^{-6}$ 
 and we perform $N_{\textrm{MC}}=100$ Monte-Carlo trials for different values of the SNR: 
\begin{equation*}
     \textrm{SNR} \,=\, \textrm{10 log}_{\textrm{10}}\Bigg(\frac{ \norm{\boldsymbol{Z}}^{2}_{\textrm{F}}}{\norm{\boldsymbol{W}}^{2}_{\textrm{F}}}\Bigg),
\end{equation*}
\noindent where $\|.\|_{\rm{F}}$ is the Frobenius norm.
We also use the normalized root MSE (NRMSE) as performance measure which defined as follows:
\begin{equation*}
    \textrm{NRMSE} ~=~\frac{1}{N_{\textrm{MC}}}\sum_{\ell=1}^{N_{\textrm{MC}}} \frac{\|\bm{Z}_{\ell}-\widehat{\bm{Z}}_{\ell}\|_{\textrm{F}}}{\|\bm{Z}_{\ell}\|_{\textrm{F}}}
\end{equation*}
where $\bm{Z}_{\ell}$ is $\ell$th realization of $\bm{\mathsf{Z}}$ and $\widehat{\bm{Z}}_{\ell}$ is its reconstruction during the $\ell$th Monte-Carlo trial. As per BiG-VAMP's initialization setting, all the initial means and covariances were set to the all-zero vector and the identity matrix, respectively. The results disclosed in the sequel demonstrate that BiG-VAMP yields  considerable improvements in reconstruction performance and robustness as compared to state-of-the-art algorithms, especially in presence of discrete priors on either $\bm{U}$ or $\bm{V}$.
%
\subsection{Noisy dictionary learning}
\begin{figure}[h!]
\centering
\includegraphics[scale=\figscale]{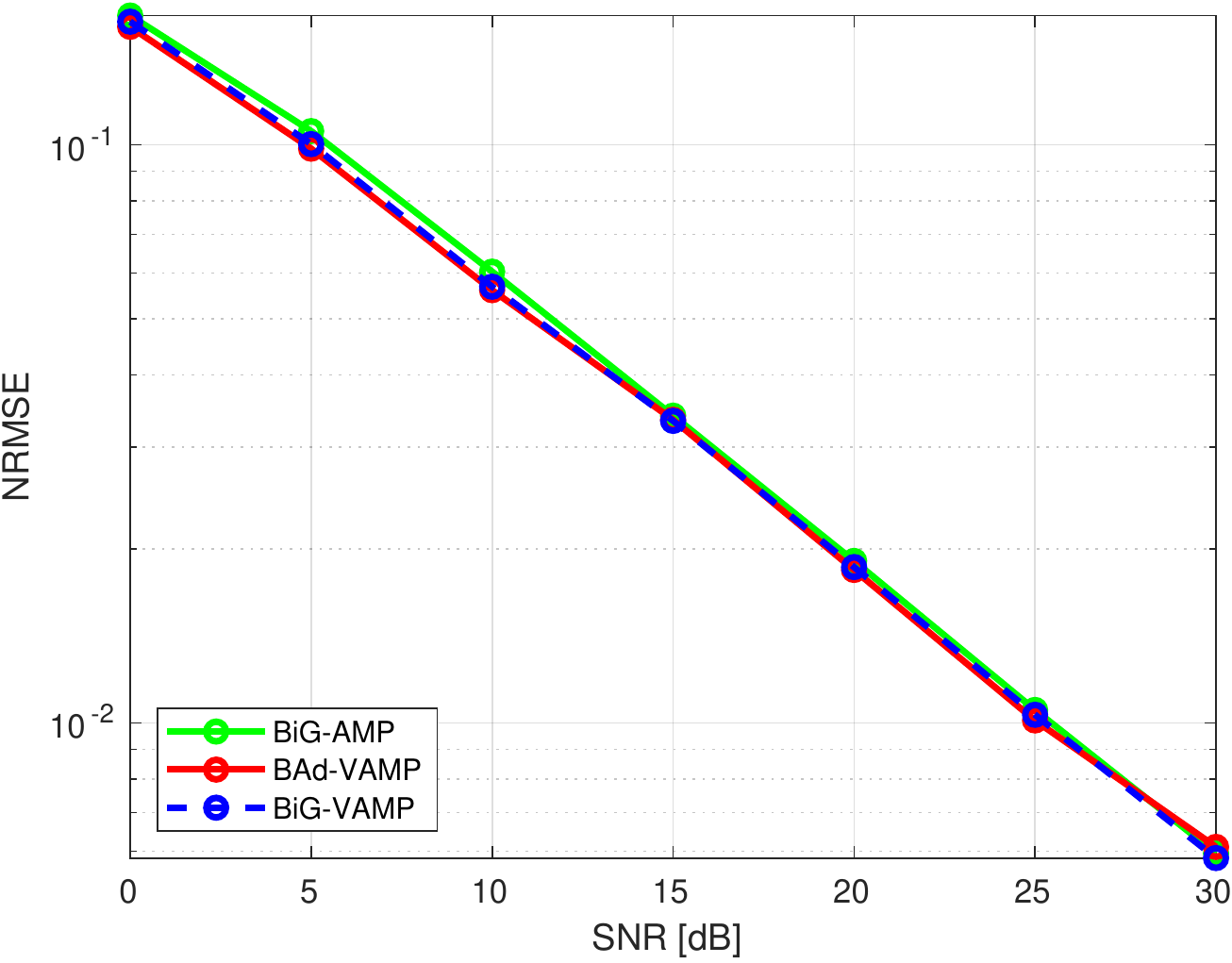}
\caption{NRMSE of BiG-VAMP and  BAd-VAMP vs. the SNR for the dictionary learning problem:  Gaussian prior on $\boldsymbol{U}$, Bernoulli-Gaussian prior on $\boldsymbol{V}$ with a sparsity level of $95\%$, $N=1000$, $M=1000$, and $r=20$.}
\label{fig:gauss-bg-benchmark}
\end{figure}

\begin{figure}[h!]
\centering
\includegraphics[scale=\figscale]{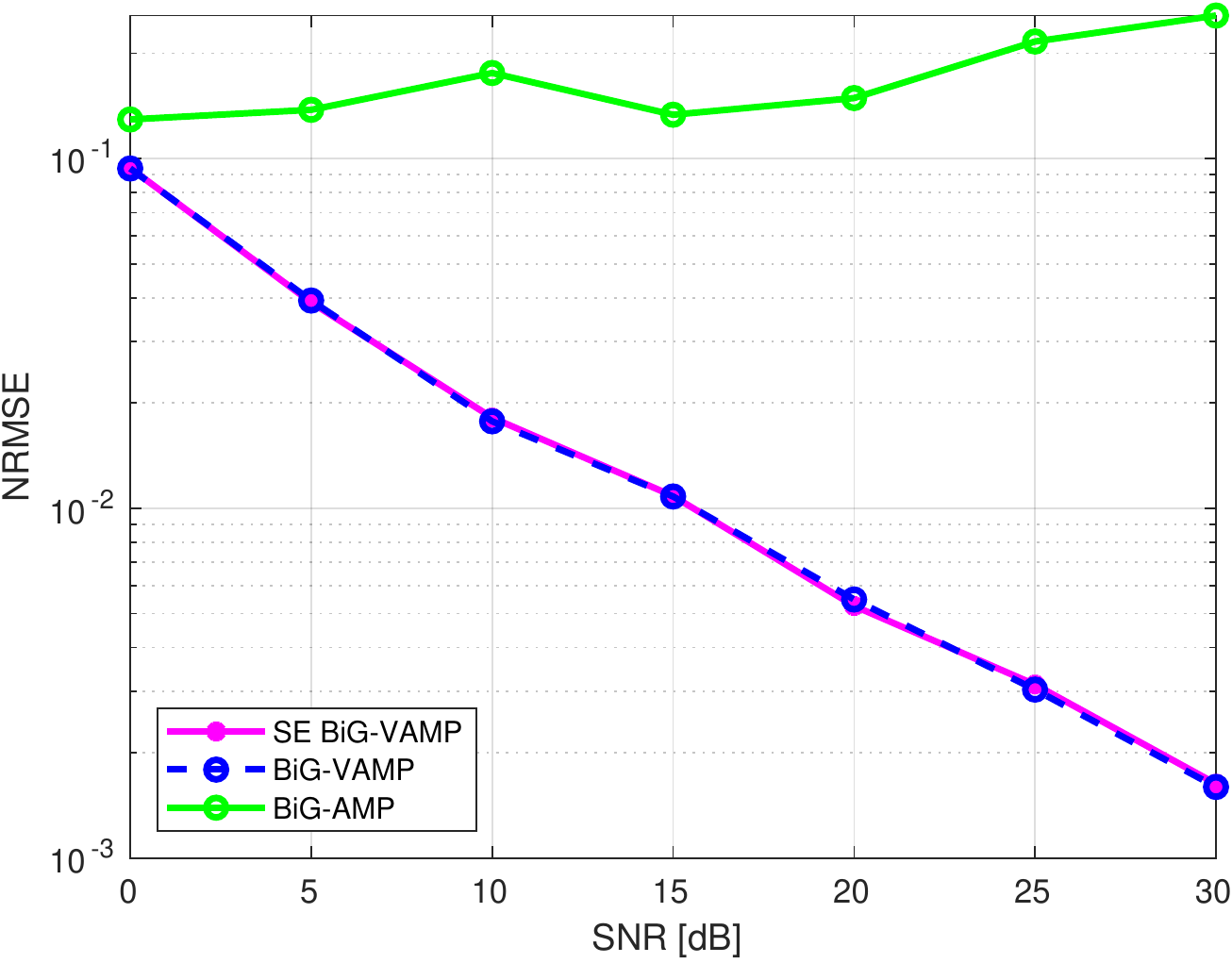}
\caption{NRMSE of BiG-VAMP and BiG-AMP vs. the SNR for the dictionary learning problem: binary prior on $\boldsymbol{U}$, Bernoulli-Gaussian prior on $\boldsymbol{V}$  with a sparsity level of $95\%$, $N=100$, $M=100$ and $r=5$.}
\label{fig:binary-bigvamp-bigamp-dictionary-learning-low-rank}
\end{figure}

\begin{figure}[h!]
\centering
    \includegraphics[scale=\figscale]{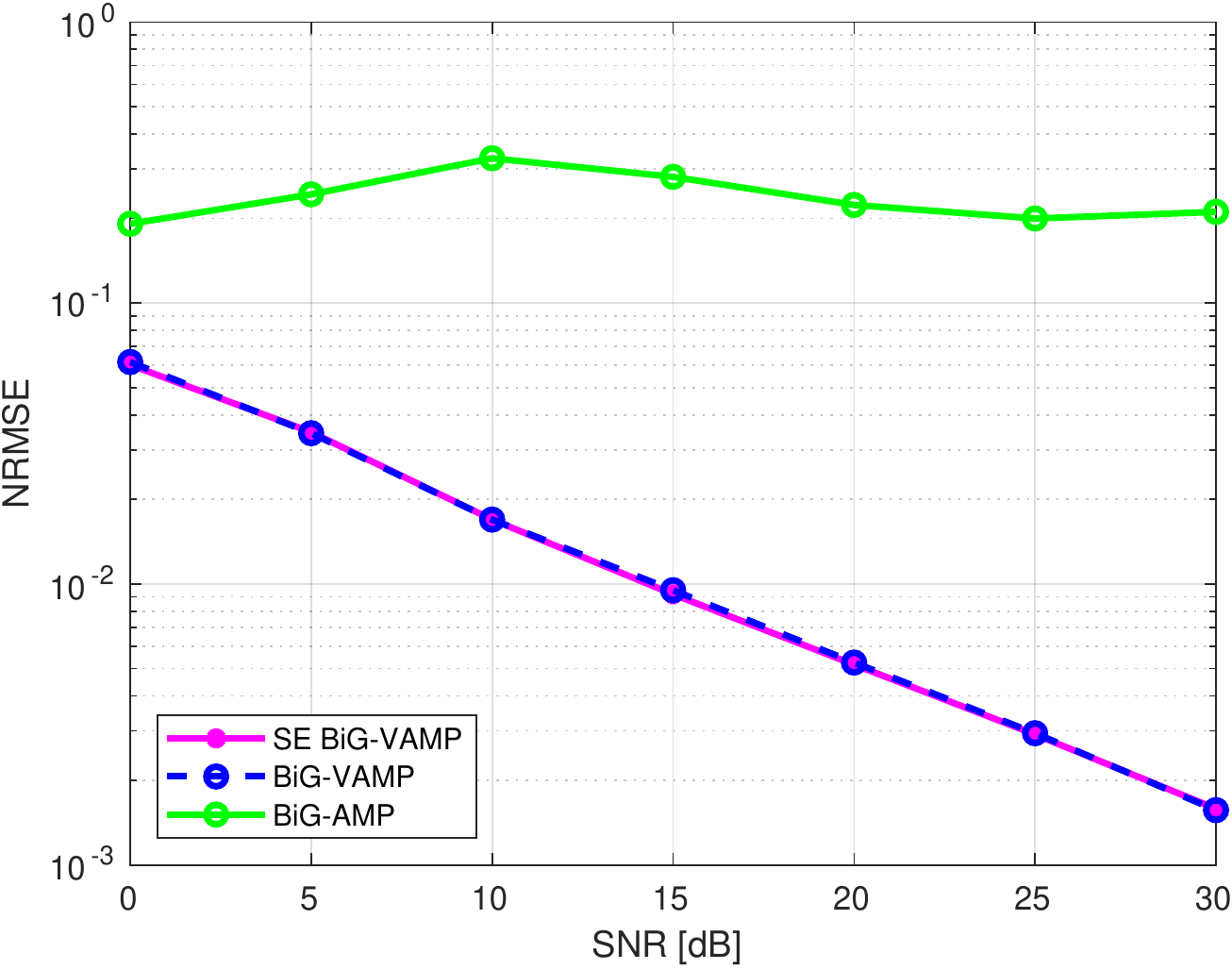}
\caption{NRMSE of BiG-VAMP and BiG-AMP vs. the SNR for the dictionary learning problem:  binary prior on  $\boldsymbol{U}$, Bernoulli-Gaussian prior on $\boldsymbol{V}$ with a sparsity level of $95\%$, $N=500$, $M=500$ and $r=25$.}
\label{fig:binary-bigvamp-bigamp-dictionary-learning-high-rank}
\end{figure}

Here, we  apply the proposed BiG-VAMP algorithm to the well-known dictionary learning problem wherein the goal is to find, from a noisy observation $\boldsymbol{Y}$, a dictionary matrix $\boldsymbol{U}$ and a sparse matrix $\boldsymbol{V}$. For that purpose, we use  Gaussian and Bernoulli-Gaussian priors on $\boldsymbol{U}$ and $\boldsymbol{V}$ matrices, respectively. We depict the NRMSE on the estimated $\boldsymbol{\widehat{Z}}$ in Fig. \ref{fig:gauss-bg-benchmark} wherin we bechchmark the proposed algorithm against BiG-AMP and BAd-VAMP.
Note that BAd-VAMP was simulated with its  defaults parameters from \cite{sarkar2019bilinear}, i.e., $\tau_{1,\textrm{max}} = 1$ and $\tau_{2,\textrm{max}} = 0$, $\zeta=0.8$ and $\gamma_{\textrm{min}}=10^{-6}$. In Fig. \ref{fig:gauss-bg-benchmark}, it is seen that BiG-AMP and BAd-VAMP\footnote{Note here that the Gaussian prior on $\bm{U}$ was incorporated during the maximization step of the EM algorithm inside BAd-VAMP. In this special case, BAd-VAMP is actually performing MAP  estimation of $\bm{U}$ (instead of maximum likelihood estimation).} exhibit the same performance as BiG-VAMP which is hardly surprising since all  algorithms are optimally exploiting the considered priors on $\bm{U}$ and $\bm{V}$ and do not suffer from any convergence issues. 
\\ 
Next, we consider a binary prior on the unknown dictionary $\boldsymbol{U}$. Note that in this case the underlying \textit{dictionary learning} problem is also known as  the $\mathbb{Z}/2$ synchronization problem in the mathematical literature \cite{Perry2016} or blind detection problem  in the communication literature \cite{Mezghani2018}. In this context, we again compare BiG-VAMP to BiG-AMP and we consider both low rank (i.e., $r=5$) and moderately high rank (i.e., $r=25$)  structures as illustrated in Figs. \ref{fig:binary-bigvamp-bigamp-dictionary-learning-low-rank} and \ref{fig:binary-bigvamp-bigamp-dictionary-learning-high-rank}, respectively.

\noindent We do not include BAd-VAMP in the subsequent simulations
since it is not designed to handlenon-Gaussian priors on  $\boldsymbol{U}$ 
This is in fact  due to the inherent limitation imposed by the combinatorial maximization step in the EM algorithm that BAd-VAMP  uses to update $\boldsymbol{U}$ at every iteration.\\
Figs. \ref{fig:binary-bigvamp-bigamp-dictionary-learning-low-rank} and \ref{fig:binary-bigvamp-bigamp-dictionary-learning-high-rank} show order of-magnitude difference in the NRMSE performance of BiG-VAMP and BiG-AMP. While BiG-VAMP outperforms BiG-AMP under both low-rank and high-rank structures, the gap between the two algorithms is not inherently related to the rank value, but rather to the inablity of BiG-AMP to converge under discrete priors. Moreover, it is seen from both figures that the empirical NRMSE of BiG-VAMP is accurately predicted by the analytical state evolution recursion thereby corroborating  the theoretical analysis we conducted in Section \ref{sec:state-evolution}. 
\subsection{Matrix  factorization}\label{sec:matrix-factorization}
\begin{figure}[h!]
\centering
\includegraphics[scale=\figscale]{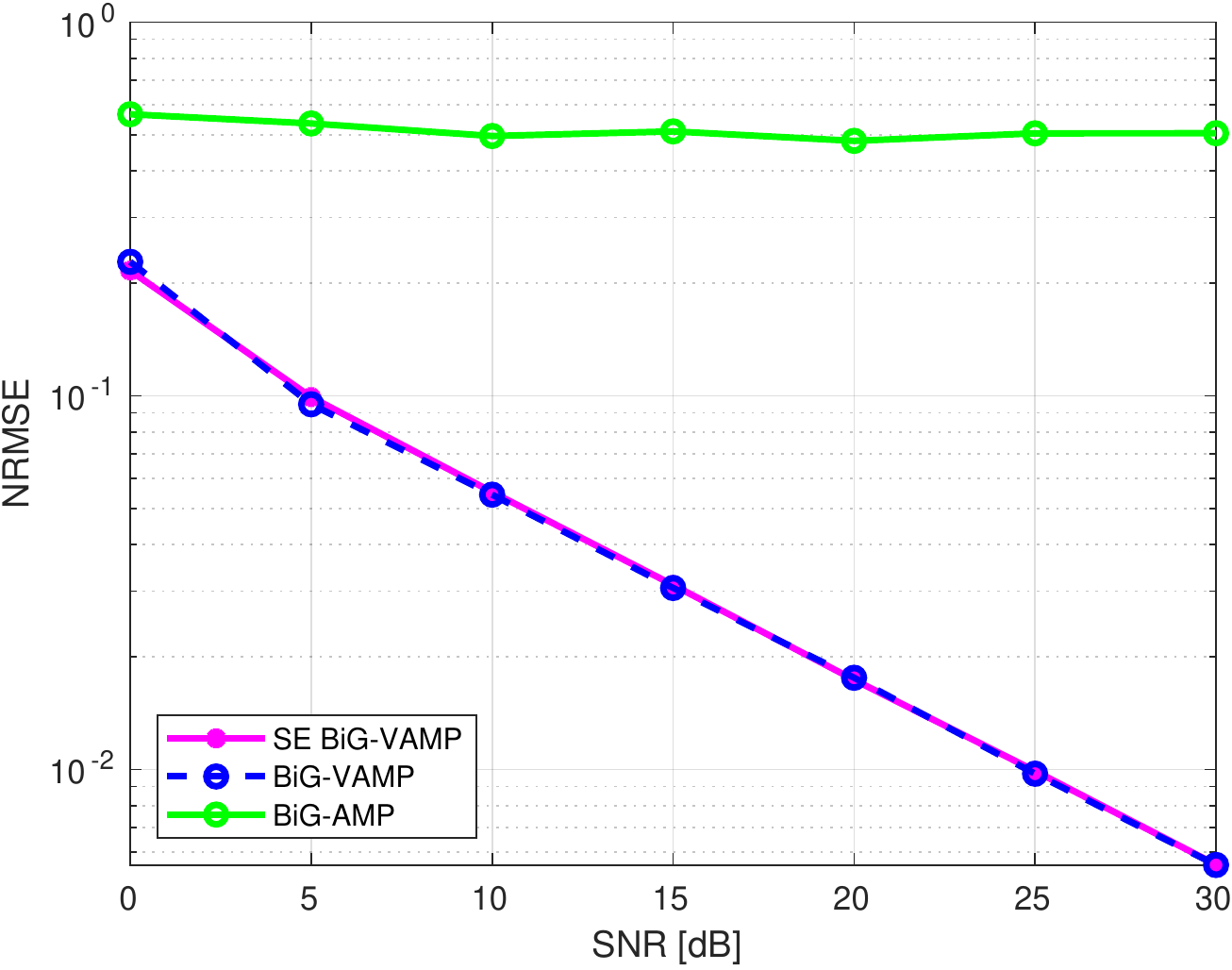}
\caption{NRMSE of BiG-VAMP and BiG-AMP vs. the SNR for the matrix factorization problem: binary prior on  $\boldsymbol{U}$, Gaussian prior on $\boldsymbol{V}$, $N=1000$, $M=200$ and $r=30$.}
\label{fig:binary-bigvamp-bigamp}
\end{figure}

In this case, we compare BiG-VAMP to BIG-AMP \cite{bigampCode} for the case where $\boldsymbol{U}$ is a binary matrix and $\bm{V}$ is Gaussian-distributed. Fig. \ref{fig:binary-bigvamp-bigamp} depicts the NRMSE of both algorithms and reveals that BiG-AMP's performance again deteriorates considerably in presence of a discrete prior either on $\bm{U}$ or $\bm{V}$ (due to the symmetry of the problem).  BiG-VAMP, however, finds an accurate factorisation of $\bm{Z}$ over the entire SNR range and its empirical NRMSE is again theoretically predicted by the established state evolution analysis.  This endows the proposed algorithm with offline design guidelines when applied to different engineering problems in practice.

\begin{figure}[h!]
\centering
\includegraphics[scale=\figscale]{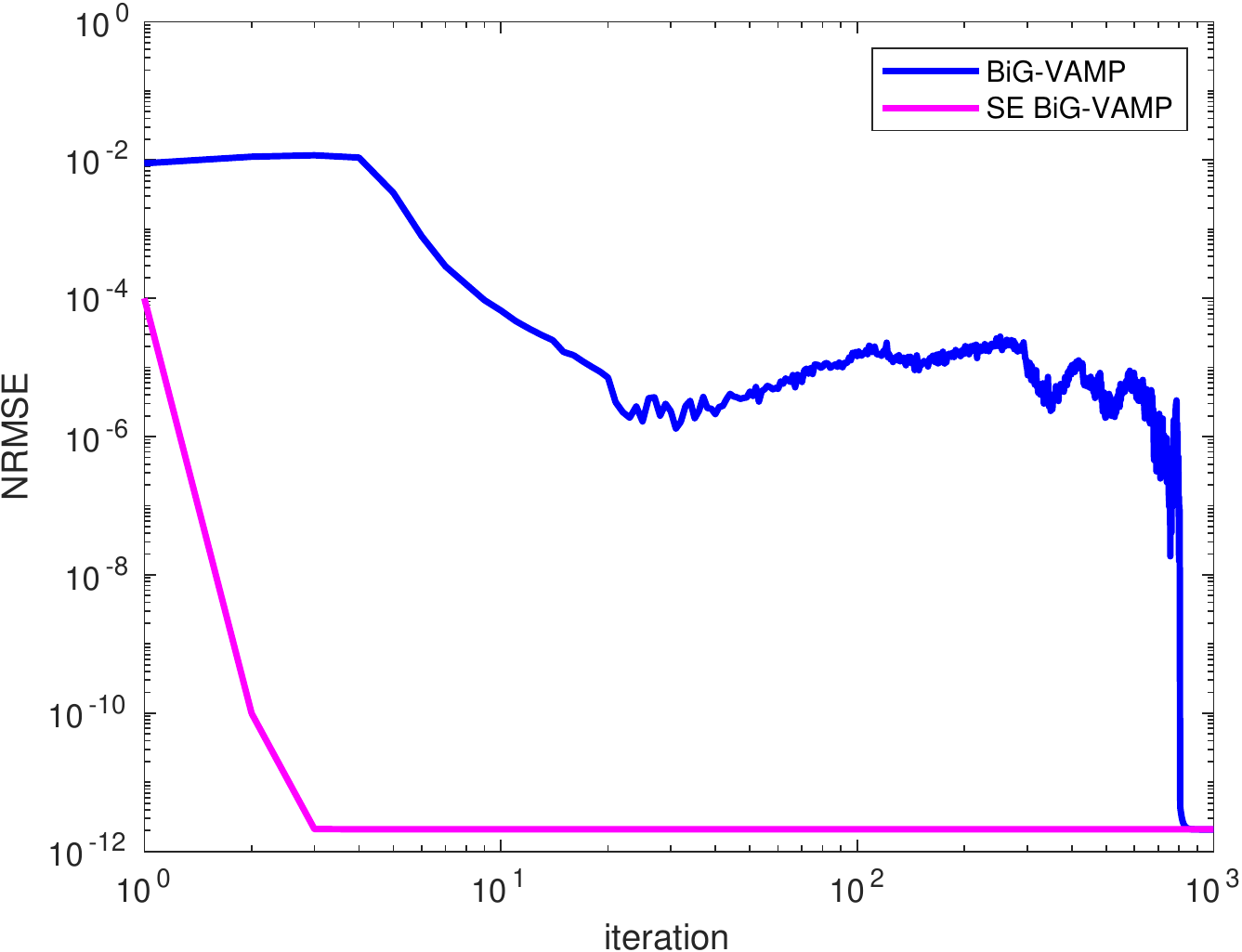}
\caption{NRMSE of BiG-VAMP and its SE for the matrix factorization problem at $\textrm{SNR}=10$ dB: binary prior on  $\boldsymbol{U}$, Gaussian prior on $\boldsymbol{V}$, $N=1000$, $M=500$ and $r=10$.}
\label{fig:binary-bigvamp-chaotic-transit-behavior}
\end{figure}

\noindent A close inspection of the NRMSE time evolution, as shown in Fig. \ref{fig:binary-bigvamp-chaotic-transit-behavior}, illustrates qualitatively a transiently chaotic trajectory that is similar to those of fluid parcels in turbulent flows studied in \cite{ercsey2011optimization}.
Such a chaotic behaviour may be a generic feature of algorithms searching for solutions in hard optimization problems with applications as diverse as protein folding and Sudoku \cite{elser2007searching}.

\subsection{Matrix completion}
\begin{figure}[h!]
\centering
\includegraphics[scale=\figscale]{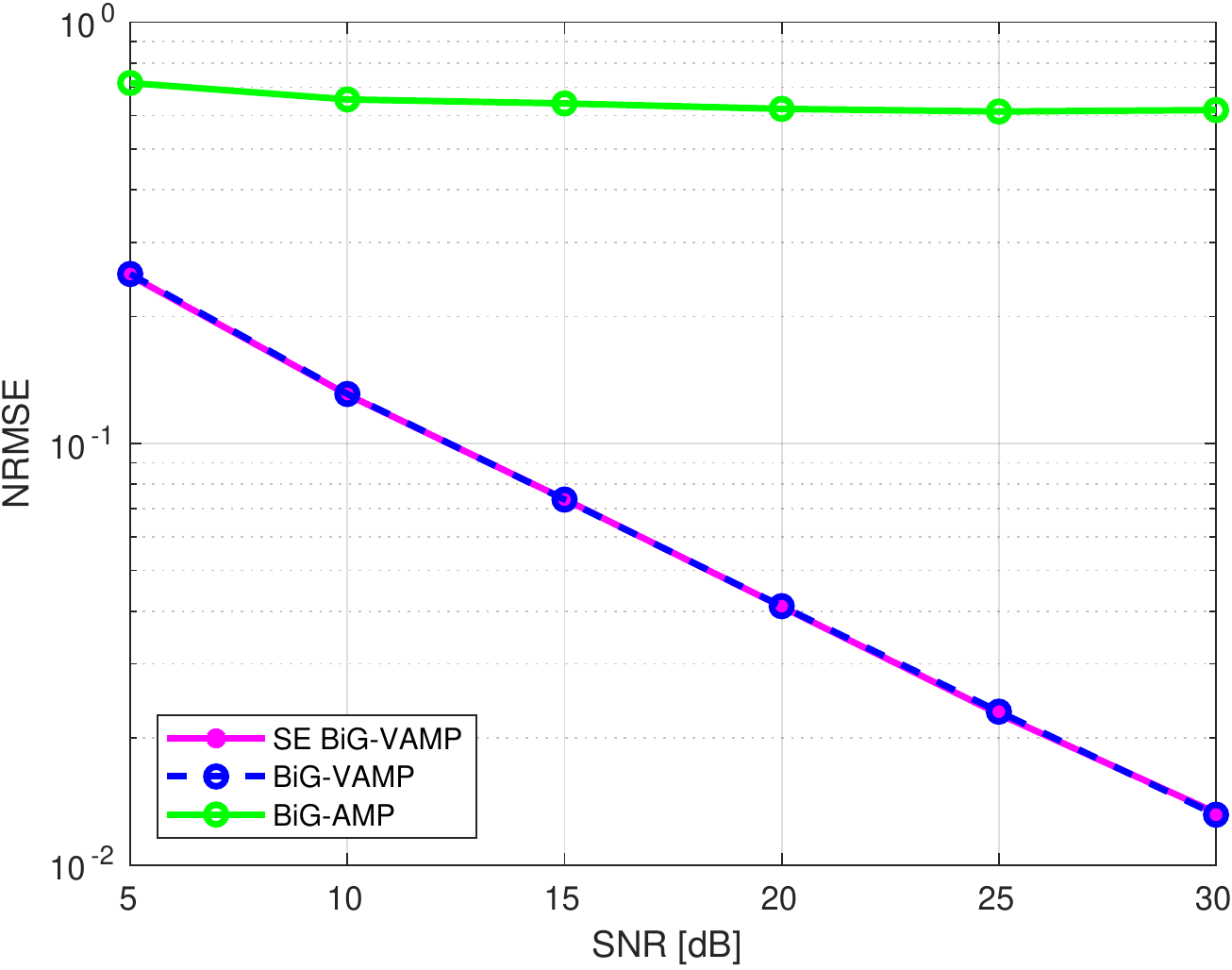}
\caption{NRMSE of BiG-VAMP and BiG-AMP vs. the SNR for the matrix completion problem with a selection rate of $20 \%$: binary prior on $\boldsymbol{U}$, Gaussian prior on $\boldsymbol{V}$,  $N=1000$, $M=500$ and $r=30$.}
\label{fig:binary-bigvamp-bigamp-matrix-completion}
\end{figure}

Unlike the previous two applications (i.e., matrix factorization and dictionary learning) 
wherein the observation model  was linear, we now turn our attention to a popular generalized bilnear recovery problem, namely the matrix completion problem. In this context,  Fig. \ref{fig:binary-bigvamp-bigamp-matrix-completion} compares BiG-VAMP to BIG-AMP under the nonlinear observation model in (\ref{eq:bilinear-model}) by taking  $\phi(\cdot)$  to be a random selection with a rate of $20\%$.
There, it is also seen that BiG-VAMP outperforms by far BiG-AMP which is in principle able to deal with any non-linearity in the observation model. But its deficiency here stems from the fact that it diverges for the considered challenging problem.
\par
\begin{figure}[h!]
\centering
\includegraphics[scale=\figscale]{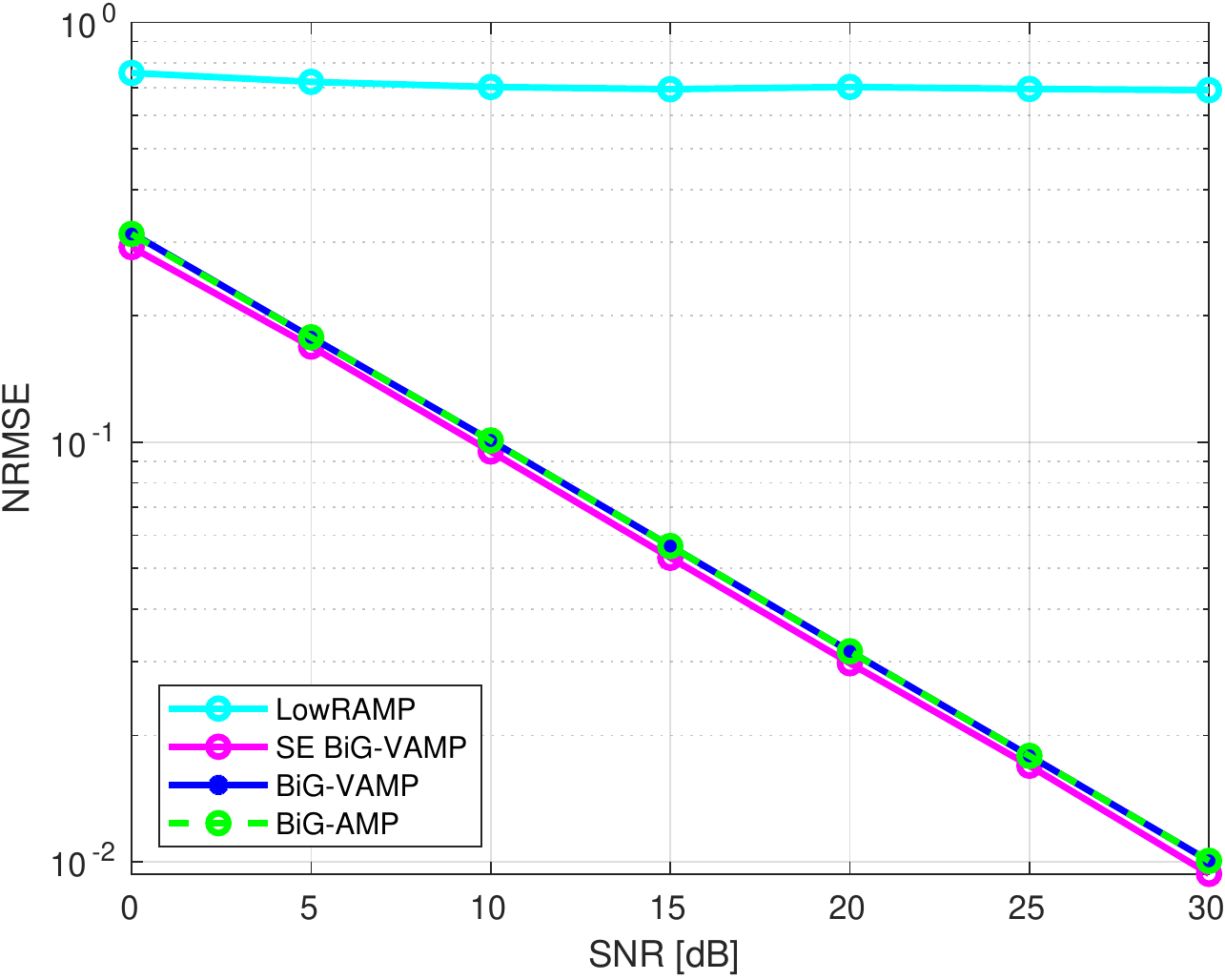}
\caption{NRMSE of BiG-VAMP, BiG-AMP, and LowRAMP vs. the SNR for the matrix completion with the selection rate of $10 \%$: Gaussian prior on $\bm{U}$, Gaussian prior on $\mathbf{V}$, $N=1000$, $M=500$ and $r=3$.}
\label{fig:lowrank-bigvamp-bigamp-lowramp}
\end{figure}

 For the  case where the only problem structure available at hand is the low rank approximation, we also benchmark  BiG-VAMP  against LowRAMP \cite{lesieur2015phase} whose MATLAB code is publicly available from \cite{lowrampCode}. The results are depicted in Fig. \ref{fig:lowrank-bigvamp-bigamp-lowramp} and as seen there  LowRAMP is not able to correctly recover $\boldsymbol{U}$ and $\boldsymbol{V}$  while BiG-VAMP exhibits the same performance as BiG-VAMP. Under the considered non linear selection function, $\phi(\cdot)$, the performance degradation of LowRAMP  is mainly due to the fact the second-order Taylor series approximation of the output channel (\ref{eq:non-linear-distribution}), around $z_{ij}= 0$, is not accurate high SNR values (which is a crucial condition to solve the matrix completion problem).\\

\begin{figure}[h!]
\centering
\includegraphics[scale=\figscale]{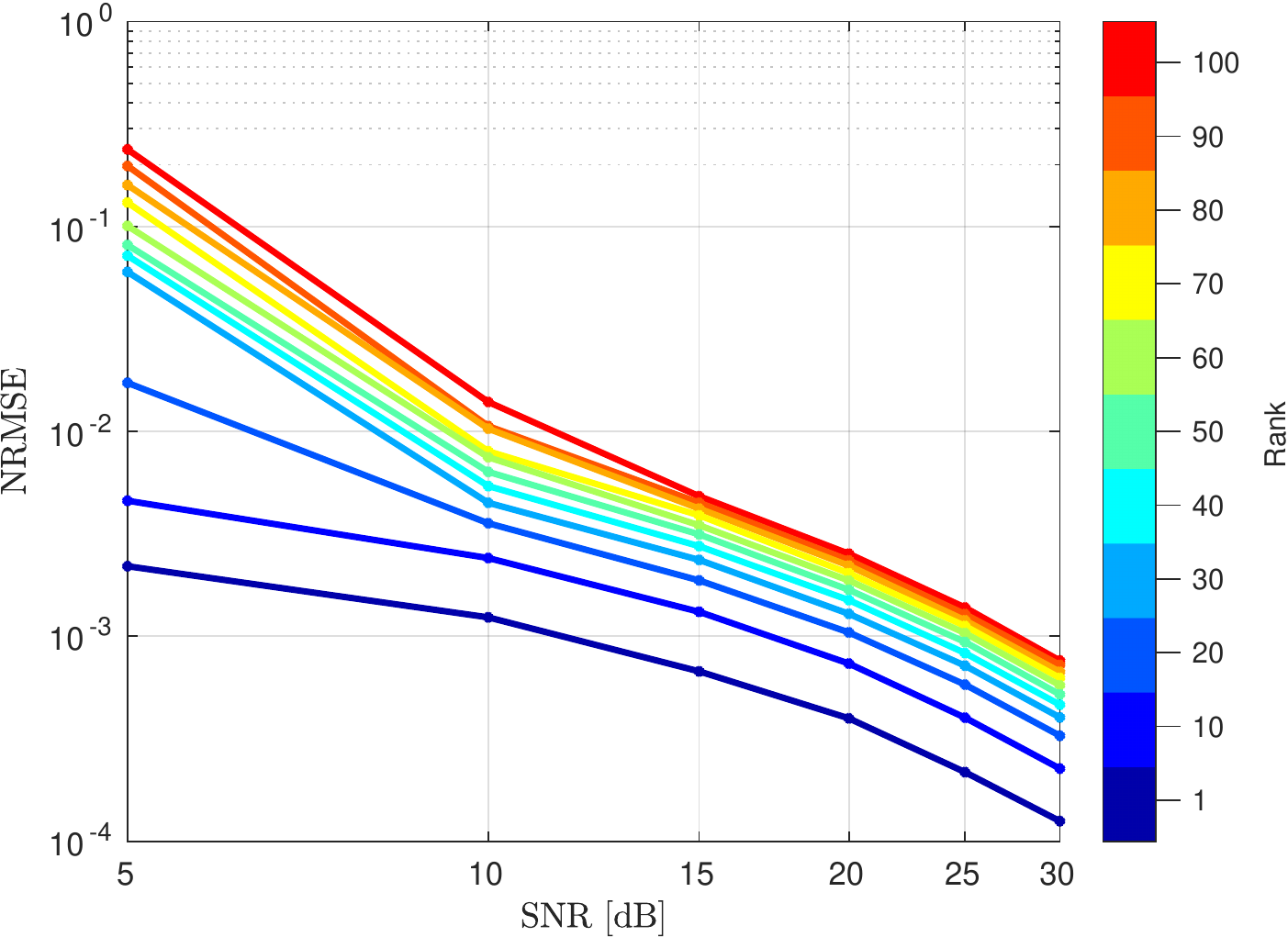}
\caption{Phase transition diagram BiG-VAMP for the matrix completion problem with $N=1000$, $M=500$, $1\leq r\leq 100$, a selection rate of $20\%$, binary matrix $\mathbf{U}$ and Gaussian matrix $\boldsymbol{V}$.}
\label{fig:phase-transition-matrix-completion}
\end{figure}

Fig. \ref{fig:phase-transition-matrix-completion} shows the phase transition diagrams in the NRMSE for both low- and high-SNR regimes at a fixed selection rate of $20\%$. In the large-rank limit, we observe a low detectability regime when the SNR is in $[0,10]$ dB. This result is consistent with the non negligible uncertainty of estimators \cite{chen2019inference} to conduct inference in the low-SNR regime for noisy matrix completion.\\ 

\section{Conclusion}
In this work, we introduced a new algorithm, dubbed BiG-VAMP, to solve the generalized bilinear  recovery problem  based on the approximate message passing paradigm while treating  both the MMSE and MAP inference problems in a unified framework. We described how BiG-VAMP provides a broader solution to the bilinear recovery problem under different structured matrices beyond the ``low rank'' structure heavily investigated in the existing literature. In particular, our numerical results for applications in matrix-factorization, dictionary learning, and matrix completion demonstrated that BiG-VAMP exhibits the best reconstruction performance under non Gaussian priors as compared to existing state-of-the-art algorithms such as BiG-AMP, BAd-VAMP, and LowRAMP. Additionally, we derived the state evolution equations of BiG-VAMP and characterized its phrase transition for the matrix completion problem. 

\section*{Acknowledgment}
This work was supported by the Discovery Grants (DG) program  from the Natural Science and Engineering Council of Canada (NSERC).


\balance
\bibliographystyle{IEEEtran}
\bibliography{IEEEabrv,references.bib}

\begin{thebibliography}{10}
\providecommand{\url}[1]{#1}
\csname url@samestyle\endcsname
\providecommand{\newblock}{\relax}
\providecommand{\bibinfo}[2]{#2}
\providecommand{\BIBentrySTDinterwordspacing}{\spaceskip=0pt\relax}
\providecommand{\BIBentryALTinterwordstretchfactor}{4}
\providecommand{\BIBentryALTinterwordspacing}{\spaceskip=\fontdimen2\font plus
\BIBentryALTinterwordstretchfactor\fontdimen3\font minus
  \fontdimen4\font\relax}
\providecommand{\BIBforeignlanguage}[2]{{%
\expandafter\ifx\csname l@#1\endcsname\relax
\typeout{** WARNING: IEEEtran.bst: No hyphenation pattern has been}%
\typeout{** loaded for the language `#1'. Using the pattern for}%
\typeout{** the default language instead.}%
\else
\language=\csname l@#1\endcsname
\fi
#2}}
\providecommand{\BIBdecl}{\relax}
\BIBdecl

\bibitem{gribonval2015sparse}
R.~Gribonval, R.~Jenatton, and F.~Bach, ``Sparse and spurious: dictionary
  learning with noise and outliers,'' \emph{IEEE Transactions on Information
  Theory}, vol.~61, no.~11, pp. 6298--6319, 2015.

\bibitem{montanari2012graphical}
A.~Montanari, Y.~Eldar, and G.~Kutyniok, ``Graphical models concepts in
  compressed sensing,'' \emph{Compressed Sensing: Theory and Applications}, pp.
  394--438, 2012.

\bibitem{zou2006sparse}
H.~Zou, T.~Hastie, and R.~Tibshirani, ``Sparse principal component analysis,''
  \emph{Journal of computational and graphical statistics}, vol.~15, no.~2, pp.
  265--286, 2006.

\bibitem{koren2009matrix}
Y.~Koren, R.~Bell, and C.~Volinsky, ``Matrix factorization techniques for
  recommender systems,'' \emph{Computer}, no.~8, pp. 30--37, 2009.

\bibitem{matsushita2013low}
R.~Matsushita and T.~Tanaka, ``Low-rank matrix reconstruction and clustering
  via approximate message passing,'' in \emph{Advances in Neural Information
  Processing Systems}, 2013, pp. 917--925.

\bibitem{wang2019symmetric}
L.~Wang, Z.~Zhang, and D.~Dunson, ``Symmetric bilinear regression for signal
  subgraph estimation,'' \emph{IEEE Transactions on Signal Processing},
  vol.~67, no.~7, pp. 1929--1940, 2019.

\bibitem{lin2010augmented}
Z.~Lin, M.~Chen, and Y.~Ma, ``The augmented lagrange multiplier method for
  exact recovery of corrupted low-rank matrices,'' \emph{arXiv preprint
  arXiv:1009.5055}, 2010.

\bibitem{boyd2011distributed}
S.~Boyd, N.~Parikh, E.~Chu, B.~Peleato, J.~Eckstein \emph{et~al.},
  ``Distributed optimization and statistical learning via the alternating
  direction method of multipliers,'' \emph{Foundations and
  Trends{\textregistered} in Machine learning}, vol.~3, no.~1, pp. 1--122,
  2011.

\bibitem{babacan2012sparse}
S.~D. {Babacan}, M.~{Luessi}, R.~{Molina}, and A.~K. {Katsaggelos}, ``Sparse
  bayesian methods for low-rank matrix estimation,'' \emph{IEEE Transactions on
  Signal Processing}, vol.~60, no.~8, pp. 3964--3977, 2012.

\bibitem{donoho2009message}
D.~L. Donoho, A.~Maleki, and A.~Montanari, ``Message-passing algorithms for
  compressed sensing,'' \emph{Proceedings of the National Academy of Sciences},
  vol. 106, no.~45, pp. 18\,914--18\,919, 2009.

\bibitem{eldar2012compressed}
Y.~C. Eldar and G.~Kutyniok, \emph{Compressed sensing: theory and
  applications}.\hskip 1em plus 0.5em minus 0.4em\relax Cambridge university
  press, 2012.

\bibitem{rangan2011generalized}
S.~{Rangan}, ``Generalized approximate message passing for estimation with
  random linear mixing,'' in \emph{2011 IEEE International Symposium on
  Information Theory Proceedings}, 2011, pp. 2168--2172.

\bibitem{bayati2011dynamics}
M.~Bayati and A.~Montanari, ``The dynamics of message passing on dense graphs,
  with applications to compressed sensing,'' \emph{IEEE Transactions on
  Information Theory}, vol.~57, no.~2, pp. 764--785, 2011.

\bibitem{barbier2019optimal}
J.~Barbier, F.~Krzakala, N.~Macris, L.~Miolane, and L.~Zdeborov{\'a}, ``Optimal
  errors and phase transitions in high-dimensional generalized linear models,''
  \emph{Proceedings of the National Academy of Sciences}, vol. 116, no.~12, pp.
  5451--5460, 2019.

\bibitem{mezard2009information}
M.~Mezard and A.~Montanari, \emph{Information, physics, and computation}.\hskip
  1em plus 0.5em minus 0.4em\relax Oxford University Press, 2009.

\bibitem{rangan2019vector}
S.~{Rangan}, P.~{Schniter}, and A.~K. {Fletcher}, ``Vector approximate message
  passing,'' in \emph{2017 IEEE International Symposium on Information Theory
  (ISIT)}, 2017, pp. 1588--1592.

\bibitem{schniter2016vector}
P.~Schniter, S.~Rangan, and A.~K. Fletcher, ``Vector approximate message
  passing for the generalized linear model,'' in \emph{2016 50th Asilomar
  Conference on Signals, Systems and Computers}.\hskip 1em plus 0.5em minus
  0.4em\relax IEEE, 2016, pp. 1525--1529.

\bibitem{rangan2019vector_IT}
S.~Rangan, P.~Schniter, and A.~K. Fletcher, ``Vector approximate message
  passing,'' \emph{IEEE Transactions on Information Theory}, vol.~65, no.~10,
  pp. 6664--6684, 2019.

\bibitem{ma2017orthogonal}
J.~{Ma} and L.~{Ping}, ``Orthogonal amp,'' \emph{IEEE Access}, vol.~5, pp.
  2020--2033, 2017.

\bibitem{lesieur2015phase}
T.~Lesieur, F.~Krzakala, and L.~Zdeborov{\'a}, ``Phase transitions in sparse
  pca,'' in \emph{2015 IEEE International Symposium on Information Theory
  (ISIT)}.\hskip 1em plus 0.5em minus 0.4em\relax IEEE, 2015, pp. 1635--1639.

\bibitem{parker2014bilinear}
J.~T. Parker, P.~Schniter, and V.~Cevher, ``Bilinear generalized approximate
  message passing—part i: Derivation,'' \emph{IEEE Transactions on Signal
  Processing}, vol.~62, no.~22, pp. 5839--5853, 2014.

\bibitem{parker2016parametric}
J.~T. Parker and P.~Schniter, ``Parametric bilinear generalized approximate
  message passing,'' \emph{IEEE Journal of Selected Topics in Signal
  Processing}, vol.~10, no.~4, pp. 795--808, 2016.

\bibitem{sarkar2019bilinear}
S.~Sarkar, A.~K. Fletcher, S.~Rangan, and P.~Schniter, ``Bilinear recovery
  using adaptive vector-amp,'' \emph{IEEE Transactions on Signal Processing},
  vol.~67, no.~13, pp. 3383--3396, 2019.

\bibitem{dempster1977maximum}
A.~P. Dempster, N.~M. Laird, and D.~B. Rubin, ``Maximum likelihood from
  incomplete data via the em algorithm,'' \emph{Journal of the Royal
  Statistical Society: Series B (Methodological)}, vol.~39, no.~1, pp. 1--22,
  1977.

\bibitem{tulino2004random}
A.~M. Tulino, S.~Verd{\'u} \emph{et~al.}, ``Random matrix theory and wireless
  communications,'' \emph{Foundations and Trends{\textregistered} in
  Communications and Information Theory}, vol.~1, no.~1, pp. 1--182, 2004.

\bibitem{bigampCode}
\BIBentryALTinterwordspacing
P.~Schniter, ``{BiGAMP}.'' [Online]. Available:
  \url{https://sourceforge.net/projects/gampmatlab/}
\BIBentrySTDinterwordspacing

\bibitem{badvampCode}
\BIBentryALTinterwordspacing
S.~Sarkar, ``{Bilinear Adaptive Vector Approximate Message Passing},'' Jul.
  2019. [Online]. Available: \url{https://github.com/sbrsarkar/BAdVAMP}
\BIBentrySTDinterwordspacing

\bibitem{Perry2016}
A.~Perry, A.~Wein, A.~Bandeira, and A.~Moitra, ``{Message-Passing Algorithms
  for Synchronization Problems over Compact Group},'' \emph{Communications on
  Pure and Applied Mathematics}, 10 2016.

\bibitem{Mezghani2018}
A.~{Mezghani} and A.~L. {Swindlehurst}, ``{Blind Estimation of Sparse Broadband
  Massive MIMO Channels With Ideal and One-bit ADCs},'' \emph{IEEE Transactions
  on Signal Processing}, vol.~66, no.~11, pp. 2972--2983, 2018.

\bibitem{lowrampCode}
\BIBentryALTinterwordspacing
F.~Krzakala, ``{Low rank Matrix Factorization with AMP},'' Aug. 2015. [Online].
  Available: \url{https://github.com/krzakala/LowRAMP}
\BIBentrySTDinterwordspacing

\bibitem{chen2019inference}
Y.~Chen, J.~Fan, C.~Ma, and Y.~Yan, ``Inference and uncertainty quantification
  for noisy matrix completion,'' \emph{Proceedings of the National Academy of
  Sciences}, vol. 116, no.~46, pp. 22\,931--22\,937, 2019.

\end{thebibliography}
 
\end{document}